\documentclass[10pt]{article}
\usepackage{amsmath,amsthm,amssymb}
\usepackage[dvips]{graphics}
\usepackage{xcolor}
\usepackage{subfigure}
\usepackage{multirow}
\usepackage{graphicx} 
\usepackage{bm}     
\usepackage{setspace} 
\usepackage{booktabs} 
\usepackage{verbatim}
\usepackage{appendix} 
\usepackage{fullpage} 
\usepackage{natbib} 
\usepackage[T1]{fontenc}
\usepackage{changepage}
\usepackage{setspace}
\usepackage[normalem]{ulem} 

\newcommand{\R}{\mathbb{R}}
\newcommand{\N}{\mathbb{N}}

\newcommand{\Pb}{\mathbb{P}}
\newcommand{\I}{\mathbb{I}}
\newcommand{\Sy}{\mathbb{S}}
\newcommand{\F}{\mathcal{F}}

\newcommand{\B}{\mathcal{B}}

\newcommand{\card}{\mathrm{card}}
\newcommand{\diff}{\mathrm{d}}

\newcommand{\Rep}{\mathrm{Rep}}
\newcommand{\NRep}{\mathrm{NRep}}

\numberwithin{equation}{section}

\theoremstyle{plain}
\newtheorem{Theo}{Theorem}[section]
\newtheorem{Prop}[Theo]{Proposition}

\newtheorem{Lem}[Theo]{Lemma}

\theoremstyle{definition}
\newtheorem{Def}{Definition}[section]

\theoremstyle{remark}

\bibpunct{(}{)}{;}{a}{}{,}
\begin{document}

\singlespace

\title{Parsimonious Hierarchical Modeling Using Repulsive Distributions}
\author{Jos\'{e} Quinlan \\ Departamento de Estad\'{i}stica \\ Pontificia Universidad Cat\'{o}lica de Chile \\  jjquinla@mat.puc.cl
        \and Fernando A. Quintana \\ Departamento de Estad\'{i}stica \\ Pontificia Universidad Cat\'{o}lica de Chile \\ quintana@mat.uc.cl
        \and Garritt L. Page \\ Department of Statistics \\ Brigham Young University \\ page@stat.byu.edu}
\maketitle


\begin{abstract}
Employing nonparametric methods for density estimation has become routine in Bayesian statistical practice. Models based on discrete nonparametric priors such as Dirichlet 
Process Mixture (DPM) models are very attractive choices due to their flexibility and tractability. However, a common problem in fitting DPMs or other discrete models to 
data is that they tend to produce a large number of (sometimes) redundant clusters. In this work we propose a method that produces parsimonious mixture models (i.e. mixtures 
that discourage the creation of redundant clusters), without sacrificing flexibility or model fit. This method is based on the idea of repulsion, that is, that any two 
mixture components are encouraged to be well separated. We propose a family of $d$-dimensional probability densities whose coordinates tend to repel each other in a smooth 
way. The induced probability measure has a close relation with Gibbs measures, graph theory and point processes. We investigate its global properties and explore its use in 
the context of mixture models for density estimation. Computational techniques are detailed and we illustrate its usefulness with some well-known data sets and a small 
simulation study.
\end{abstract}


\doublespace


\section{Introduction}\label{Introduction}

Hierarchical mixture models have been very successfully employed in a myriad of applications of Bayesian modeling. A typical formulation for such models adopts the basic 
form
\begin{equation}
  \boldsymbol{y}_{i}\mid\boldsymbol{\theta}_{i}\stackrel{ind.}{\sim}k(\boldsymbol{y}_{i};\boldsymbol{\theta}_{i}),\qquad
  \boldsymbol{\theta}_{1},\ldots,\boldsymbol{\theta}_{n}\stackrel{i.i.d.}{\sim}\sum_{k=1}^{N}\pi_{k}\delta_{\boldsymbol{\phi}_{k}},\qquad
  \boldsymbol{\phi}_{1},\ldots,\boldsymbol{\phi}_{N}\stackrel{i.i.d.}{\sim}G_{0},
  \label{General.Mixture.Model}
\end{equation}
where $k(\,\cdot\,;\boldsymbol{\theta})$ is a suitable kernel density indexed by $\boldsymbol{\theta}$, $1\leq N\leq\infty$, component weights $\pi_{1},\dots,\pi_{N}$ are 
nonnegative and $\sum_{k=1}^{N}\pi_{k}=1$ with probability 1, and $G_{0}$ is a suitable probability distribution. Here $N$ could be regarded as fixed or random and in the 
latter case a prior $p(N)$ would need to be specified. Depending on the modeling goals and data particularities, the model could have additional parameters and levels in the 
hierarchy. The generic model \eqref{General.Mixture.Model} includes, as special cases, finite mixture models \citep{fruhwirth:mixt} and species sampling mixture models 
\citep{pitman:96,quintana:06}, in turn including several well-known particular examples such as the Dirichlet Process (DP) \citep{ferguson:73} and the Pitman-Yor Process 
\citep{PY97}.

A common feature of models like \eqref{General.Mixture.Model} is the use of i.i.d. atoms $\boldsymbol{\phi}_{1},\ldots,\boldsymbol{\phi}_{N}$. This choice seems to have been 
largely motivated by the resulting tractability of the models, specially in the nonparametric case ($N=\infty$). There is also a substantial body of literature concerning 
important properties such as wide support, posterior consistency, and posterior convergence rates, among others. See, for instance, \cite{GhosalVanDerVaart:07} and
\cite{ShenTokdarGhosal:13}.

\begin{figure}[htb]
  \begin{center}
  \begin{tabular}{ccc}
    \includegraphics[width=8cm,height=8cm]{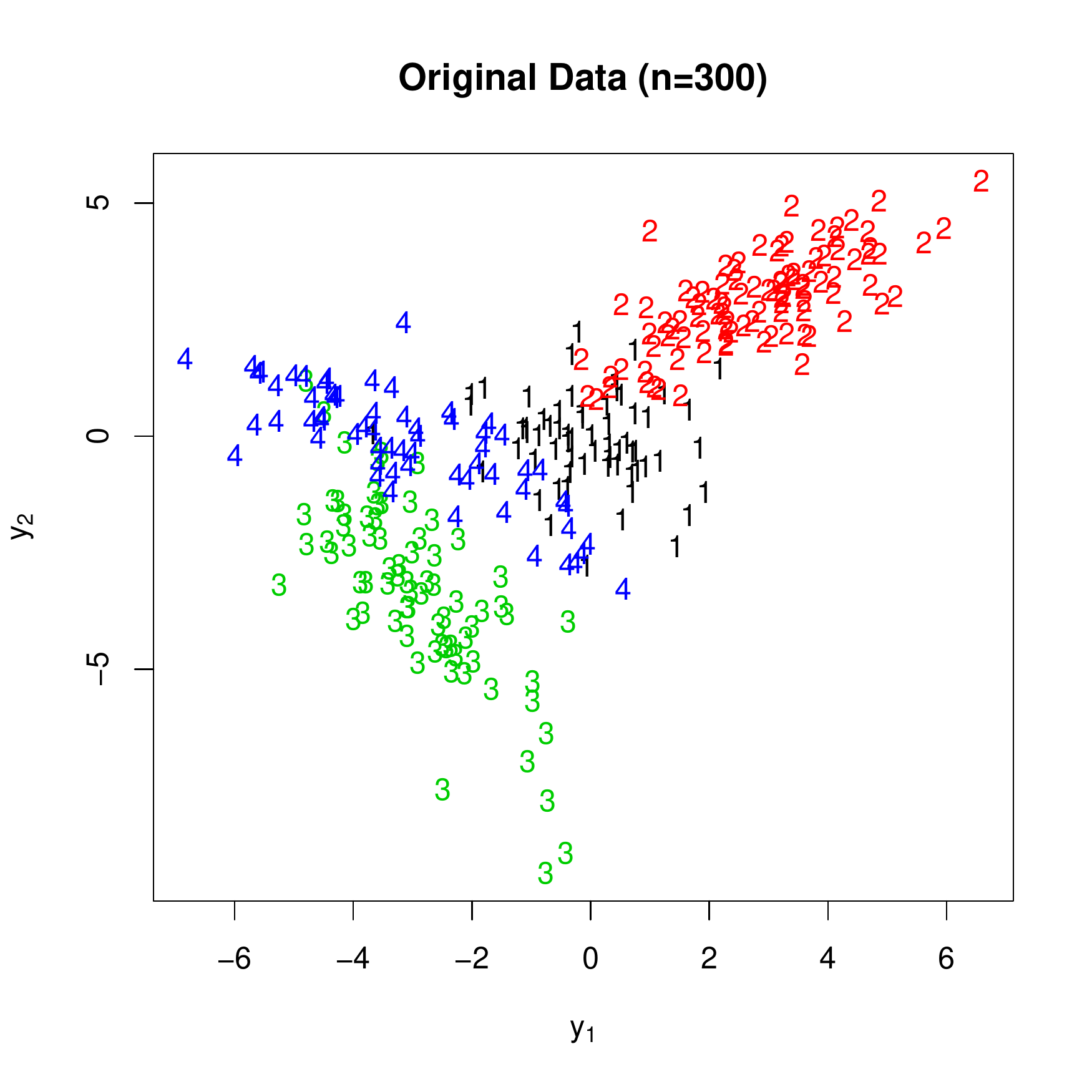} & \hspace{-1cm} & \includegraphics[width=8cm,height=8cm]{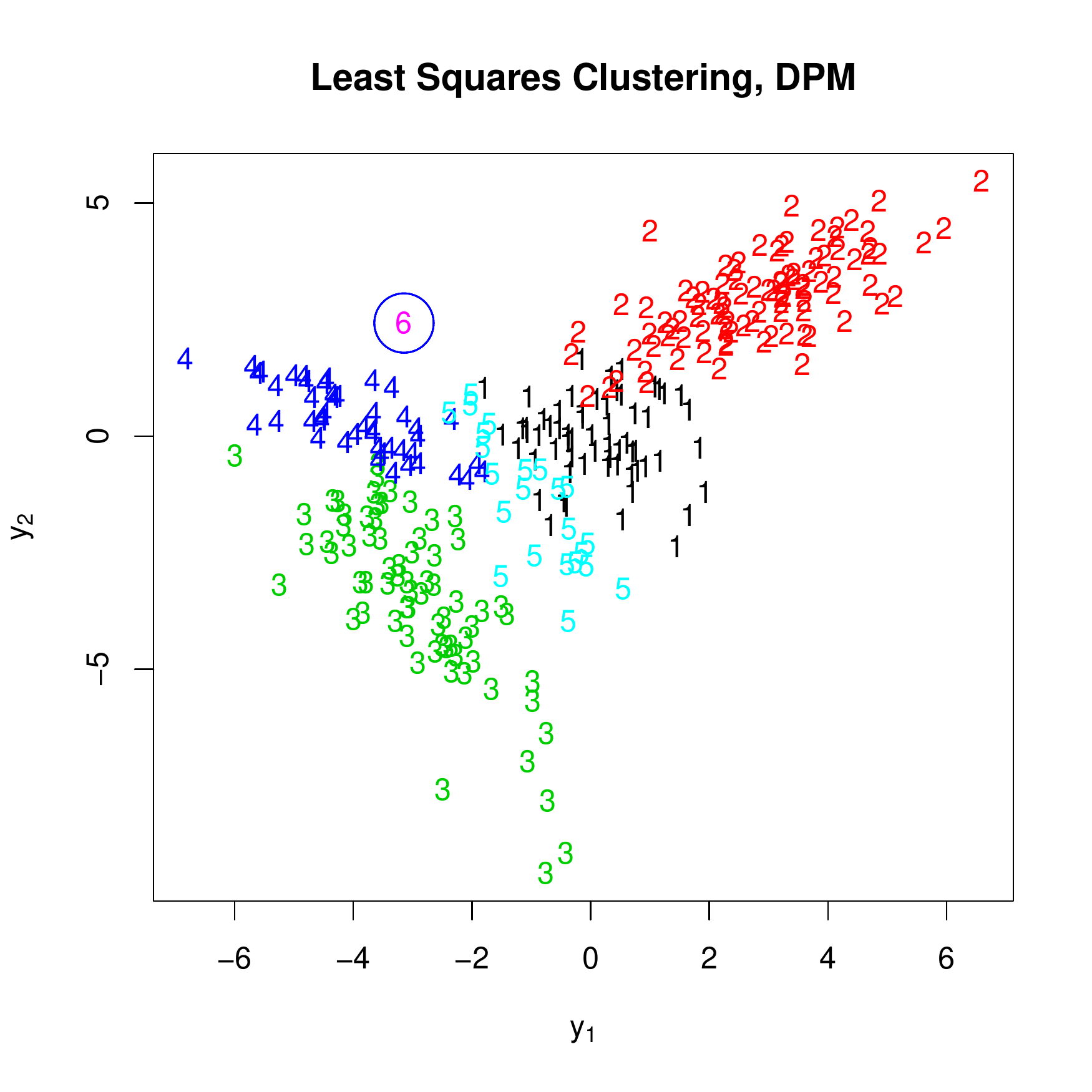}
  \end{tabular}
  \end{center}
  \caption{Data simulated from the mixture of 4 bivariate normal densities in \eqref{Bivariate.Normal.Simulation}. The left panel shows the original $n=300$ data points with 
  colors and numbers indicating the original cluster. The right panel shows the clustering resulting from applying Dahl's least squares clustering algorithm to a DPM.}
  \label{Example.Introduction}
\end{figure}

While the use of i.i.d. atoms in \eqref{General.Mixture.Model} is technically (and practically) convenient, a typical summary of the induced posterior clustering will 
usually contain a number of very small clusters or even some singletons. As a specific example, we considered a synthetic data set of $n=300$ independent observations 
simulated from the following mixture of 4 bivariate normal distributions:
\begin{equation}
  \boldsymbol{y}\sim
  0.2\mathrm{N}_{2}(\boldsymbol{\mu}_{1},\boldsymbol{\Sigma}_{1})
  +0.3\mathrm{N}_{2}(\boldsymbol{\mu}_{2},\boldsymbol{\Sigma}_{2})
  +0.3\mathrm{N}_{2}(\boldsymbol{\mu}_{3},\boldsymbol{\Sigma}_{3})
  +0.2\mathrm{N}_{2}(\boldsymbol{\mu}_{4},\boldsymbol{\Sigma}_{4}),
  \label{Bivariate.Normal.Simulation}
\end{equation}
with
\begin{align*}
  &\boldsymbol{\mu}_{1}=(0,0)^{\top},\quad
  \boldsymbol{\mu}_{2}=(3,3)^{\top},\quad
  \boldsymbol{\mu}_{3}=(-3,-3)^{\top},\quad
  \boldsymbol{\mu}_{4}=(-3,0)^{\top} \\
  &\boldsymbol{\Sigma}_{1}=
  \Bigg(
  \begin{array}{ccc}
  1 & & 0 \\
  0 & & 1
  \end{array}
  \Bigg),\quad
  \boldsymbol{\Sigma}_{2}=
  \Bigg(
  \begin{array}{ccc}
  2 & & 1 \\
  1 & & 1
  \end{array}
  \Bigg),\quad
  \boldsymbol{\Sigma}_{3}=
  \Bigg(
  \begin{array}{rcc}
  1 & & 1 \\
  -1 & & 3
  \end{array}
  \Bigg),\quad
  \boldsymbol{\Sigma}_{4}=
  \Bigg(
  \begin{array}{rcr}
  3 & & -2 \\
  -2 & & 2 \end{array}
  \Bigg).
\end{align*}
The left panel in Figure~\ref{Example.Introduction} shows the original data and clusters, labeled with different numbers and colors. We fit to these data the variation of 
model \eqref{General.Mixture.Model} implemented in the function \texttt{DPdensity} of \texttt{DPpackage} \citep{DPpackage}, which is the bivariate version of the DP-based 
model discussed in \cite{escobar&west:95}. The right panel of Figure~\ref{Example.Introduction} shows the same data but now displays the cluster configuration resulting 
from the least squares algorithm described in \cite{dahl:2006}. The estimated partition can be thought of as a particular yet useful summary of the posterior distribution of 
partitions for this model. What we observe is a common situation in the application of models like \eqref{General.Mixture.Model}: we find 6 clusters (the simulation truth 
involved 4 clusters), one of which is a singleton. Such small clusters are very hard to interpret and a natural question arises, is it possible to limit and ideally, avoid 
such occurrences?

In an example like what is described above, our main motivation is not pinning down the ``true'' number of simulated clusters. What we actually want to accomplish is to 
develop a model that encourages joining such small clusters with other larger ones. This would certainly facilitate interpretation of the resulting clusters. Doing so has 
another conceptual advantage, which is sparsity. The non-sparse behavior shown in the right panel of Figure~\ref{Example.Introduction} is precisely facilitated by the fact 
that the atoms in the mixture are i.i.d. and therefore, can move freely with respect to each other. Thus to achieve our desired goal, we need atoms that mutually 
\textit{repel} each other.

Colloquially, the concept of \textit{repulsion} among a set of objects implies that the objects tend to separate rather than congregate. This notion of repulsion has been 
studied in the context of Point Processes. For example, Determinantal Point Processes \citep{LavancierMollerRubak:2015}, Strauss Point Processes 
\citep{MateuMontes:2000,OgataTanemura:1985} and Mat\'ern-type Point Processes \citep{RaoAdamsDunson:2016} are all able to generate point patterns that exhibit more repulsion 
than that expected from a Poisson Point Process \citep{DaleyVereJones:2002}. Given a fixed number of points within a bounded (Borel) set, the Poisson Point Process can 
generate point configurations such that two points can be very close together simply by chance. The repulsion in Determinantal, Strauss and Mat\'ern-type Processes 
discourages such behavior and is controlled by a set of parameters that inform pattern configurations. Among these, to our knowledge, only Determinantal Point Processes have 
been employed to introduce the notion of repulsion in statistical modeling (see \cite{XuMullerTelesca:2016}).

An alternative way to incorporate the notion of repulsion in modeling is to construct a probability distribution that explicitly parameterizes repulsion. Along these lines 
\cite{Steel:2016} develop a family of probability densities called Non-Local Priors that incorporates repulsion by penalizing small relative distances between 
coordinates. Our approach to incorporating repulsion is to model coordinate interactions through potentials (functions that describe the ability to interact) found in so 
called (second order) Gibbs measures. As will be shown, this allows us to control the strength of repulsion and also consider a large variety of types of repulsion.

Gibbs measures have been widely studied and used for describing phenomena from Mechanical Statistics \citep{DaleyVereJones:2002}. Essentially, they are used to model the 
average macroscopic behavior of particle systems through a set of probability and physical laws that are imposed over the possible microscopic states of the system. Through 
the action of potentials, Gibbs measures can induce attraction or repulsion between particles. A number of authors have approached repulsive distributions by specifying a 
particular potential in a Gibbs measure (though the connections to Gibbs measures was not explicitly stated). For example, \cite{NIPS2012_4589} use a Lennard-Jones type 
potential \citep{LennardJones:1924} to introduce repulsion. Interestingly, there is even a connection between Gibbs measures and Determinantal Point Processes via versions 
of Papangelou intensities \citep{Papangelou:1974}. See \cite{GeorgiiYoo:2005} for more details. It is worth noting that in each of the works just cited, the particles 
(following the language in Mechanical Statistics) represent location parameters in mixture models.

Similar to the works just mentioned, we focus on a particular potential specification that introduces repulsion via a joint distribution. There are at least three benefits 
to employing the class of repulsive distributions we develop for statistical modeling:
\begin{itemize}
  \item [(i)] The repulsion is explicitly parameterized in the model and produces a flexible and smooth repulsion effect.

  \item [(ii)] The normalizing constant and induced probability distribution have closed forms, they are (almost) tractable and provide intuition regarding the presence of 
  repulsion.

  \item [(iii)] The computational aspects related to simulation are fairly simple to implement.
\end{itemize}

In what follows, we discuss theoretical and applied aspects of the proposed class of repulsive distributions and in particular we emphasize how the repulsive class of 
distributions achieves the three properties just listed.

The remainder of this chapter will be organized as follows. In Section~\ref{Probability.Repulsive.Distributions} we formally introduce the notion of repulsion in the context 
of a probability distribution and discuss several resulting properties. In Section~\ref{Gaussian.Mixture.Model}, we detail how the repulsive probability distributions can be 
employed in hierarchical mixture modeling for density estimation. Section~\ref{Simulation.Study} contains results from a small simulation study that compares the repulsive 
mixture model we develop to DPM and finite mixture models. In Section~\ref{Data.Illustrations} we apply the methodology to two well known datasets. Proofs of all technical 
results and computational strategies are provided in Appendix~\ref{Algorith.RGMM}--\ref{Proof.Posterior.Convergence.Rate}.


\section{Probability Repulsive Distributions}\label{Probability.Repulsive.Distributions}
We start by providing contextual background and introducing notation that will be used throughout.

\subsection{Background and Preliminaries}\label{Preliminaries}

We will use the $k$-fold product space of $\R^{d}$ denoted by $\R^{d}_{k}=\prod_{i=1}^{k}\R^{d}$ and $\B(\R^{d}_{k})$ its associated $\sigma$-algebra as the reference space 
on which the class of distributions we derive will be defined. Here, $k\in\N$ $(k\geq2)$ and $d\in\N$. Let 
$\boldsymbol{x}_{k,d}=(\boldsymbol{x}_{1},\ldots,\boldsymbol{x}_{k})$ with $\boldsymbol{x}_{1},\ldots,\boldsymbol{x}_{k}\in\R^{d}$. The $k$-tuple $\boldsymbol{x}_{k,d}$ can 
be thought of as $k$ ordered objects of dimension $d$ jointly allocated in $\R^{d}_{k}$. We add to the measurable space $(\R^{d}_{k},\B(\R^{d}_{k}))$ a $\sigma$-finite 
measure $\lambda^{k}_{d}$, that is the $k$-fold product of the $d$-dimensional Lebesgue measure $\lambda_{d}$. To represent integrals with respect to $\lambda^{k}_{d}$, we 
will use $\diff\boldsymbol{x}_{k,d}$ instead of $\diff\lambda^{k}_{d}(\boldsymbol{x}_{k,d})$. Also, given two metric spaces $(\Omega_{1},d_{1})$ and $(\Omega_{2},d_{2})$ we 
denote by $C(\Omega_{1};\Omega_{2})$ the class of all continuous functions $f:\Omega_{1}\to\Omega_{2}$. In what follows we use the term \textit{repulsive distribution} to 
reference a distribution that formally incorporates the notion of repulsion.

As mentioned previously, our construction of non-i.i.d. distributions depends heavily on Gibbs measures where dependence (and hence repulsion) between the coordinates of 
$\boldsymbol{x}_{k,d}$ is introduced via functions that model interactions between them. More formally, consider $\varphi_{1}:\R^{d}\to[-\infty,\infty]$ a measurable 
function and $\varphi_{2}:\R^{d}\times\R^{d}\to[-\infty,\infty]$ a measurable and symmetric function. Define
\begin{equation}
  \nu_{\mathrm{G}}\Bigg(\prod_{i=1}^{k}A_{i}\Bigg)=
  \int_{\prod_{i=1}^{k}A_{i}}
  \exp\Bigg\{-\sum_{i=1}^{k}\varphi_{1}(\boldsymbol{x}_{i})-\sum_{r<s}^{k}\varphi_{2}(\boldsymbol{x}_{r},\boldsymbol{x}_{s})\Bigg\}\diff\boldsymbol{x}_{k,d},
  \label{Gibbs.Measure}
\end{equation}
where $\prod_{i=1}^{k}A_{i}$ is the cartesian product of Borel sets $A_{1},\ldots,A_{k}$ in $\R^{d}$. Here, $\varphi_{1}$ can be thought of as a physical force that controls 
the influence that the environment has on each coordinate $\boldsymbol{x}_{i}$ while $\varphi_{2}$ controls the interaction between pairs of coordinates $\boldsymbol{x}_{r}$ 
and $\boldsymbol{x}_{s}$. If $\varphi_{1}$ and $\varphi_{2}$ are selected so that $\nu_{\mathrm{G}}(\R^{d}_{k})$ is finite, then by Caratheodory's Theorem $\nu_{\mathrm{G}}$ 
defines a unique finite measure on $(\R^{d}_{k},\B(\R^{d}_{k}))$. The induced probability measure corresponding to the normalized version of \eqref{Gibbs.Measure}, is called 
a (second order) Gibbs measure. The normalizing constant (total mass of $\R^{d}_{k}$ under $\nu_{\mathrm{G}}$)
\begin{equation*}
  \nu_{\mathrm{G}}(\R^{d}_{k})=
  \int_{\R^{d}_{k}}
  \exp\Bigg\{-\sum_{i=1}^{k}\varphi_{1}(\boldsymbol{x}_{i})-\sum_{r<s}^{k}\varphi_{2}(\boldsymbol{x}_{r},\boldsymbol{x}_{s})\Bigg\}\diff\boldsymbol{x}_{k,d}
\end{equation*}
is commonly known as partition function \citep{Pathria2011299} and encapsulates important qualitative information about the interactions and the degree of disorder present 
in the coordinates of $\boldsymbol{x}_{k,d}$. In general, $\nu_{\mathrm{G}}(\R^{d}_{k})$ is (almost) intractable mainly because of the presence of $\varphi_{2}$.

Note that symmetry of $\varphi_{2}$ (i.e., $\varphi_{2}(\boldsymbol{x}_{r},\boldsymbol{x}_{s})=\varphi_{2}(\boldsymbol{x}_{s},\boldsymbol{x}_{r})$) means that 
$\nu_{\mathrm{G}}$ defines a symmetric measure. This implies that the order of coordinates is immaterial. If $\varphi_{2}=0$ then $\nu_{\mathrm{G}}$ reduces to a structure 
where coordinates do not interact and are only subject to environmental influence through $\varphi_{1}$. When $\varphi_{2}\neq0$, it is common that 
$\varphi_{2}(\boldsymbol{x},\boldsymbol{y})$ only depends on the relative distance between $\boldsymbol{x}$ and $\boldsymbol{y}$ \citep{DaleyVereJones:2002}. More formally, 
let $\rho:\R^{d}\times\R^{d}\to[0,\infty)$ be a metric on $\R^{d}$ and $\phi:[0,\infty)\to[-\infty,\infty]$ a measurable function. To avoid pathological or degenerate cases, 
we consider metrics that do not treat singletons as open sets in the topology induced by $\rho$. Then letting 
$\varphi_{2}(\boldsymbol{x},\boldsymbol{y})=\phi\{\rho(\boldsymbol{x},\boldsymbol{y})\}$, interactions will be smooth if, for example, 
$\phi\in C([0,\infty);[-\infty,\infty])$. Following this general idea, \cite{NIPS2012_4589} use $\phi(r)=\tau(1/r)^{\nu}:\tau,\nu\in(0,\infty)$ to construct repulsive 
probability densities, which is a particular case of the Lennard-Jones type potential \citep{LennardJones:1924} that appears in Molecular Dynamics. Another potential that 
can be used to define repulsion is the (Gibbs) hard-core potential $\phi(r)=+\infty\I_{[0,b]}(r):b\in(0,\infty)$ \citep{IllianPenttinen:2008}, which is a particular case of 
the Strauss potential \citep{Strauss:1975}. Here, $\I_{A}(r)$ is the indicator function over a Borel set $A$ in $\R$. This potential, used in the context of Point Processes, 
generates disperse point patterns whose points are all separated by a distance greater than $b$ units. However, the threshold of separation $b$ prevents the repulsion from 
being smooth \citep{DaleyVereJones:2002}. Other examples of repulsive potentials can be found in \cite{OgataTanemura:1981,OgataTanemura:1985}. The key characteristic that 
differentiates the behavior of the potentials provided above is the action near 0; the faster the potential function goes to infinity as relative distance between 
coordinates goes to zero, the stronger the repulsion that the coordinates of $\boldsymbol{x}_{k,d}$ will experiment when they are separated by small distances. Even though 
\cite{Steel:2016} do not employ a potential to model repulsion, the repulsion that results from their model is very similar to that found in \cite{NIPS2012_4589} and tends 
to push coordinates far apart.

It is often the case that $\varphi_{1}$ and $\varphi_{2}$ are indexed by a set of parameters which inform the types of patterns produced. It would therefore be natural to 
estimate these parameters using observed data. However, $\nu_{\mathrm{G}}(\R^{d}_{k})$ is typically a function of the unknown parameters which makes deriving closed form 
expressions of $\nu_{\mathrm{G}}(\R^{d}_{k})$ practically impossible and renders Bayesian or frequentist estimation procedures intractable. To avoid this complication, 
pseudo-maximum likelihood methods have been proposed to approximate $\nu_{\mathrm{G}}(\R^{d}_{k})$ when carrying out estimation \citep{OgataTanemura:1981,Penttinen:1984}. We 
provide details of a Bayesian approach in subsequent sections.


\subsection{$\mathrm{Rep}_{k,d}(f_{0},C_{0},\rho)$ Distribution}

As mentioned, our principal objective is to construct a family of probability densities for $\boldsymbol{x}_{k,d}$ that relaxes the i.i.d. assumption associated with its 
coordinates and we will do this by employing Gibbs measures that include an interaction function that mutually separates the $k$ coordinates. Of all the potentials that 
might be considered in a Gibbs measure, we seek one that permits modeling repulsion flexibly so that a soft type of repulsion is available which avoids forcing large 
distances among the coordinates. As noted by \cite{DaleyVereJones:2002} and \cite{OgataTanemura:1981} the following potential
\begin{align}
  \phi(r)=-\log\{1-\exp(-cr^{2})\}:c\in(0,\infty)
  \label{Very.Soft-core.Potential}
\end{align}
produces smoother repulsion compared to other types of potentials in terms of ``repelling strength'' and for this reason we employ it as an example of interaction function 
in a Gibbs measure. A question that naturally arises at this point relates to the possibility of specifying a tractable class of repulsive distributions that incorporates 
the features discussed above. Note first that connecting \eqref{Very.Soft-core.Potential} with $\nu_{\mathrm{G}}$ is straightforward: if we take
\begin{equation*}
  \varphi_{2}(\boldsymbol{x},\boldsymbol{y})=-\log[1-C_{0}\{\rho(\boldsymbol{x},\boldsymbol{y})\}],\qquad C_{0}(r)=\exp(-cr^{2}):c\in(0,\infty)
\end{equation*}
then $\nu_{\mathrm{G}}$ will have a ``pairwise-interaction term'' given by
\begin{equation}
  \exp\Bigg\{-\sum_{r<s}^{k}\varphi_{2}(\boldsymbol{x}_{r},\boldsymbol{x}_{s})\Bigg\}=
  \prod_{r<s}^{k}[1-C_{0}\{\rho(\boldsymbol{x}_{r},\boldsymbol{x}_{s})\}].
  \label{Interaction.Potential}
\end{equation}
The right-hand side of \eqref{Interaction.Potential} induces a particular interaction structure that separates the coordinates of $\boldsymbol{x}_{k,d}$, thus introducing a 
notion of repulsion. The degree of separation is regulated by the speed at which $C_{0}$ decays to 0. The answer to the question posed earlier can then be given by focusing 
on functions $C_{0}:[0,\infty)\to(0,1]$ that satisfy the following properties:
\begin{itemize}
  \item [A1.] $C_{0}\in C([0,\infty);(0,1])$.
  \item [A2.] $C_{0}(0)=1$.
  \item [A3.] $C_{0}(r)\to0$ (right-side limit) when $x\to\infty$.
  \item [A4.] For all $r_{1},r_{2}\in[0,\infty)$, if $r_{1}<r_{2}$ then $C_{0}(r_{1})>C_{0}(r_{2})$.
\end{itemize}
For future reference we will call A1 to A4 the $C_{0}$-properties. The following Lemma guarantees that the type of repulsion induced by the $C_{0}$-properties is smooth in 
terms of $\boldsymbol{x}_{k,d}$.

\begin{Lem}\label{Smoothness.Repulsive.Component}
  Given a metric $\rho:\R^{d}\times\R^{d}\to[0,\infty)$ such that singletons are not open sets in the topology induced by $\rho$, the function 
  $\mathrm{R}_{\mathrm{C}}:\R^{d}_{k}\to[0,1)$ defined by
  \begin{equation}
    \mathrm{R}_{\mathrm{C}}(\boldsymbol{x}_{k,d})=
    \prod_{r<s}^{k}[1-C_{0}\{\rho(\boldsymbol{x}_{r},\boldsymbol{x}_{s})\}]
    \label{Repulsive.Component}
  \end{equation}
  belongs to $C(\R^{d}_{k};[0,1))$ for all $d\in\N$ and $k\in\N$ $(k\geq2)$.
\end{Lem}

Through out the article we will refer to \eqref{Repulsive.Component} as the \textit{repulsive component}. We finish the construction of repulsive probability measures by 
specifying a distribution supported on $\R^{d}$ which will be common for all the coordinates of $\boldsymbol{x}_{k,d}$. Let $f_{0}\in C(\R^{d};(0,\infty))$ be a probability 
density function, then under $\varphi_{1}(\boldsymbol{x})=-\log\{f_{0}(\boldsymbol{x})\}$, $\nu_{\mathrm{G}}$ will have a ``base component term'' given by
\begin{equation}
  \exp\Bigg\{-\sum_{i=1}^{k}\varphi_{1}(\boldsymbol{x}_{i})\Bigg\}=\prod_{i=1}^{k}f_{0}(\boldsymbol{x}_{i}).
  \label{Global.Potential}
\end{equation}
Incorporating \eqref{Interaction.Potential} and \eqref{Global.Potential} into \eqref{Gibbs.Measure} we get
\begin{equation*}
  \nu_{\mathrm{G}}\Bigg(\prod_{i=1}^{k}A_{i}\Bigg)=
  \int_{\prod_{i=1}^{k}A_{i}}
  \Bigg\{\prod_{i=1}^{k}f_{0}(\boldsymbol{x}_{i})\Bigg\}\mathrm{R}_{\mathrm{C}}(\boldsymbol{x}_{k,d})\diff\boldsymbol{x}_{k,d}.
\end{equation*}
The following Proposition ensures that the repulsive probability measures just constructed are well defined.

\begin{Prop}\label{Existence.Repulsive.Distribution}
  Let $f_{0}\in C(\R^{d};(0,\infty))$ be a probability density function. The function
  \begin{equation}
    g_{k,d}(\boldsymbol{x}_{k,d})=
    \Bigg\{\prod_{i=1}^{k}f_{0}(\boldsymbol{x}_{i})\Bigg\}\mathrm{R}_{\mathrm{C}}(\boldsymbol{x}_{k,d})
    \label{Unnormalized.Repulsive.Density}
  \end{equation}
  is measurable and integrable for all $d\in\N$ and $k\in\N$ $(k\geq2)$.
\end{Prop}

With Proposition~\ref{Existence.Repulsive.Distribution} it is now straightforward to construct a probability measure with the desired repulsive structure; small relative 
distances are penalized in a smooth way. Notice that the support of \eqref{Unnormalized.Repulsive.Density} is determined by the shape of the ``baseline distribution'' 
$f_{0}$ and then subsequently distorted (i.e. contracted) by the repulsive component. The normalized version of \eqref{Unnormalized.Repulsive.Density} defines a valid joint 
probability density function which we now provide.

\begin{Def}\label{Repulsive.Distribution}
  The probability distribution $\Rep_{k,d}(f_{0},C_{0},\rho)$ has probability density function
  \begin{align}
    &\Rep_{k,d}(\boldsymbol{x}_{k,d})=\frac{1}{c_{k,d}}\Bigg\{\prod_{i=1}^{k}f_{0}(\boldsymbol{x}_{i})\Bigg\}\mathrm{R}_{\mathrm{C}}(\boldsymbol{x}_{k,d}),
    \label{Repulsive.Density} \\ 
    &c_{k,d}=\int_{\R^{d}_{k}}
    \Bigg\{\prod_{i=1}^{k}f_{0}(\boldsymbol{x}_{i})\Bigg\}\mathrm{R}_{\mathrm{C}}(\boldsymbol{x}_{k,d})\diff\boldsymbol{x}_{k,d}
    \label{Repulsive.Constant}.
  \end{align}
  Here $\boldsymbol{x}_{k,d}\in\R^{d}_{k}$, $f_{0}\in C(\R^{d};(0,\infty))$ is a probability density function, $C_{0}:[0,\infty)\to(0,1]$ is a function that satisfies the 
  $C_{0}$-properties and $\rho:\R^{d}\times\R^{d}\to[0,\infty)$ is a metric such that singletons are not open sets in the topology induced by it.
\end{Def}


\subsection{$\mathrm{Rep}_{k,d}(f_{0},C_{0},\rho)$ Properties}\label{Rep.Properties}

In this section we will investigate a few general properties of the $\Rep_{k,d}(f_{0},C_{0},\rho)$ class. The distributional results are provided to further understanding 
regarding characteristics of \eqref{Repulsive.Density} from a qualitative and analytic point of view. As a first observation, because of symmetry, 
$\Rep_{k,d}(\boldsymbol{x}_{k,d})$ is an exchangeable distribution in $\boldsymbol{x}_{1},\ldots,\boldsymbol{x}_{k}$. This facilitates the study of computational techniques 
motivated by $\Rep_{k,d}(f_{0},C_{0},\rho)$. However, it is worth noting that $\{\Rep_{k,d}(f_{0},C_{0},\rho)\}_{k\geq2}$ does not induce a sample-size consistent sequence 
of finite-dimensional distributions, meaning that
\begin{equation*}
  \int_{\R^{d}}\Rep_{k+1,d}(\boldsymbol{x}_{k+1,d})\diff\boldsymbol{x}_{k+1}\neq\Rep_{k,d}(\boldsymbol{x}_{k,d}).
\end{equation*}
This makes predicting locations of new coordinates problematic. In Section~\ref{Gaussian.Mixture.Model} we address how this may be accommodated in modeling contexts. To 
simplify notation, in what follows we will use $[m]=\{1,\ldots,m\}$, with $m\in\N$.


\subsubsection{Normalizing Constant}\label{Normalizing.Constant}

Because $\mathrm{R}_{\mathrm{C}}(\boldsymbol{x}_{k,d})$ is invariant under permutations of the coordinates of $\boldsymbol{x}_{k,d}$, an interaction's direction is 
immaterial to whether it is present or absent (i.e., $\boldsymbol{x}_{r}$ interacts with $\boldsymbol{x}_{s}$ if and only if $\boldsymbol{x}_{s}$ interacts with 
$\boldsymbol{x}_{r}$). Therefore it is sufficient to represent the interaction between $\boldsymbol{x}_{r}$ and $\boldsymbol{x}_{s}$ as $(r,s)\in I_{k}$ where 
$I_{k}=\{(r,s):1\leq r<s\leq k\}$. In this setting, $I_{k}$ reflects the set of all pairwise interactions between the $k$ coordinates of $\boldsymbol{x}_{k,d}$ and 
$\ell_{k}=\card(I_{k})=\frac{k(k-1)}{2}$, where $\card(E)$ is the cardinality of a set $E$. Now, expanding \eqref{Repulsive.Component} term-by-term results in
\begin{equation}
  \mathrm{R}_{\mathrm{C}}(\boldsymbol{x}_{k,d})=
  1+\sum_{l=1}^{\ell_{k}}(-1)^{l}\sum_{\substack{A\subseteq I_{k} \\ \card(A)=l}}\Bigg[\prod_{(r,s)\in A}C_{0}\{\rho(\boldsymbol{x}_{r},\boldsymbol{x}_{s})\}\Bigg]
  \label{Expanded.Repulsive.Component}.
\end{equation}
The right-side of \eqref{Expanded.Repulsive.Component} is connected to graph theory in the following way: $A\subseteq I_{k}$ can be interpreted as a non-directed graph whose 
edges are $(r,s)\in A$.

\begin{figure}[htb]
  \begin{center}
  \begin{tabular}{c}
    \includegraphics[width=15cm,height=8cm]{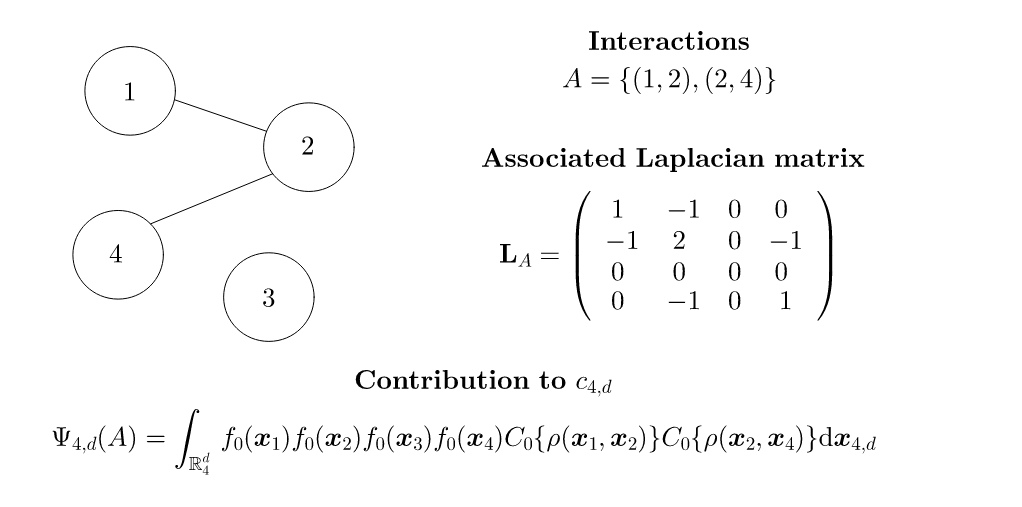}
  \end{tabular}
  \end{center}
  \caption{The graph and Laplacian matrix for a possible interaction for $k=4$ coordinates.}
\label{Example.Laplacian}
\end{figure}

Using \eqref{Expanded.Repulsive.Component}, it can be shown that expression \eqref{Repulsive.Constant} in Definition~\ref{Repulsive.Distribution} has the following form:
\begin{align}
  &c_{k,d}=1+\sum_{l=1}^{\ell_{k}}(-1)^{l}\sum_{\substack{A\subseteq I_{k} \\ \card(A)=l}}\Psi_{k,d}(A)
  \label{Expanded.Repulsive.Constant} \\
  &\Psi_{k,d}(A)=\int_{\R^{d}_{k}}
  \Bigg\{\prod_{i=1}^{k}f_{0}(\boldsymbol{x}_{i})\Bigg\}\Bigg[\prod_{(r,s)\in A}C_{0}\{\rho(\boldsymbol{x}_{r},\boldsymbol{x}_{s})\}\Bigg]\diff\boldsymbol{x}_{k,d}
  \label{Repulsive.Integral}.
\end{align}
Note that representing $A$ as a graph or Laplacian matrix can help develop intuition on how each summand contributes to the expression \eqref{Expanded.Repulsive.Constant}. 
Figure~\ref{Example.Laplacian} shows one particular case of how 3 of $k=4$ coordinates in $\R^{d}$ might interact by providing the respective Laplacian matrix together with 
the contribution that \eqref{Repulsive.Integral} brings to calculating $c_{4,d}$ according to \eqref{Expanded.Repulsive.Constant}.

Equation~\eqref{Expanded.Repulsive.Constant} retains connections with the probabilistic version of the Inclusion-Exclusion Principle. This result, which is very useful in 
Enumerative Combinatorics, says that in any probability space $(\Omega,\F,\Pb)$
\begin{equation*}
  \Pb\Bigg(\bigcap_{i=1}^{k}A_{i}^{c}\Bigg)=1+\sum_{l=1}^{k}(-1)^{l}\sum_{\substack{I\subseteq[k] \\ \card(I)=l}}\Pb\Bigg(\bigcap_{i\in I}A_{i}\Bigg),
\end{equation*}
with $A_{1},\ldots,A_{k}$ events on $\F$ and $A_{i}^{c}$ denoting the complement of $A_{i}$. With this in mind, $c_{k,d}$ is the result of adding/substracting all the 
contributions $\Psi_{k,d}(A)$ that emerge for every non-empty set $A\subseteq I_{k}$. If we think of $c_{k,d}$ as an indicator of the strength of repulsion, 
$\Psi_{k,d}(A)$ provides the specific contribution from the interactions $(r,s)\in A$. Moreover, it quantifies how distant a $\Rep_{k,d}(f_{0},C_{0},\rho)$ distribution is 
from the (unattainable) extreme case $C_{0}=0$ (i.e., the coordinates $\boldsymbol{x}_{1},\ldots,\boldsymbol{x}_{k}$ are mutually independent and share a common probability 
law $f_{0}$).

The tractability of $c_{k,d}$ depends heavily on the number of coordinates $k$ since the cost of evaluating \eqref{Repulsive.Integral} becomes prohibitive as it requires 
carrying out (at least) $2^{\ell_{k}}-1$ numerical calculations. In Subsection~\ref{Particular.Case} we highlight a particular choice of $f_{0}$, $C_{0}$ and $\rho$ that 
produces a closed form expression for \eqref{Repulsive.Integral}.


\section{Gaussian Mixture Models and $\mathrm{NRep}_{k,d}(\boldsymbol{\mu},\boldsymbol{\Sigma},\tau)$ Distribution}\label{Gaussian.Mixture.Model}

In this section we will briefly introduce Gaussian Mixture Models, which are very popular in the context of density estimation \citep{escobar&west:95} because of their 
flexibility and computational tractability. Then we show that repulsion can be incorporated by modeling location parameters with the repulsion distribution described 
previously. 


\subsection{Repulsive Gaussian Mixture Models (RGMM)}\label{Particular.Case}

Consider $n\in\N$ experimental units whose responses $\boldsymbol{y}_{1},\ldots,\boldsymbol{y}_{n}$ are $d$-dimensional and assumed to be exchangeable. Gaussian mixtures can 
be thought of as a way of grouping the $n$ units into several clusters, say $k\in\N$, each having its own specific characteristics. In this context, the $j$th cluster 
($j\in[k]$) is modeled through a Gaussian density $\mathrm{N}_{d}(\,\cdot\,;\boldsymbol{\theta}_{j},\boldsymbol{\Lambda}_{j})$ with location 
$\boldsymbol{\theta}_{j}\in\R^{d}$ and scale $\boldsymbol{\Lambda}_{j}\in\Sy^{d}$. Here, $\Sy^{d}$ is the space of real, symmetric and positive-definite matrices of 
dimension $d\times d$. We let $\boldsymbol{\theta}_{k,d}=(\boldsymbol{\theta}_{1},\ldots,\boldsymbol{\theta}_{k})\in\R^{d}_{k}$ and 
$\boldsymbol{\Lambda}_{k,d}=(\boldsymbol{\Lambda}_{1},\ldots,\boldsymbol{\Lambda}_{k})\in\Sy^{d}_{k}$ where $\Sy^{d}_{k}$ is the $k$-fold product space of $\Sy^{d}$. Next 
let $\boldsymbol{\pi}_{k,1}=(\pi_{1},\ldots,\pi_{k})\in\Delta_{k-1}$, where $\Delta_{k-1}$ is the standard $(k-1)$-simplex $(\Delta_{0}=\{1\})$, denote a set of weights that 
reflect the probability of allocating $\boldsymbol{y}_{i}:i\in[n]$ to a cluster. Then the standard Gaussian Mixture Model is
\begin{equation}
  \boldsymbol{y}_{i}\mid\boldsymbol{\pi}_{k,1},\boldsymbol{\theta}_{k,d},\boldsymbol{\Lambda}_{k,d}
  \stackrel{i.i.d.}{\sim}
  \sum_{j=1}^{k}\pi_{j}\mathrm{N}_{d}(\boldsymbol{y}_{i};\boldsymbol{\theta}_{j},\boldsymbol{\Lambda}_{j}).
  \label{Likelihood.GMM}
\end{equation}
It is common to restate \eqref{Likelihood.GMM} by introducing latent cluster membership indicators $z_{1},\ldots,z_{n}\in[k]$ such that $\boldsymbol{y}_{i}$ is drawn from 
the $j$th mixture component if and only if $z_{i}=j$:
\begin{align}
  &\boldsymbol{y}_{i}\mid z_{i},\boldsymbol{\theta}_{k,d},\boldsymbol{\Lambda}_{k,d}
  \stackrel{ind.}{\sim}
  \mathrm{N}_{d}(\boldsymbol{y}_{i};\boldsymbol{\theta}_{z_{i}},\boldsymbol{\Lambda}_{z_{i}})
  \label{Response.given.Cluster.Label} \\
  &z_{i}\mid\boldsymbol{\pi}_{k,1}
  \stackrel{i.i.d.}{\sim}
  \Pb(z_{i}=j)=\pi_{j}.
  \label{Cluster.Label}
\end{align}
after marginalizing over the $z_{i}$ indicators. The model is typically completed with conjugate-style priors for all parameters.

Specifying a prior distribution for $k\in\N$ is possible. For example, DPM models by construction induce a prior distribution on the number of clusters $k$. Alternatively, 
Reversible Jump MCMC \citep{RJMCMC,Richardson:1997} or Birth-Death Chains \citep{Stephens:2000} could be employed after assigning a particular prior for $k$. These methods 
do not translate well to the non-i.i.d. case and so we employ a case-specific upper bound $k\geq2$.

In the above mixture model, the location parameters associated with each mixture component are typically assumed to be independent a priori. This is precisely the assumption 
that facilitates the presence of redundant mixture components. In contrast, our work focuses on employing $\Rep_{k,d}(f_{0},C_{0},\rho)$ as a model for location parameters 
in \eqref{Likelihood.GMM} which promotes reducing redundant mixture components without sacrificing goodness-of-fit, i.e, more parsimony relative to alternatives with 
independent locations. Moreover, the responses will be allocated to a few well-separated clusters. This desired behavior can be easily incorporated in the mixture model by 
assuming
\begin{align}
  &\boldsymbol{\theta}_{k,d}\sim\Rep_{k,d}(f_{0},C_{0},\rho)\nonumber \\
  &f_{0}(\boldsymbol{x})=\mathrm{N}_{d}(\boldsymbol{x};\boldsymbol{\mu},\boldsymbol{\Sigma}):\boldsymbol{\mu}\in\R^{d},\boldsymbol{\Sigma}\in\Sy^{d}
  \label{Normal.Baseline.Density} \\
  &C_{0}(r)=\exp(-0.5\tau^{-1}r^{2}):\tau\in(0,\infty)
  \label{Exponential.Repulsion} \\
  &\rho(\boldsymbol{x},\boldsymbol{y})=\{(\boldsymbol{x}-\boldsymbol{y})^{\top}\boldsymbol{\Sigma}^{-1}(\boldsymbol{x}-\boldsymbol{y})\}^{1/2}.
  \label{Generalized.Euclidean.Distance}
\end{align}
The specific forms of $f_{0}$, $C_{0}$ and $\rho$ are admissible according to Definition~\ref{Repulsive.Distribution}. The repulsive distribution parameterized by 
\eqref{Normal.Baseline.Density}--\eqref{Generalized.Euclidean.Distance} will be denoted by $\NRep_{k,d}(\boldsymbol{\mu},\boldsymbol{\Sigma},\tau)$. Because
$\NRep_{k,d}(\boldsymbol{\mu},\boldsymbol{\Sigma},\tau)$ introduces dependence a priori (in particular, repulsion) between the coordinates of $\boldsymbol{\theta}_{k,d}$, 
they are no longer conditionally independent given $(\boldsymbol{y}_{n,d},\boldsymbol{z}_{n,1},\boldsymbol{\Lambda}_{k,d})$, with 
$\boldsymbol{y}_{n,d}=(\boldsymbol{y}_{1},\ldots,\boldsymbol{y}_{n})\in\R^{d}_{n}$ and $\boldsymbol{z}_{n,1}=(z_{1},\ldots,z_{n})\in[k]^{n}$. The parameter $\tau$ in 
\eqref{Exponential.Repulsion} controls the strength of repulsion associated with coordinates in $\boldsymbol{\theta}_{k,d}$ via \eqref{Generalized.Euclidean.Distance}: as 
$\tau\to0$ (right-side limit), the repulsion becomes weaker. The selection of \eqref{Normal.Baseline.Density} mimics the usual i.i.d. multivariate normal assumption.

To facilitate later reference we state the ``repulsive mixture model'' in its entirety:
\begin{align}
  &\boldsymbol{y}_{i}\mid z_{i},\boldsymbol{\theta}_{k,d},\boldsymbol{\Lambda}_{k,d}
  \stackrel{ind.}{\sim}
  \mathrm{N}_{d}(\boldsymbol{y}_{i};\boldsymbol{\theta}_{z_{i}},\boldsymbol{\Lambda}_{z_{i}})
  \label{NRep.Response.given.Cluster.Label} \\
  &z_{i}\mid \boldsymbol{\pi}_{k,1}
  \stackrel{i.i.d.}{\sim}
  \Pb(z_{i}=j)=\pi_{j}
  \label{NRep.Cluster.Label}
\end{align}
together with the following mutually independent prior distributions:
\begin{align}
  &\boldsymbol{\pi}_{k,1}
  \sim
  \mathrm{Dir}(\boldsymbol{\alpha}_{k,1}):\boldsymbol{\alpha}_{k,1}\in(0,\infty)^{k}
  \label{Prior.Pi.NRep} \\
  &\boldsymbol{\theta}_{k,d}
  \sim
  \NRep_{k,d}(\boldsymbol{\mu},\boldsymbol{\Sigma},\tau):\boldsymbol{\mu}\in\R^{d},\boldsymbol{\Sigma}\in\Sy^{d},\tau\in(0,\infty)
  \label{Prior.Theta.NRep} \\
  &\boldsymbol{\Lambda}_{j}\stackrel{i.i.d.}{\sim}\mathrm{IW}_{d}(\boldsymbol{\Psi},\nu):\boldsymbol{\Psi}\in\Sy^{d},\nu\in(0,\infty).
  \label{Prior.Lambda.NRep}
\end{align}
In what follows we will refer to the model in \eqref{NRep.Response.given.Cluster.Label}--\eqref{Prior.Lambda.NRep} as the (Bayesian) Repulsive Gaussian Mixture Model 
(abbreviated as RGMM).


\subsubsection{Parameter Calibration}\label{Parameter.Calibration}

We briefly discuss stategies of selecting values for parameters that control the prior distributions in \eqref{Prior.Pi.NRep}--\eqref{Prior.Lambda.NRep}. We select values
for $\boldsymbol{\mu}$, $\boldsymbol{\Sigma}$ and $\tau$ of the RGMM instead of treating them as unknown and assigning them hyperprior distributions because of computational
cost. First notice that $(\boldsymbol{\mu},\boldsymbol{\Sigma})$ acts as a location/scale parameter: if $\boldsymbol{\Sigma}=\mathbf{C}\mathbf{C}^{\top}$ is the 
corresponding Cholesky decomposition for $\boldsymbol{\Sigma}$, then $\boldsymbol{\theta}_{k,d}\sim\NRep_{k,d}(\mathbf{0}_{d},\mathbf{I}_{d},\tau)$ implies that
\begin{equation*}
  \mathbf{1}_{k}\otimes\boldsymbol{\mu}+(\mathbf{I}_{k}\otimes\mathbf{C})\boldsymbol{\theta}_{k,d}\sim\NRep_{k,d}(\boldsymbol{\mu},\boldsymbol{\Sigma},\tau),
\end{equation*}
where $\mathbf{I}_{d}$ is the $d\times d$ identity matrix and $\mathbf{0}_{d},\mathbf{1}_{d}\in\R^{d}$ are $d$-dimensional vectors of zeroes and ones, respectively. Although 
a Gaussian hyperprior for $\boldsymbol{\mu}$ is a reasonable candidate (the full conditional distribution is also Gaussian), it is not straightforward how to select its 
associated hyperparameters. A slightly more complicated problem occurs with $\boldsymbol{\Sigma}$, since this parameter participates in the repulsive component and no closed 
form is available for its posterior distribution. Even more problematic, the induced full conditional distribution for $\tau$ turns out to be doubly-intractable 
\citep{murray2006} and as a result the standard MCMC algorithms do not apply. To see this, it can be shown using \eqref{Expanded.Repulsive.Constant}, 
\eqref{Repulsive.Integral} and the Gaussian integral that the normalizing constant of $\NRep_{k,d}(\boldsymbol{\mu},\boldsymbol{\Sigma},\tau)$ is
\begin{equation*}
  c_{k,d}=1+\sum_{l=1}^{\ell_{k}}(-1)^{l}\sum_{\substack{A\subseteq I_{k} \\ \card(A)=l}}
  \det(\mathbf{I}_{k}\otimes\mathbf{I}_{d}+\mathbf{L}_{A}\otimes\tau^{-1}\mathbf{I}_{d})^{-1/2},
\end{equation*}
where $\mathbf{I}_{k}$ is the $k\times k$ identity matrix, $\mathbf{L}_{A}$ denotes the Laplacian matrix associated to the set of interactions $A\subseteq I_{k}$ (see 
Subsection~\ref{Normalizing.Constant}) and $\otimes$ is the matrix Kronecker product, making it a function of $\tau$. 

To facilitate hyperparameter selection we standardize the $\boldsymbol{y}_i$'s (a common practice in mixture models see, e.g. \citealt{BDA3}). Upon standardizing the 
response, it is reasonable to assume that $\boldsymbol{\mu}=\mathbf{0}_{d}$ and $\boldsymbol{\Sigma}=\mathbf{I}_{d}$. Further \cite{BDA3} argue that setting 
$\boldsymbol{\alpha}_{k,1}=k^{-1}\mathbf{1}_{d}$ produces a weakly informative prior for $\boldsymbol{\pi}_{k,1}$. Selecting $\nu$ and $\boldsymbol{\Psi}$ is particularly 
important as they can dominate the repulsion effect. Setting $\nu=d+4$ and $\boldsymbol{\Psi}=3\psi\mathbf{I}_{d}$ with $\psi\in(0,\infty)$ guarantees that each scale matrix 
$\boldsymbol{\Lambda}_{j}$ is centered on $\psi\mathbf{I}_{d}$ and that their entries possess finite variances. The value of $\psi$ can be set to a value that accommodates 
the desired variability.

To calibrate $\tau$, we follow the strategy outlined in \cite{Steel:2016}. Their approach consists of first specifying the probability that the coordinates of 
$\boldsymbol{\theta}_{k,d}$ are separated by a certain distance $u$ and then set $\tau$ to the value that achieves the desired probability. To formalize this idea, suppose 
first that $\boldsymbol{\theta}_{1},\ldots,\boldsymbol{\theta}_{k}$ are a random sample coming from $\mathrm{N}_{d}(\mathbf{0}_{d},\mathbf{I}_{d})$. To favor separation 
among these random vectors we can use \eqref{Exponential.Repulsion} and \eqref{Generalized.Euclidean.Distance} with $\boldsymbol{\Sigma}=\mathbf{I}_{d}$ to choose $\tau$ 
such that for all $r\neq s\in[k]$
\begin{equation*}
  \Pb[1-\exp\{-0.5\tau^{-1}(\boldsymbol{\theta}_{r}-\boldsymbol{\theta}_{s})^{\top}(\boldsymbol{\theta}_{r}-\boldsymbol{\theta}_{s})\}\leq u]=p,
\end{equation*}
for fixed values $u,p\in(0,1)$. Letting $w(u)=-\log(1-u)$ for $u\in(0,1)$, standard properties of the Gaussian distribution guarantee that the previous relation is 
equivalent to
\begin{equation}
  \Pb\{G\leq w(u)\tau\}=p,\qquad
  G=\frac{1}{2}(\boldsymbol{\theta}_{r}-\boldsymbol{\theta}_{s})^{\top}(\boldsymbol{\theta}_{r}-\boldsymbol{\theta}_{s})\sim\mathrm{G}(d/2,1/2).
  \label{Tau.Calibration}
\end{equation}
Creating a grid of points in $(0,\infty)$ it is straightforward to find a $\tau$ that fulfills criterion \eqref{Tau.Calibration}. This criterion allows the repulsion to be 
small (according to $u$), while at the same time preventing it with probability $p$ from being too strong. This has the added effect of avoiding degeneracy of 
\eqref{Prior.Theta.NRep}, thus making computation numerically more stable. In practice, we apply the procedure outlined above to the vectors coming from the repulsive
distribution \eqref{Prior.Theta.NRep}, treating them as if they were sampled from a multivariate Gaussian distribution. This gives us a simple procedure to approximately 
achieve the desired goal of prior separation with a pre-specified probability.


\subsection{Theoretical Properties}\label{Theoretical.Properties}

In this section we explore properties associated with the support and posterior consistency of \eqref{Likelihood.GMM} under \eqref{Prior.Pi.NRep}--\eqref{Prior.Lambda.NRep}. 
These results are based on derivations found in \cite{NIPS2012_4589}. However, we highlight extensions and generalizations that we develop here. Consider for $k\in\N$ the 
family of probability densities $\F_{k}=\{f(\,\,\cdot\,\,;\boldsymbol{\xi}_{k}):\boldsymbol{\xi}_{k}\in\boldsymbol{\Theta}_{k}\}$, where 
$\boldsymbol{\xi}_{k}=\boldsymbol{\pi}_{k,1}\times\boldsymbol{\theta}_{k,1}\times\{\lambda\}=(\pi_{1},\ldots,\pi_{k})\times(\theta_{1},\ldots,\theta_{k})\times\{\lambda\}$,
$\boldsymbol{\Theta}_{k}=\Delta_{k-1}\times\R^{1}_{k}\times(0,\infty)$ and
\begin{equation*}
  f(\,\,\cdot\,\,;\boldsymbol{\xi}_{k})=\sum_{j=1}^{k}\pi_{j}\mathrm{N}(\,\,\cdot\,\,;\theta_{j},\lambda).
\end{equation*}
Let $B_{p}(\boldsymbol{x},r)$ with $\boldsymbol{x}\in\R^{1}_{k}$ and $r\in(0,\infty)$ denote an open ball centered on $\boldsymbol{x}$, and with radius $r$, and 
$D_{p}(\boldsymbol{x},r)$ its closure relative to the Euclidean $L_{p}$-metric ($p\geq1$) on $\R^{1}_{k}$.

The following four conditions will be assumed to prove the results stated afterwards.
\begin{itemize}
  \item [B1.] The true data generating density $f_{0}(\,\,\cdot\,\,;\boldsymbol{\xi}^{0}_{k_{0}})$ belongs to $\F_{k_{0}}$ for some fixed $k_{0}\geq2$, where
  $\boldsymbol{\xi}^{0}_{k_{0}}=\boldsymbol{\pi}^{0}_{k_{0},1}\times\boldsymbol{\theta}^{0}_{k_{0},1}\times\{\lambda_{0}\}=
  (\pi^{0}_{1},\ldots,\pi^{0}_{k_{0}})\times(\theta^{0}_{1},\ldots,\theta^{0}_{k_{0}})\times\{\lambda_{0}\}$.

  \item [B2.] The true locations $\theta^{0}_{1},\ldots,\theta^{0}_{k_{0}}$ satisfy $\min(|\theta^{0}_{r}-\theta^{0}_{s}|:r\neq s\in[k_{0}])\geq v$ for some $v>0$.

  \item [B3.] The number of components $k\in\N$ follows a discrete distribution $\kappa$ on the measurable space $(\N,2^{\N})$ such that $\kappa(k_{0})>0$.

  \item [B4.] For $k\geq2$ we have $\boldsymbol{\xi}_{k}\sim\mathrm{Dir}(k^{-1}\mathbf{1}_{k})\times\NRep_{k,1}(\mu,\sigma^{2},\tau)\times\mathrm{IG}(a,b)$. In the case that 
  $k=1$, $\boldsymbol{\xi}_{k}\sim\delta_{1}\times\mathrm{N}(\mu,\sigma^{2})\times\mathrm{IG}(a,b)$ with $\delta_{1}$ a Dirac measure centred on 1. In both scenarios 
  $\mu\in\R$ and $\sigma^{2},\tau,a,b\in(0,\infty)$ are fixed values.
\end{itemize}
Condition B2 requires that the true locations are separated by a minimum (Euclidian) distance $v$, which favors disperse mixture component centroids within the range of the 
response. For condition B4, the sequence $\{\boldsymbol{\xi}_{k}:k\in\N\}$ can be constructed (via the Kolmogorov's Extension Theorem) in a way that the elements are 
mutually independent upon adding to each $\boldsymbol{\Theta}_{k}$ an appropriate $\sigma$-algebra. This guarantees the existence of a prior distribution $\Pi$ defined on 
$\F=\bigcup_{k=1}^{\infty}\F_{k}$ which correspondingly connects the elements of $\F$ with $\boldsymbol{\xi}=\prod_{k=1}^{\infty}\boldsymbol{\xi}_{k}$. To calculate 
probabilities with respect to $\Pi$, the following stochastic representation will be useful
\begin{equation}
  \boldsymbol{\xi}\mid K=k\sim\boldsymbol{\xi}_{k},\qquad K\sim\kappa.
  \label{Distribution.Xi.given.K}
\end{equation}
Our study of the support of $\Pi$ employs the Kullback-Leibler (KL) divergence to measure the similarity between probability distributions. We will say that 
$f_{0}\in\F_{k_{0}}$ belongs to the KL support with respect to $\Pi$ if, for all $\varepsilon>0$
\begin{equation}
  \Pi\Bigg\{\Bigg(f\in\F:
  \int_{\R}\log\Bigg\{\frac{f_{0}(x;\boldsymbol{\xi}^{0}_{k_{0}})}{f(x;\boldsymbol{\xi}_{\star})}\Bigg\}f_{0}(x;\boldsymbol{\xi}^{0}_{k_{0}})\diff x<\varepsilon
  \Bigg)\Bigg\}>0,
  \label{Definition.Kullback.Leibler.Support}
\end{equation}
where $\boldsymbol{\xi}_{\star}\in\bigcup_{k=1}^{\infty}\boldsymbol{\Theta}_{k}$. Condition \eqref{Definition.Kullback.Leibler.Support} can be understood as $\Pi$'s ability 
to assign positive mass to arbitrarily small neighborhoods around the true density $f_{0}$. A fundamental step to proving that $f_{0}$ lies in the KL support of $\Pi$ is 
based on the following Lemmas.
\begin{Lem}\label{Dominated.Convergence}
  Under condition $\mathrm{B1}$, let $\varepsilon>0$. Then there exists $\delta>0$ such that
  \begin{equation*}
    \int_{\R}\log\Bigg\{\frac{f_{0}(x;\boldsymbol{\xi}^{0}_{k_{0}})}{f(x;\boldsymbol{\xi}_{k_{0}})}\Bigg\}f_{0}(x;\boldsymbol{\xi}^{0}_{k_{0}})\diff x<\varepsilon
  \end{equation*}
  for all $\boldsymbol{\xi}_{k_{0}}\in B_{1}(\boldsymbol{\theta}^{0}_{k_{0},1},\delta)\times B_{1}(\boldsymbol{\pi}^{0}_{k_{0},1},\delta)\times
  (\lambda_{0}-\delta,\lambda_{0}+\delta)$.
\end{Lem}
\begin{Lem}\label{Repulsive.Probability.Property}
  Assume condition $\mathrm{B2}$ and let $\boldsymbol{\theta}_{k_{0},1}\sim\NRep_{k_{0},1}(\mu,\sigma^{2},\tau)$. Then there exists $\delta_{0}>0$ such that
  \begin{equation*}
    \Pb\{\boldsymbol{\theta}_{k_{0},1}\in B_{1}(\boldsymbol{\theta}^{0}_{k_{0},1},\delta)\}>0.
  \end{equation*}
  for all $\delta\in(0,\delta_{0}]$. This result remains valid even when replacing $B_{1}(\boldsymbol{\theta}^{0}_{k_{0},1},\delta)$ with 
  $D_{1}(\boldsymbol{\theta}^{0}_{k_{0},1},\delta)$.
\end{Lem}
\noindent Using Lemmas \ref{Dominated.Convergence} and \ref{Repulsive.Probability.Property} we are able to prove the following Proposition.
\begin{Prop}\label{Kullback.Leibler.Support}
  Assume that conditions $\mathrm{B1}$--$\mathrm{B4}$ hold. Then $f_{0}$ belongs to the KL support of $\Pi$.
\end{Prop}
We next study the rate of convergence of the posterior distribution corresponding to a particular prior distribution (under suitable regularity conditions). To do this, we 
will use arguments that are similar to those employed in Theorem 3.1 of \cite{Scricciolo:2011}, to show that the posterior rates derived there are the same here when 
considering univariate Gaussian Mixture Models and cluster-location parameters that follow condition B4. First, we need the following two Lemmas.
\begin{Lem}\label{Repulsive.Probability.Inequality}
  For each $k\geq2$ the coordinates of $\boldsymbol{\theta}_{k,1}\sim\NRep_{k,1}(\mu,\sigma^{2},\tau)$ share the same functional form. Moreover, there exists 
  $\gamma\in(0,\infty)$ such that
  \begin{equation*}
  \Pb(|\theta_{i}|>t)\leq\frac{2}{(2\pi)^{1/2}}\frac{c_{k-1}}{c_{k}}\sigma(|\mu|+1)^{-1}\exp\Big\{-(4\sigma^{2})^{-1}t^{2}\Big\}
  \end{equation*}
  for all $t\in[\gamma,\infty)$ and $i\in[k]$. Here, $c_{k}=c_{k,1}$ is the normalizing constant of $\NRep_{k,1}(\mu,\sigma^{2},\tau)$ with $c_{1}=1$.
\end{Lem}
\begin{Lem}\label{Repulsive.Constant.Property}
  The sequence $\{c_{k}:k\in\N\}$ defined in Lemma \ref{Repulsive.Probability.Inequality} satisfies
  \begin{equation*}
    0<\frac{c_{k-1}}{c_{k}}\leq A_{1}\exp(A_{2}k)
  \end{equation*}
  for all $k\in\N$ $(k\geq2)$ and some constants $A_{1},A_{2}\in(0,\infty)$.
\end{Lem}
These results permit us to adapt certain arguments found in \cite{Scricciolo:2011} that are applicable when the location parameters of each mixture component are independent 
and follow a common distribution that is absolutely continuous with respect to the Lebesgue measure, whose support is $\R$ and with tails that decay exponentially. Using
Lemmas \ref{Repulsive.Probability.Inequality} and \ref{Repulsive.Constant.Property}, we now state the following
\begin{Prop}\label{Posterior.Convergence.Rate}
  Assume that conditions $\mathrm{B1}$, $\mathrm{B2}$ and $\mathrm{B4}$ hold. Replace condition $\mathrm{B3}$ with:
  \begin{itemize}
    \item [$\mathrm{B3'}$.] There exists $B_{1}\in(0,\infty)$ such that for all $k\in\N$, $0<\kappa(k)\leq B_{1}\exp\{-B_{2}k\}$, where $B_{2}>A_{2}$ and 
    $A_{2}\in(0,\infty)$ is given by Lemma~\ref{Repulsive.Constant.Property}.
  \end{itemize}
  Then, the posterior rate of convergence relative to the Hellinger metric is $\varepsilon_{n}=n^{-1/2}\log(n)$.
\end{Prop}


\subsection{Sampling From $\NRep_{k,d}(\boldsymbol{\mu},\boldsymbol{\Sigma},\tau)$}\label{Computational.Implementation}

Here we describe an algorithm that can be used to sample from $\NRep_{k,d}(\boldsymbol{\mu},\boldsymbol{\Sigma},\tau)$. Upon introducing component labels, sampling 
marginally from the joint posterior distribution of $\boldsymbol{\theta}_{k,d}$, $\boldsymbol{\Lambda}_{k,d}$, $\boldsymbol{\pi}_{k,1}$ and $\boldsymbol{z}_{n,1}$ can be 
done with a Gibbs sampler. However, the full conditionals of each coordinate of $\boldsymbol{\theta}_{k,d}$ are not conjugate but they are all functionally similar. Because 
of this, evaluating these densities is computationally cheap making it straightforward to carry out sampling from $\NRep_{k,d}(\boldsymbol{\mu},\boldsymbol{\Sigma},\tau)$ 
via a Metropolis--Hastings step inside the Gibbs sampling scheme. In Appendix~\ref{Algorith.RGMM} we detail the entire MCMC algorithm (Algorithm RGMM), but here we focus 
on the nonstandard aspects.

To begin, the distribution $(\boldsymbol{\theta}_{k,d}\mid\cdots)$ is given by
\begin{align*}
  (\boldsymbol{\theta}_{k,d}\mid\cdots)&\propto
  \Bigg\{\prod_{j=1}^{k}\mathrm{N}_{d}(\boldsymbol{\theta}_{j};\boldsymbol{\mu}_{j},\boldsymbol{\Sigma}_{j})\Bigg\}
  \prod_{r<s}^{k}
  [1-\exp\{-0.5\tau^{-1}(\boldsymbol{\theta}_{r}-\boldsymbol{\theta}_{s})^{\top}\boldsymbol{\Sigma}^{-1}(\boldsymbol{\theta}_{r}-\boldsymbol{\theta}_{s})\}]
\end{align*}
where
$\boldsymbol{\mu}_{j}=\boldsymbol{\Sigma}_{j}(\boldsymbol{\Sigma}^{-1}\boldsymbol{\mu}+\boldsymbol{\Lambda}^{-1}_{j}\boldsymbol{s}_{j})$,
$\boldsymbol{s}_{j}=\sum_{i=1}^{n}\I_{\{j\}}(z_{i})\boldsymbol{y}_{i}$, $\boldsymbol{\Sigma}_{j}=(\boldsymbol{\Sigma}^{-1}+n_{j}\boldsymbol{\Lambda}^{-1}_{j})^{-1}$ 
and $n_{j}=\card(i\in[n]:z_{i}=j)$. Now, the complete conditional distributions $(\boldsymbol{\theta}_{j}\mid\boldsymbol{\theta}_{-j},\cdots)$ for $j\in[k]$ and
$\boldsymbol{\theta}_{-j}=(\boldsymbol{\theta}_{l}:l\neq j)\in\R^{d}_{k-1}$, have the following form
\begin{equation*}
  f(\boldsymbol{\theta}_{j}\mid\boldsymbol{\theta}_{-j},\cdots)\propto
  \mathrm{N}_{d}(\boldsymbol{\theta}_{j};\boldsymbol{\mu}_{j},\boldsymbol{\Sigma}_{j})
  \prod_{l\neq j}^{k}
  [1-\exp\{-0.5\tau^{-1}(\boldsymbol{\theta}_{j}-\boldsymbol{\theta}_{l})^{\top}\boldsymbol{\Sigma}^{-1}(\boldsymbol{\theta}_{j}-\boldsymbol{\theta}_{l})\}].
\end{equation*}
 
The following pseudo-code describes how to sample from $f(\boldsymbol{\theta}_{k,d}\mid\cdots)$ by way of $(\boldsymbol{\theta}_{j}\mid\boldsymbol{\theta}_{-j},\cdots)$ via 
a random walk Metropolis--Hastings step within a Gibbs sampler:
\begin{enumerate}
  \item Let $\boldsymbol{\theta}^{(0)}_{k,d}=(\boldsymbol{\theta}^{(0)}_{1},\ldots,\boldsymbol{\theta}^{(0)}_{k})\in\R^{d}_{k}$ be the actual state for
  $\boldsymbol{\theta}_{k,d}$.

  \item For $j=1,\ldots,k$:
  \begin{enumerate}
    \item Generate a candidate $\boldsymbol{\theta}^{(1)}_{j}$ from $\mathrm{N}_{d}(\boldsymbol{\theta}^{(0)}_{j},\boldsymbol{\Gamma}_{j})$ with 
    $\boldsymbol{\Gamma}_{j}\in\Sy^{d}$.

    \item Set $\boldsymbol{\theta}^{(0)}_{j}=\boldsymbol{\theta}^{(1)}_{j}$ with probability $\min(1,\beta_{j})$, where
    \begin{equation*}
      \beta_{j}=
      \frac{\mathrm{N}_{d}(\boldsymbol{\theta}^{(1)}_{j};\boldsymbol{\mu}_{j},\boldsymbol{\Sigma}_{j})}
      {\mathrm{N}_{d}(\boldsymbol{\theta}^{(0)}_{j};\boldsymbol{\mu}_{j},\boldsymbol{\Sigma}_{j})}
      \prod_{l\neq j}^{k}
      \Bigg[
      \frac
      {
      1-\exp\{-0.5
      \tau^{-1}(\boldsymbol{\theta}^{(1)}_{j}-\boldsymbol{\theta}^{(0)}_{l})^{\top}\boldsymbol{\Sigma}^{-1}(\boldsymbol{\theta}^{(1)}_{j}-\boldsymbol{\theta}^{(0)}_{l})\}
      }
      {1-\exp\{-0.5
      \tau^{-1}(\boldsymbol{\theta}^{(0)}_{j}-\boldsymbol{\theta}^{(0)}_{l})^{\top}\boldsymbol{\Sigma}^{-1}(\boldsymbol{\theta}^{(0)}_{j}-\boldsymbol{\theta}^{(0)}_{l})\}
      }
      \Bigg].
    \end{equation*}
 \end{enumerate}
\end{enumerate}

The selection of $\boldsymbol{\Gamma}_{j}$ can be carried out using adaptive MCMC methods \citep{RobertsRosenthal:2009} so that the acceptance rate of the 
Metropolis--Hastings algorithm is approximately 50\% within the burn-in period for each $j\in[k]$. One approach that works well for the RGMM is to take
\begin{equation}
  \boldsymbol{\Gamma}_{j}=
  \frac{1}{B}
  \sum_{t=1}^{B}\{\boldsymbol{\Sigma}^{-1}+n^{(t)}_{j}(\boldsymbol{\Lambda}^{(t)}_{j})^{-1}\}^{-1}:n^{(t)}_{j}=\card(i\in[n]:z^{(t)}_{i}=j),
  \label{Gamma.j.Specification}
\end{equation}
where $t\in[B]$ is the $t$th iteration of the burn-in period with length $B\in\N$.


\section{Simulation Study}\label{Simulation.Study}

To provide context regarding the proposed method's performance in density estimation, we conduct a small simulation study. In the simulation we compare density estimates 
from the RGMM to what is obtained using an i.i.d. Gaussian Mixture Model (GMM) and a Dirichlet Process Gaussian Mixture Model (DPMM). This is done by treating the following
as a data generating mechanism:
\begin{equation}
  y\sim f_{0}=0.3\mathrm{N}(-5,1.0^{2})+0.05\mathrm{N}(0,0.3^{2})+0.25\mathrm{N}(1,0.3^{2})+0.4\mathrm{N}(4,0.8^{2}).
  \label{Data.Generating.Mechanism}
\end{equation}
Using \eqref{Data.Generating.Mechanism} we simulate 100 data sets with sample sizes 500, 1000 and 5000. For each of these scenarios, we compare the following 4 models 
(abbreviated by M1, M2, M3 y M4) to estimate $f_{0}$: 
\begin{itemize}
  \item [$\mathrm{M1.}$] GMM corresponding to \eqref{NRep.Response.given.Cluster.Label}--\eqref{NRep.Cluster.Label} with prior distributions given by 
  \eqref{Prior.Pi.NRep}--\eqref{Prior.Lambda.NRep}, replacing \eqref{Prior.Theta.NRep} by 
  $\boldsymbol{\theta}_{1},\ldots,\boldsymbol{\theta}_{k}\stackrel{i.i.d.}{\sim}\textrm{N}_{d}(\boldsymbol{\mu},\boldsymbol{\Sigma})$. In this case:
  \begin{itemize}
    \item [$\bullet$] $k=10$, $d=1$, $\boldsymbol{\alpha}_{k,1}=10^{-1}\mathbf{1}_{10}$, $\boldsymbol{\mu}=0$, $\boldsymbol{\Sigma}=1$, $\boldsymbol{\Psi}=0.06$ and $\nu=5$.
  \end{itemize}
  We collected 10000 MCMC iterates after discarding the first 1000 as burn-in and thinning by 10.

  \item [$\mathrm{M2.}$] RGMM with $\tau=5.45$. This value came from employing the calibration criterion from Section~\ref{Parameter.Calibration} and setting $u=0.5$ and 
  $p=0.95$. The remaining prior parameters are:
  \begin{itemize}
    \item [$\bullet$] $k=10$, $d=1$, $\boldsymbol{\alpha}_{k,1}=10^{-1}\mathbf{1}_{10}$, $\boldsymbol{\mu}=0$, $\boldsymbol{\Sigma}=1$, $\tau=5.45$, $\boldsymbol{\Psi}=0.06$ 
    and $\nu=5$.
  \end{itemize}
  We collected 10000 MCMC iterates after discarding the first 5000 as burn-in and thinning by 20.

  \item [$\mathrm{M3.}$] RGMM with $\tau=17.17$. This value came from employing the calibration criterion from Section~\ref{Parameter.Calibration} and setting $u=0.2$ and 
  $p=0.95$. Since $\tau$ is bigger here than in M2, M3 has more repulsion than M2. The remaining prior parameters are the same as in M2:
  \begin{itemize}
    \item [$\bullet$] $k=10$, $d=1$, $\boldsymbol{\alpha}_{k,1}=10^{-1}\mathbf{1}_{10}$, $\boldsymbol{\mu}=0$, $\boldsymbol{\Sigma}=1$, $\tau=17.17$, 
    $\boldsymbol{\Psi}=0.06$ and $\nu=5$.
  \end{itemize}
  We collected 10000 MCMC iterates after discarding the first 5000 as burn-in and thinning by 20.

  \item [$\mathrm{M4.}$] DPMM given by:
  \begin{align}
    &\boldsymbol{y}_{i}\mid\boldsymbol{\mu}_{i},\boldsymbol{\Sigma}_{i}\stackrel{ind.}{\sim}\mathrm{N}_{d}(\boldsymbol{\mu}_{i},\boldsymbol{\Sigma}_{i})
    \label{DPM.Response.given.Cluster} \\
    &(\boldsymbol{\mu}_{i},\boldsymbol{\Sigma}_{i})\mid H\stackrel{i.i.d.}{\sim}H
    \label{DPM.Location.Scale.Cluster} \\
    &H\mid\alpha,H_{0}\sim\mathrm{DP}(\alpha,H_{0})
    \label{DPM.DP.Location.Scale}
  \end{align}
  where the baseline distribution $H_{0}$ is the conjugate Gaussian-Inverse Wishart
  \begin{equation}
    H_{0}(\boldsymbol{\mu},\boldsymbol{\Sigma})=
    \mathrm{N}_{d}(\boldsymbol{\mu};\boldsymbol{m}_{1},k_{0}^{-1}\boldsymbol{\Sigma})\,
    \mathrm{IW}_{d}(\boldsymbol{\Sigma};\boldsymbol{\Psi}_{1},\nu_{1}):\nu_{1}\in(0,\infty).
    \label{DPM.Baseline.Distribution}
  \end{equation}
  To complete the model specification given by \eqref{DPM.Response.given.Cluster}--\eqref{DPM.Baseline.Distribution}, the following independent hyperpriors are assumed:
  \begin{align}
    &\alpha\mid a_{0},b_{0}\sim\mathrm{G}(a_{0},b_{0}):a_{0},b_{0}\in(0,\infty)
    \label{DPM.Precision.Parameter} \\
    &\boldsymbol{m}_{1}\mid\boldsymbol{m}_{2},\mathbf{S}_{2}\sim\mathrm{N}_{d}(\boldsymbol{m}_{2},\mathbf{S}_{2}):\boldsymbol{m}_{2}\in\R^{d},\mathbf{S}_{2}\in\Sy^{d}
    \label{DPM.Hyperprior.Mean.Locations} \\
    &k_{0}\mid\tau_{1},\tau_{2}\sim\mathrm{G}(\tau_{1}/2,\tau_{2}/2):\tau_{1},\tau_{2}\in(0,\infty)
    \label{DPM.Hyperprior.Precision.Locations} \\
    &\boldsymbol{\Psi}_{1}\mid\boldsymbol{\Psi}_{2},\nu_{2}\sim\mathrm{IW}_{d}(\boldsymbol{\Psi}_{2},\nu_{2}):\boldsymbol{\Psi}_{2}\in\Sy^{d},\nu_{2}\in(0,\infty).
    \label{DPM.Hyperprior.Scale.Scales}
  \end{align}
  In the simulation study we set $d=1$. The selection of hyperparameters found in \eqref{DPM.Precision.Parameter}--\eqref{DPM.Hyperprior.Scale.Scales} was based on similar 
  strategies as outlined in \cite{escobar&west:95} which produced:
  \begin{itemize}
    \item [$\bullet$] $a_{0}=2$, $b_{0}=5$, $\nu_{1}=4$, $\nu_{2}=4$, $\boldsymbol{m}_{2}=0$, $\mathbf{S}_{2}=1$, $\boldsymbol{\Psi}_{2}=1$, $\tau_{1}=2.01$ and 
    $\tau_{2}=1.01$.
  \end{itemize}
  We collected 10000 MCMC iterates after discarding the first 1000 as burn-in and thinning by 10.
\end{itemize}
Models M2 and M3 were fit using the Algorithm RGMM which was implemented in $\verb"Fortran"$. For model M4, density estimates were obtained using the function 
\verb"DPdensity" which is available in the \verb"DPpackage" of \verb"R" \citep{DPpackage}.

To compare density estimation associated with the four procedures just detailed we employ the following metrics:
\begin{itemize}
  \item Log Pseudo Marginal Likelihood (LPML) (\citealt{WesJohnsonBook}) which is a model fit metric that takes into account model complexity. This was computed by first 
  estimating all the corresponding conditional predictive ordinates \citep{Gelfand:1992} using the method in \cite{Chen:2000}.

  \item Mean Square Error (MSE).

  \item $L_{1}$-metric between the estimated posterior predictive density and $f_{0}$.
\end{itemize}

Additionally, to explore how the repulsion influences model parsimony in terms of the number of occupied mixture components, we wecorded the following numeric indicators:
\begin{itemize}
  \item Average number of occupied mixture components.

  \item Standard deviation of the average number of occupied mixture components.
\end{itemize}

\begin{figure}[htb!]
  \begin{center}
  \begin{tabular}{ccc}
    \hspace{-1.5cm}
    \includegraphics[scale=0.40]{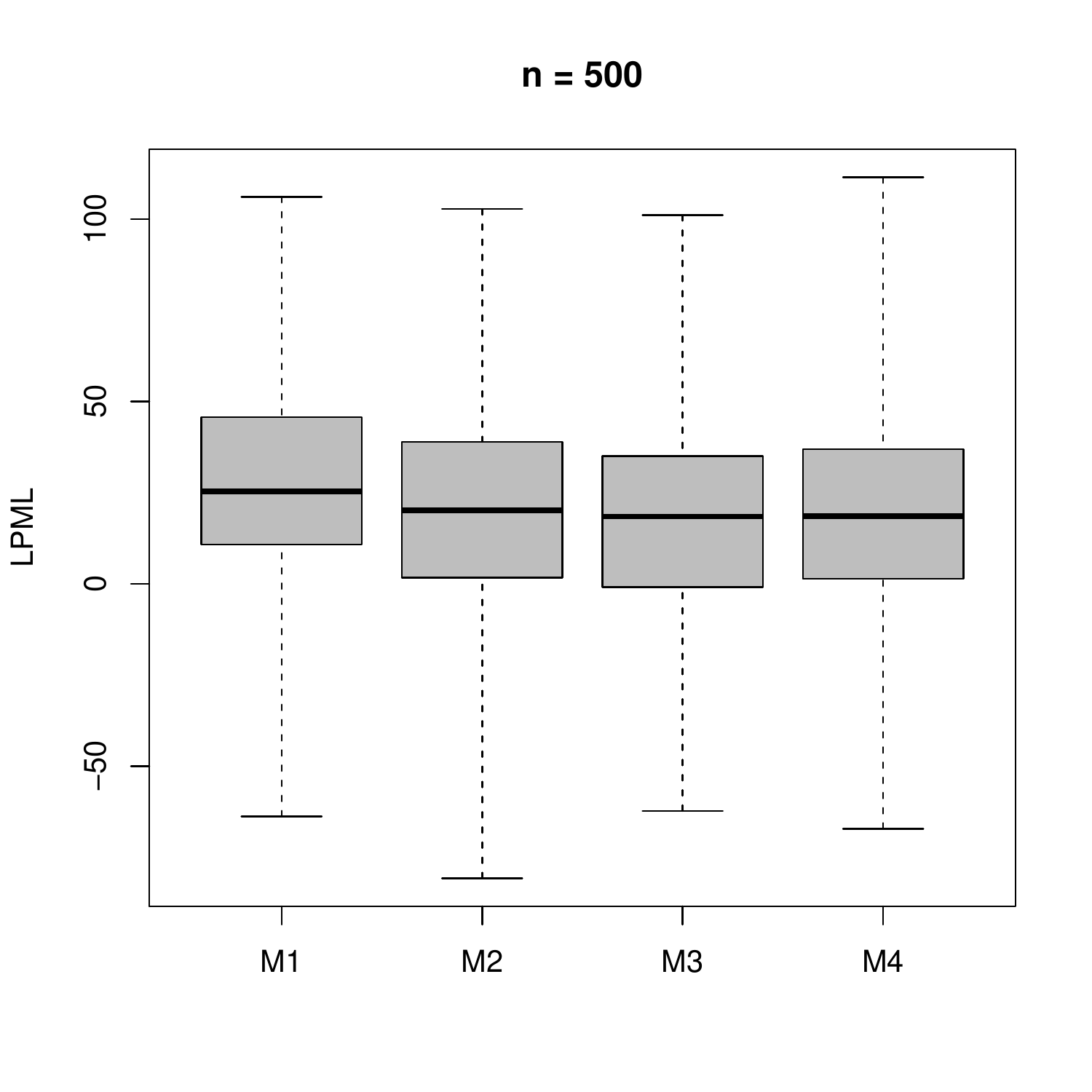} & \includegraphics[scale=0.40]{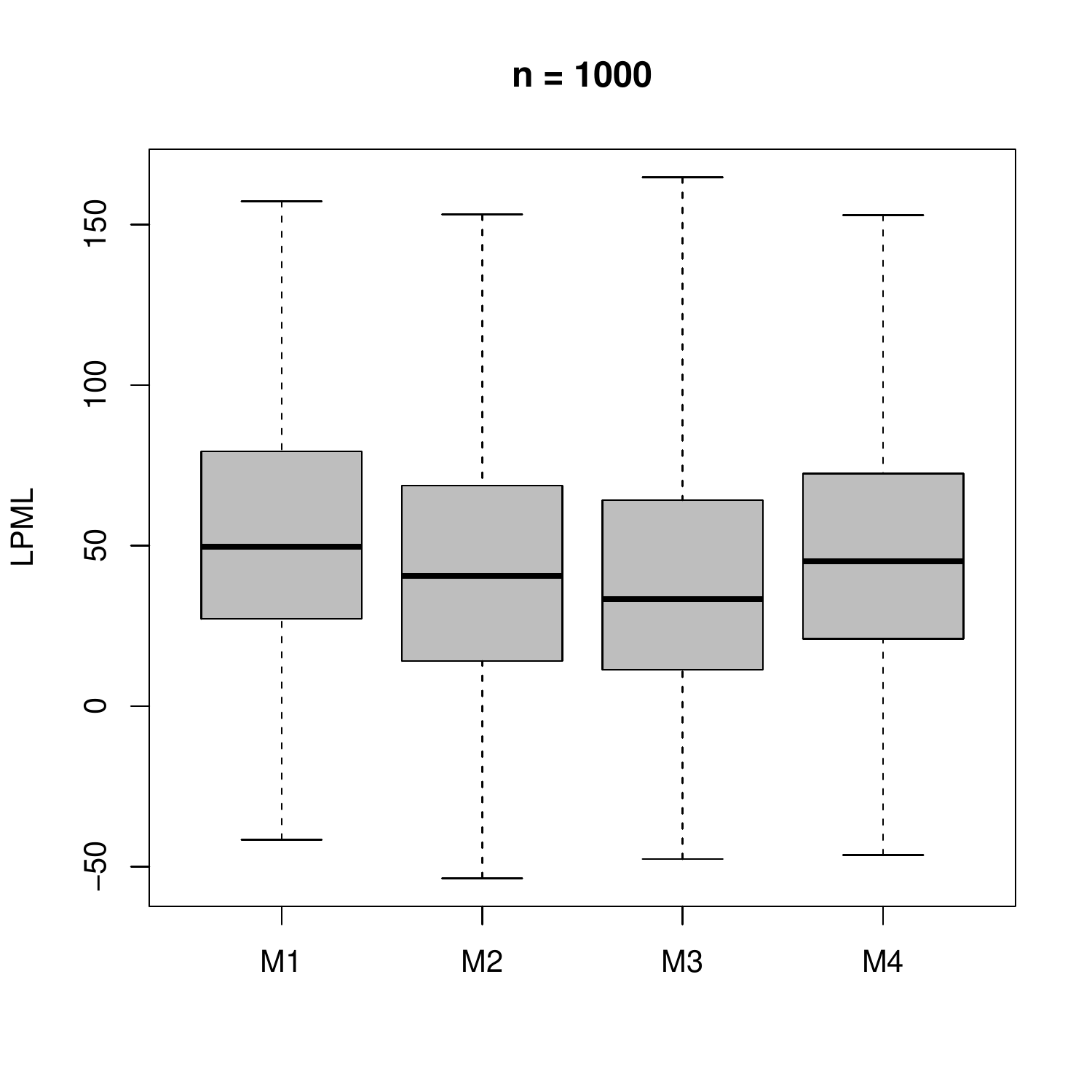} & \includegraphics[scale=0.40]{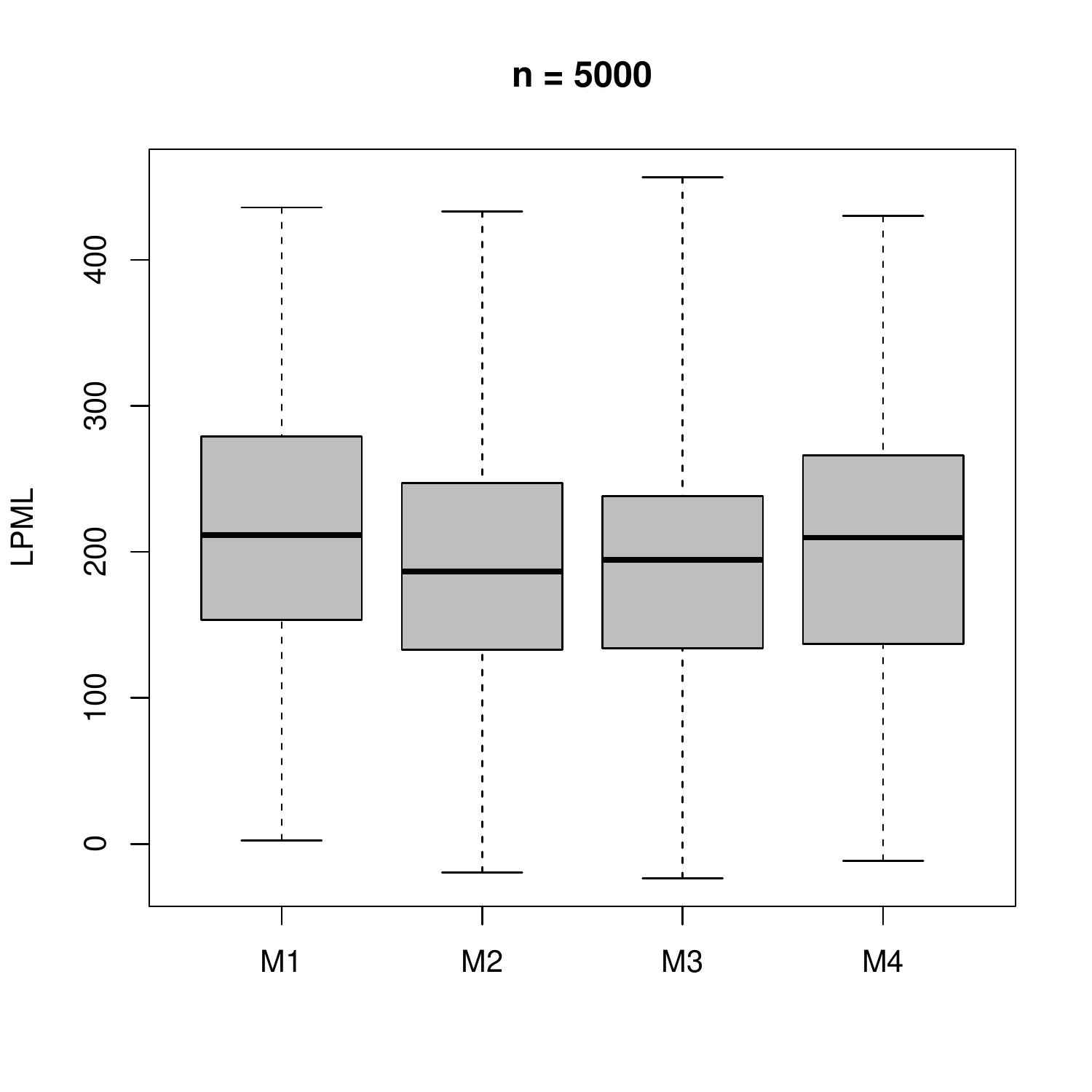}
  \end{tabular}
  \end{center}
  \caption{Boxplots that resume the behavior of LPML for each of the four models.}
  \label{Boxplots.Simulation.LPML}
\end{figure}

\begin{figure}[htb!]
  \begin{center}
  \begin{tabular}{ccc}
    \hspace{-1.5cm}
    \includegraphics[scale=0.40]{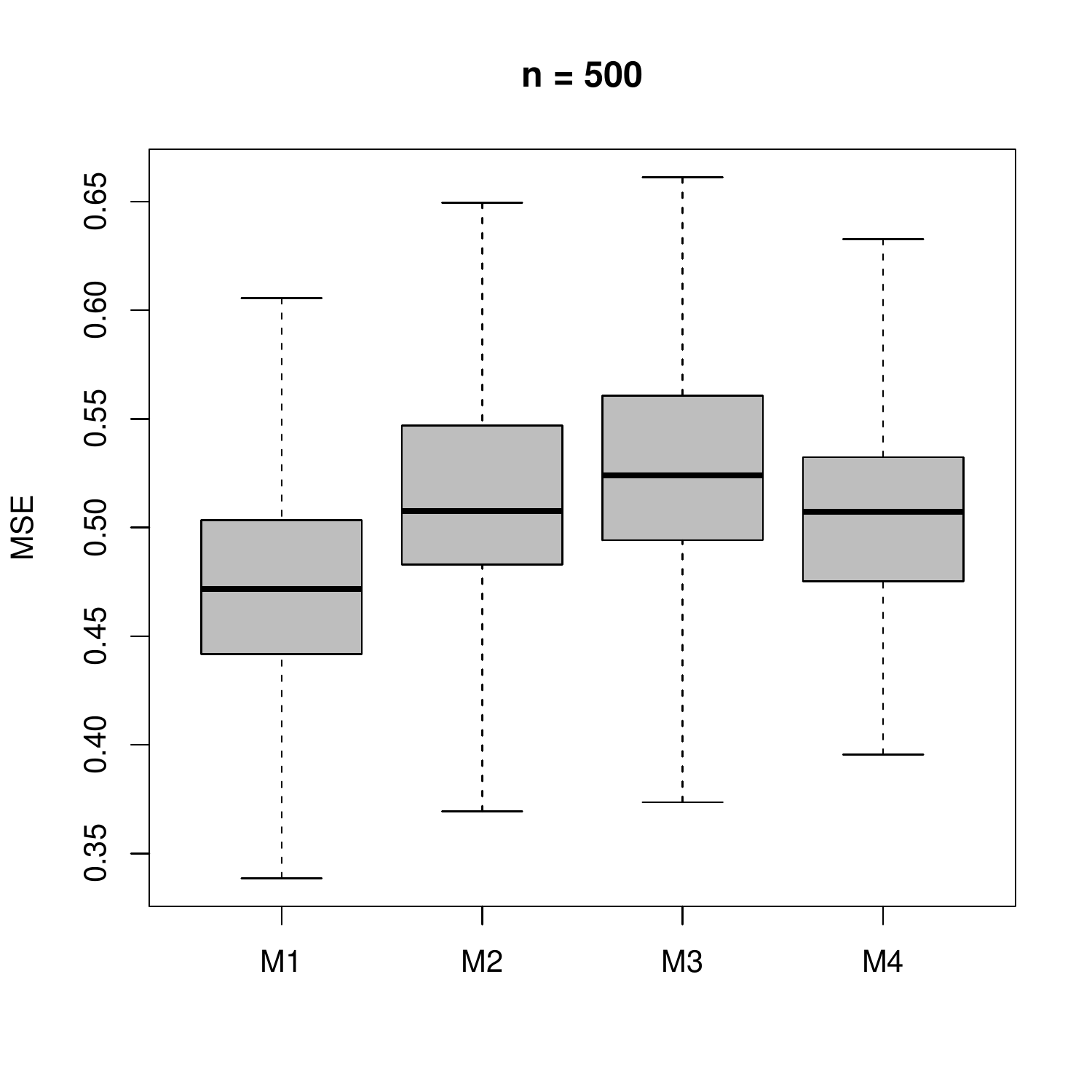} & \includegraphics[scale=0.40]{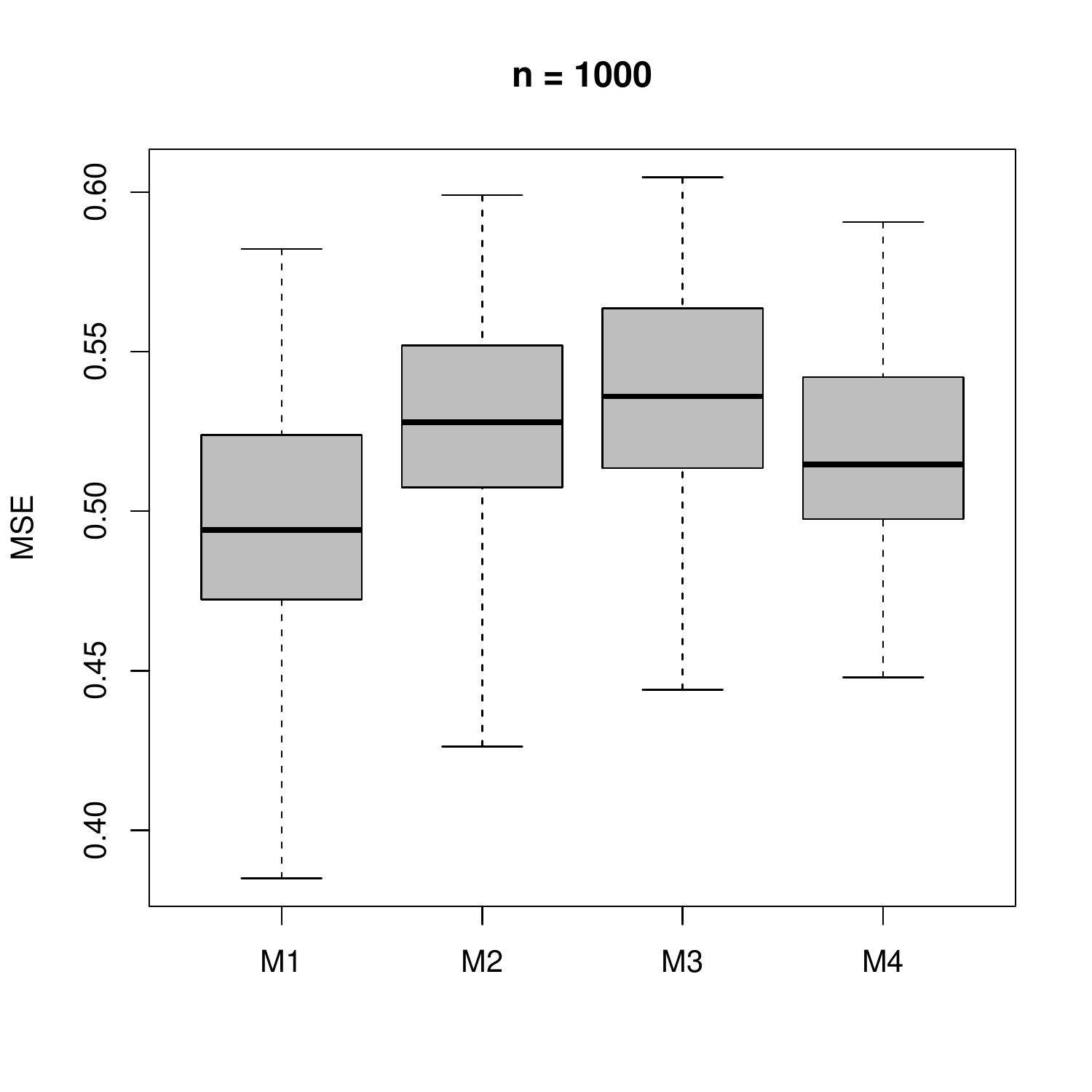} & \includegraphics[scale=0.40]{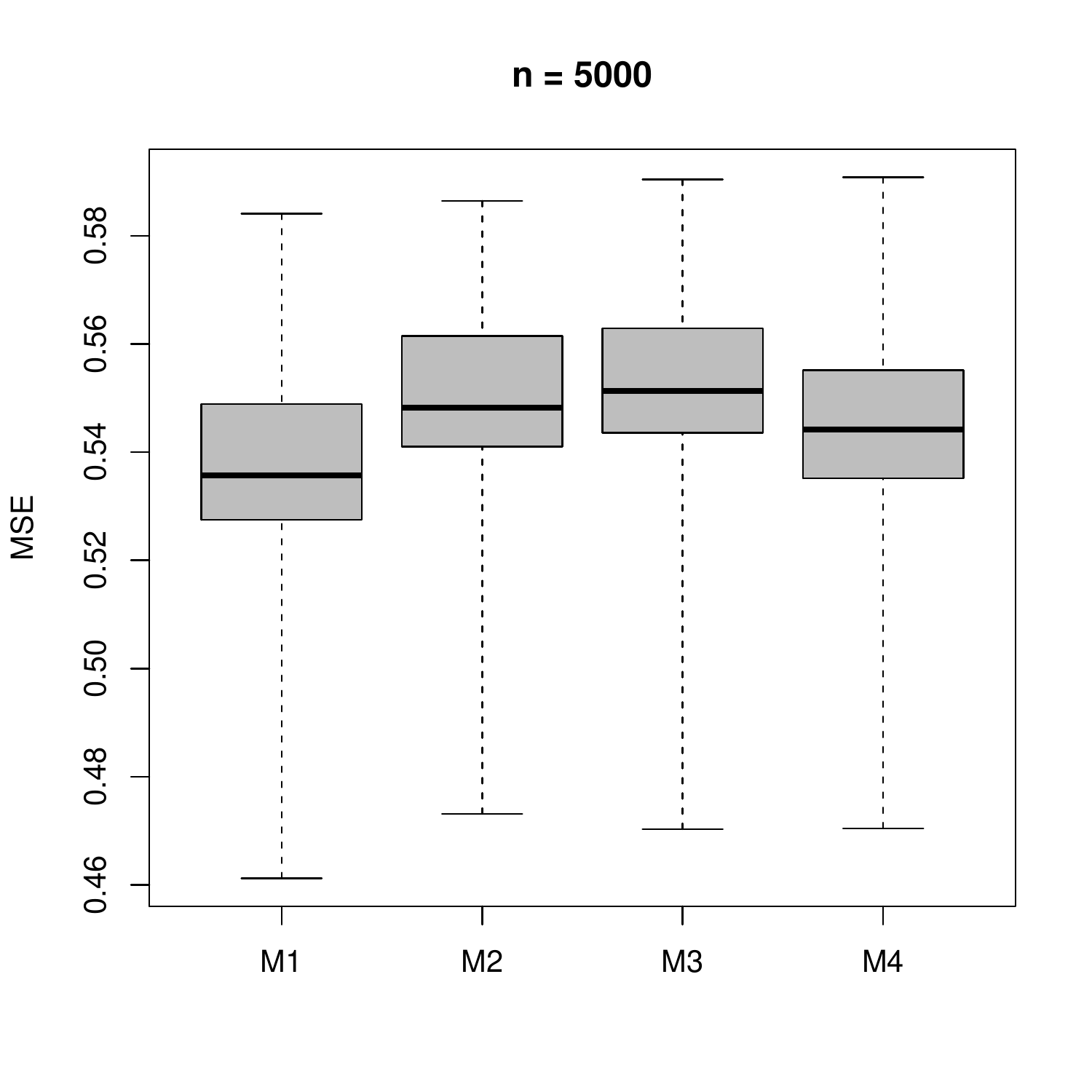}
  \end{tabular}
  \end{center}
  \caption{Boxplots that resume the behavior of MSE for each of the four models.}
  \label{Boxplots.Simulation.MSE}
\end{figure}

\begin{figure}[htb!]
  \begin{center}
  \begin{tabular}{ccc}
    \hspace{-1.5cm}
    \includegraphics[scale=0.40]{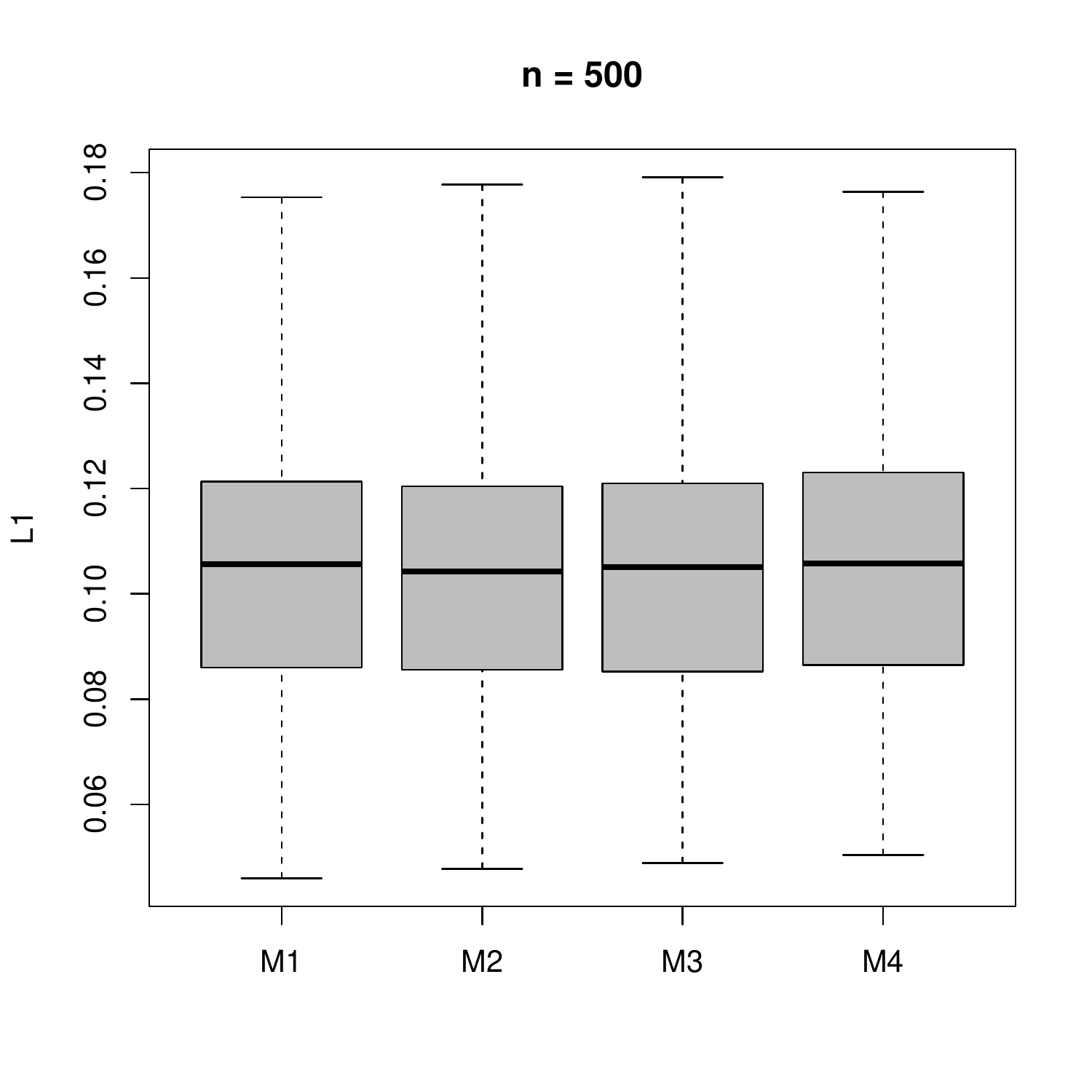} & \includegraphics[scale=0.40]{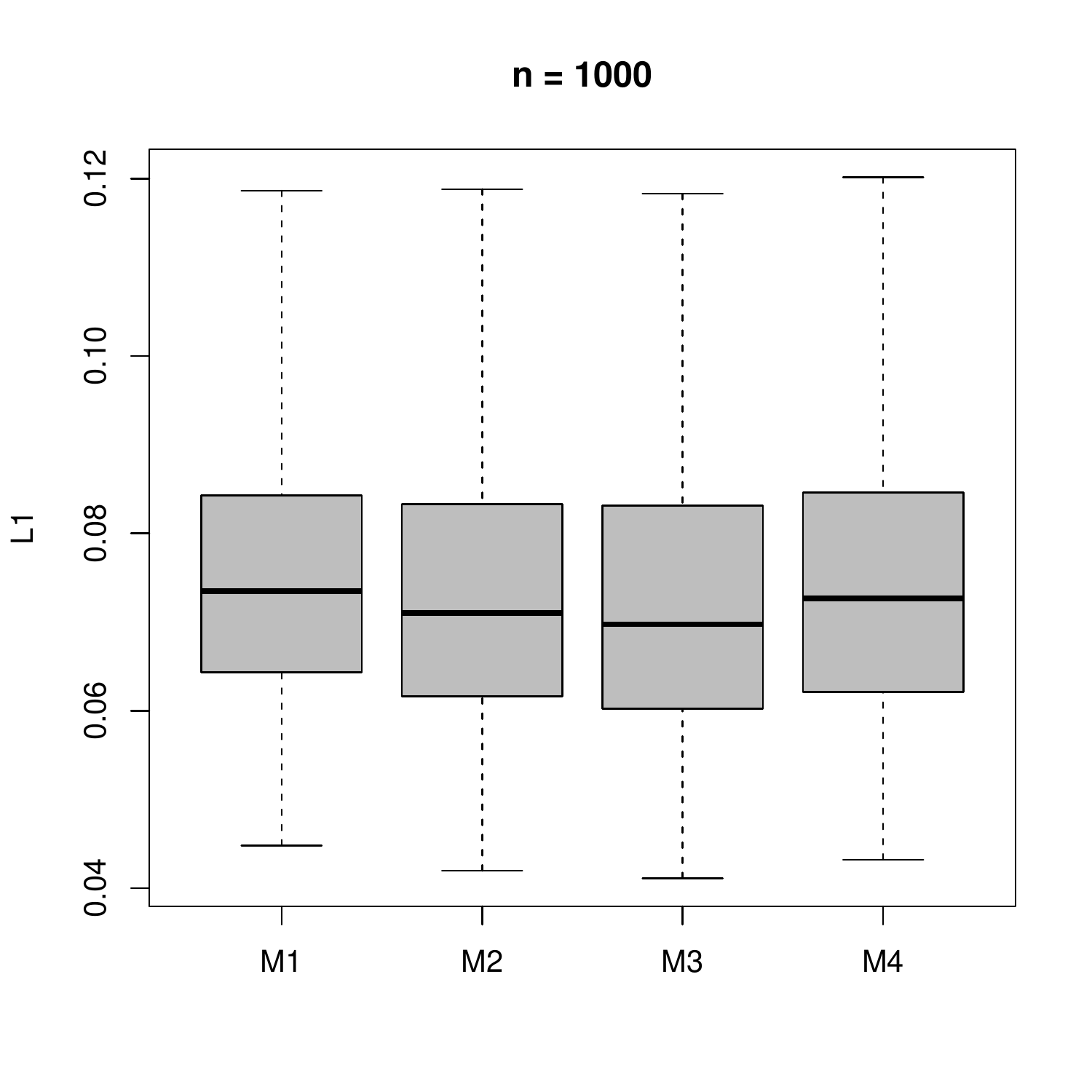} & \includegraphics[scale=0.40]{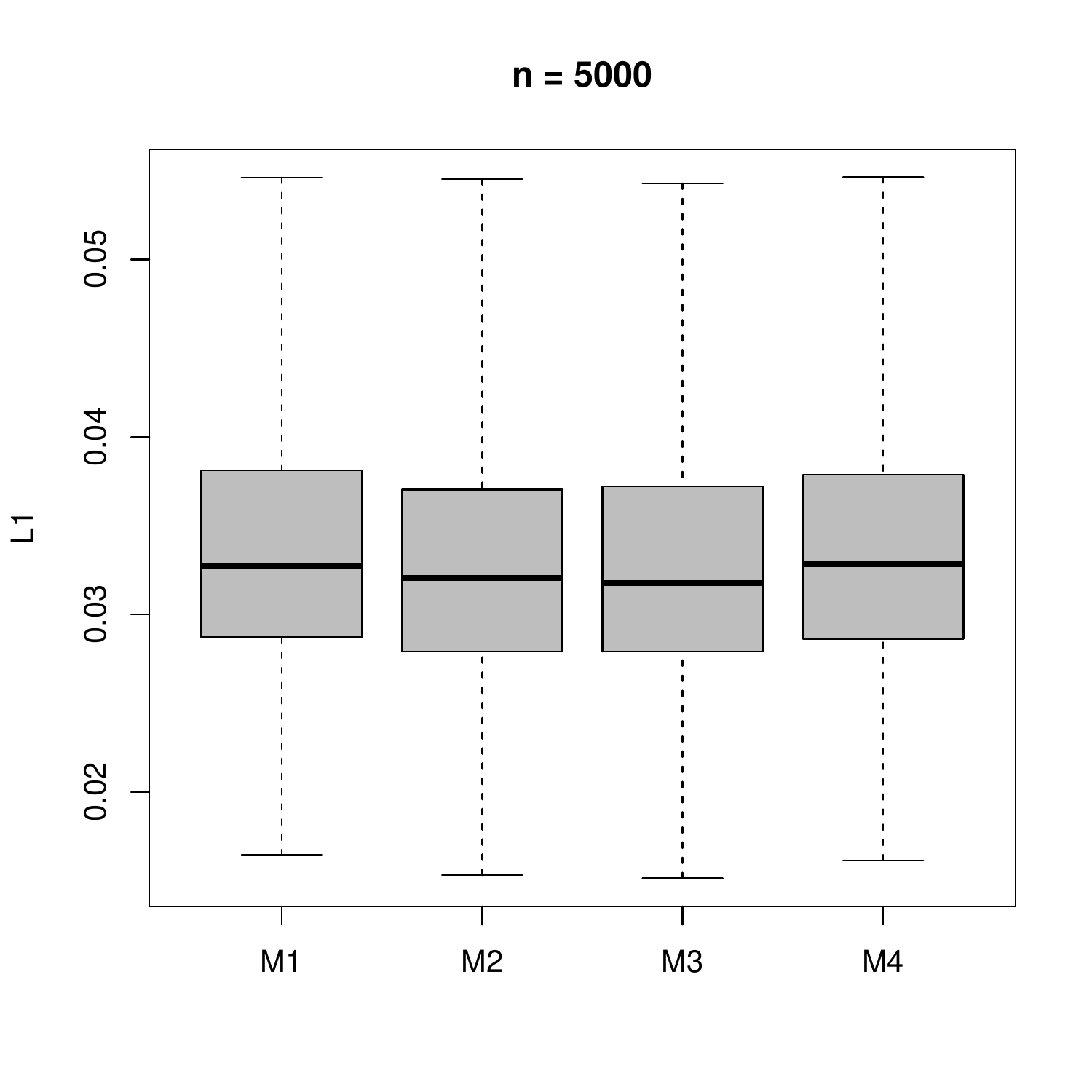}
  \end{tabular}
  \end{center}
  \caption{Boxplots that resume the behavior of $L_{1}$-metric for each of the four models.}
  \label{Boxplots.Simulation.L1}
\end{figure}

Figures \ref{Boxplots.Simulation.LPML}, \ref{Boxplots.Simulation.MSE} and \ref{Boxplots.Simulation.L1} contain side-by-side boxplots of the LPML, MSE and $L_{1}$-metric 
respectively as the sample size grows. Notice that trends seen here indicate that M1 and M4 tend to fit better, but M2 and M3 are very competitive with the advantage of 
being more parsimonious. In other words, very little model fit was sacrificed for the sake of parsimony.

\begin{figure}[htb!]
  \begin{center}
  \begin{tabular}{ccc}
    \hspace{-1.5cm}
    \includegraphics[scale=0.40]{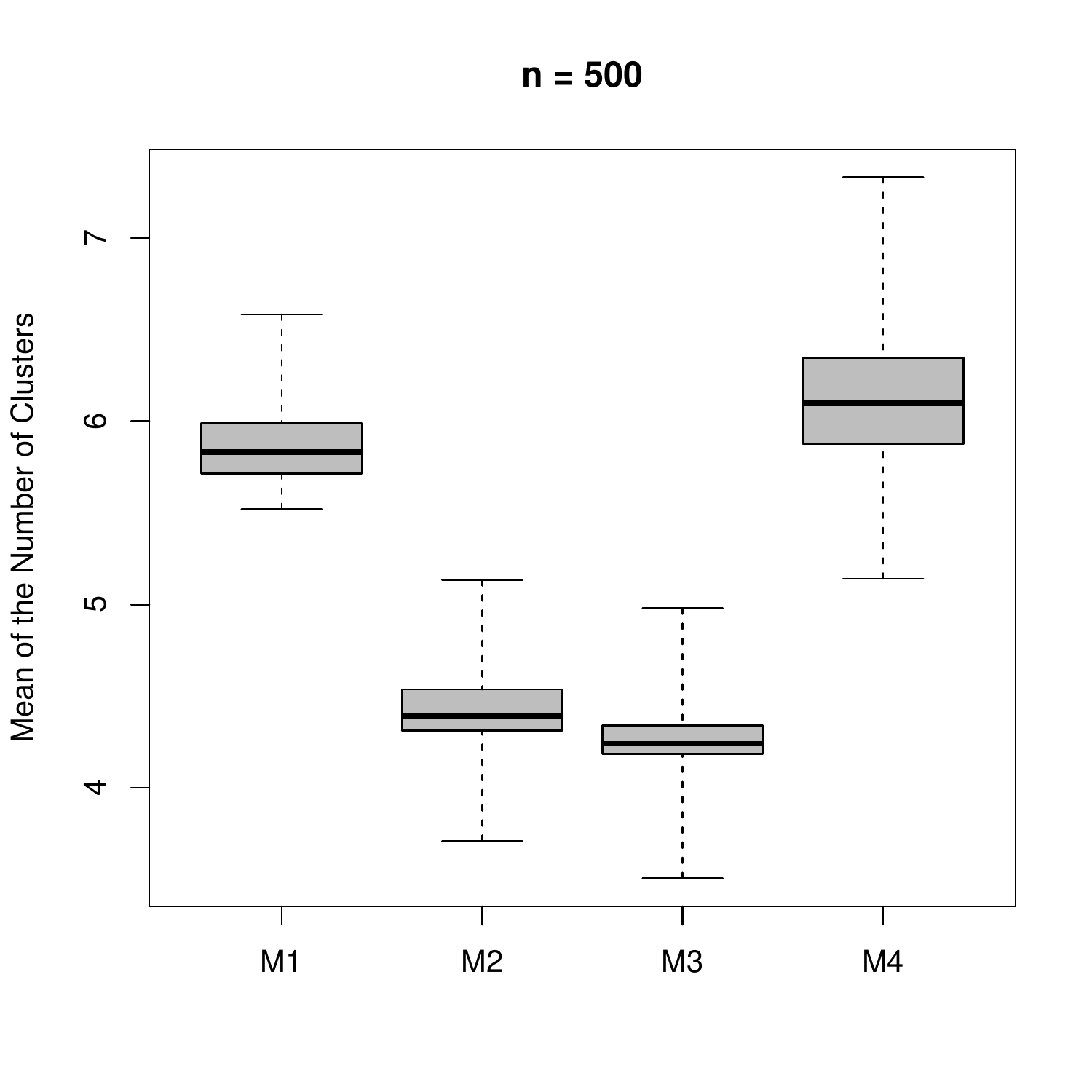} & \includegraphics[scale=0.40]{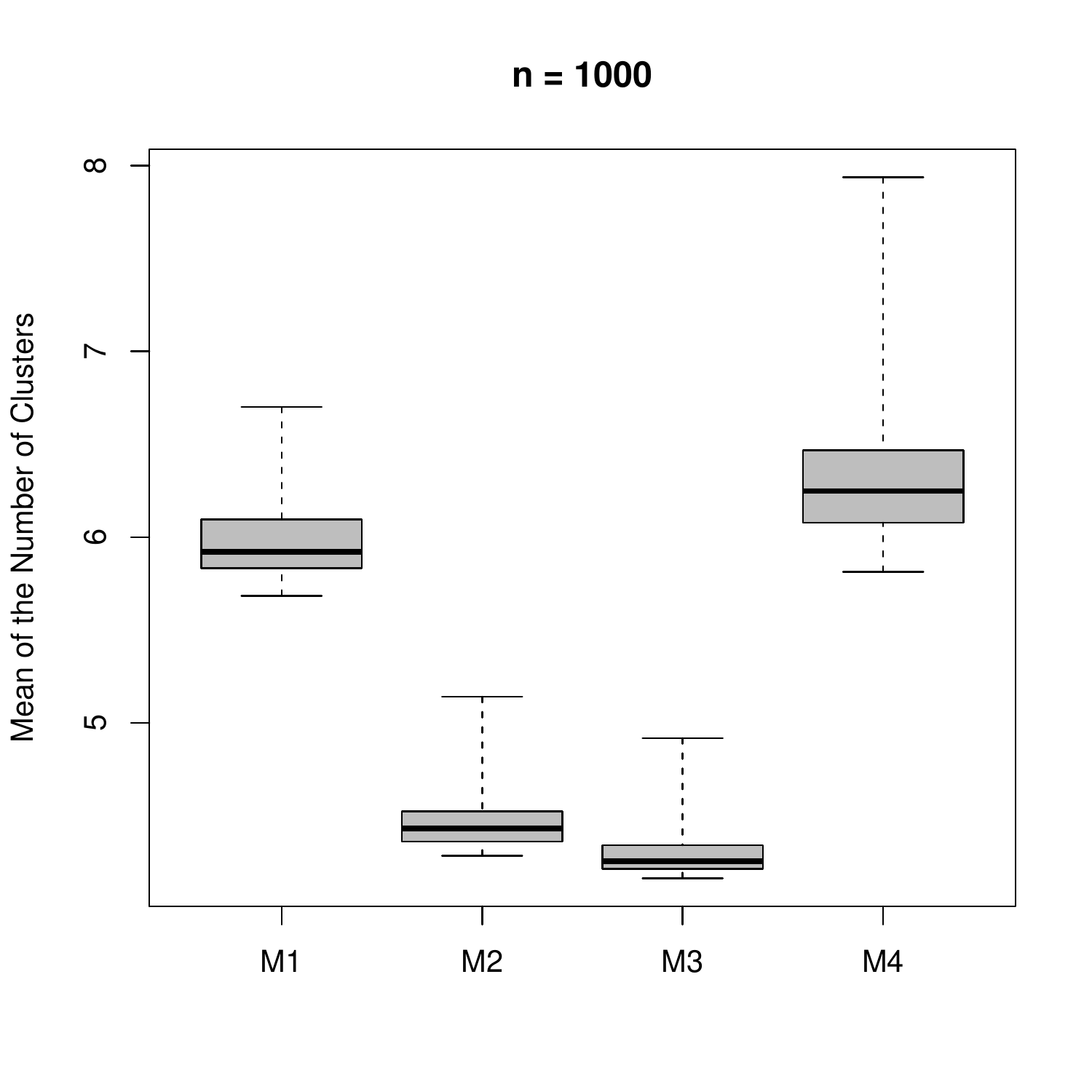} & \includegraphics[scale=0.40]{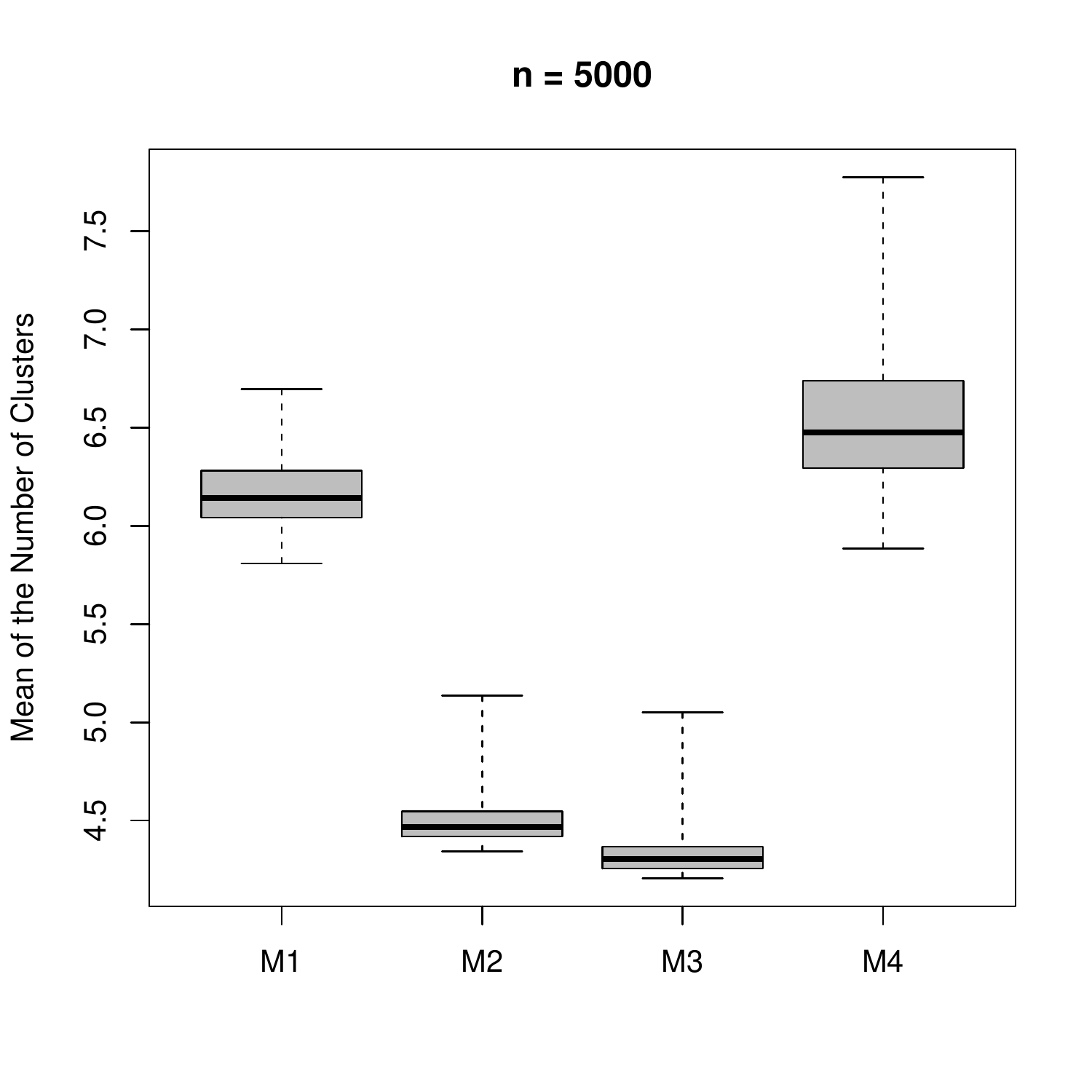}
  \end{tabular}
  \end{center}
  \caption{Side-by-side boxplots of the average number of occupied mixture components for each of the procedure.}
  \label{Boxplots.Simulation.PANOC}
\end{figure}

\begin{figure}[htb!]
  \begin{center}
  \begin{tabular}{ccc}
    \hspace{-1.5cm}
    \includegraphics[scale=0.40]{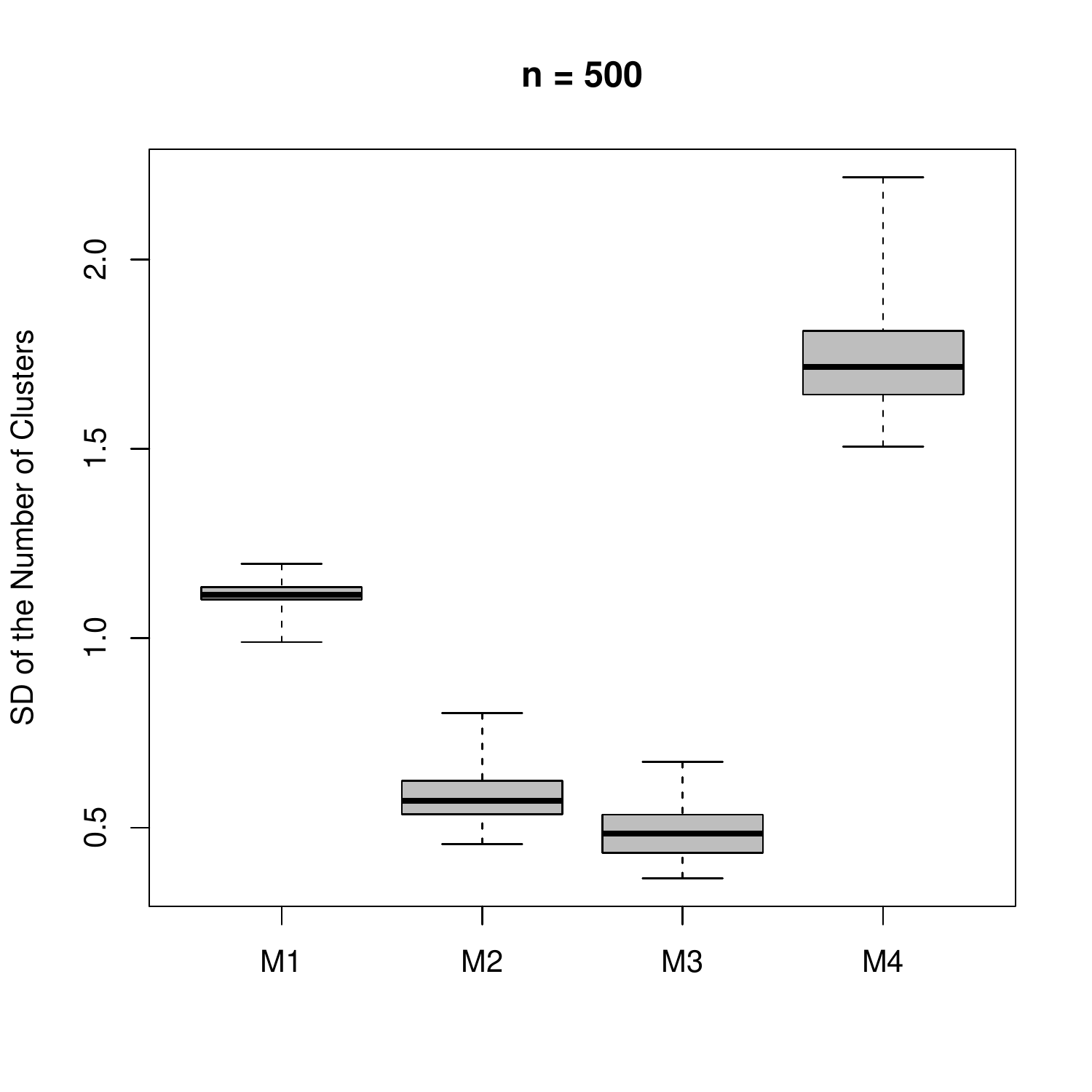} & \includegraphics[scale=0.40]{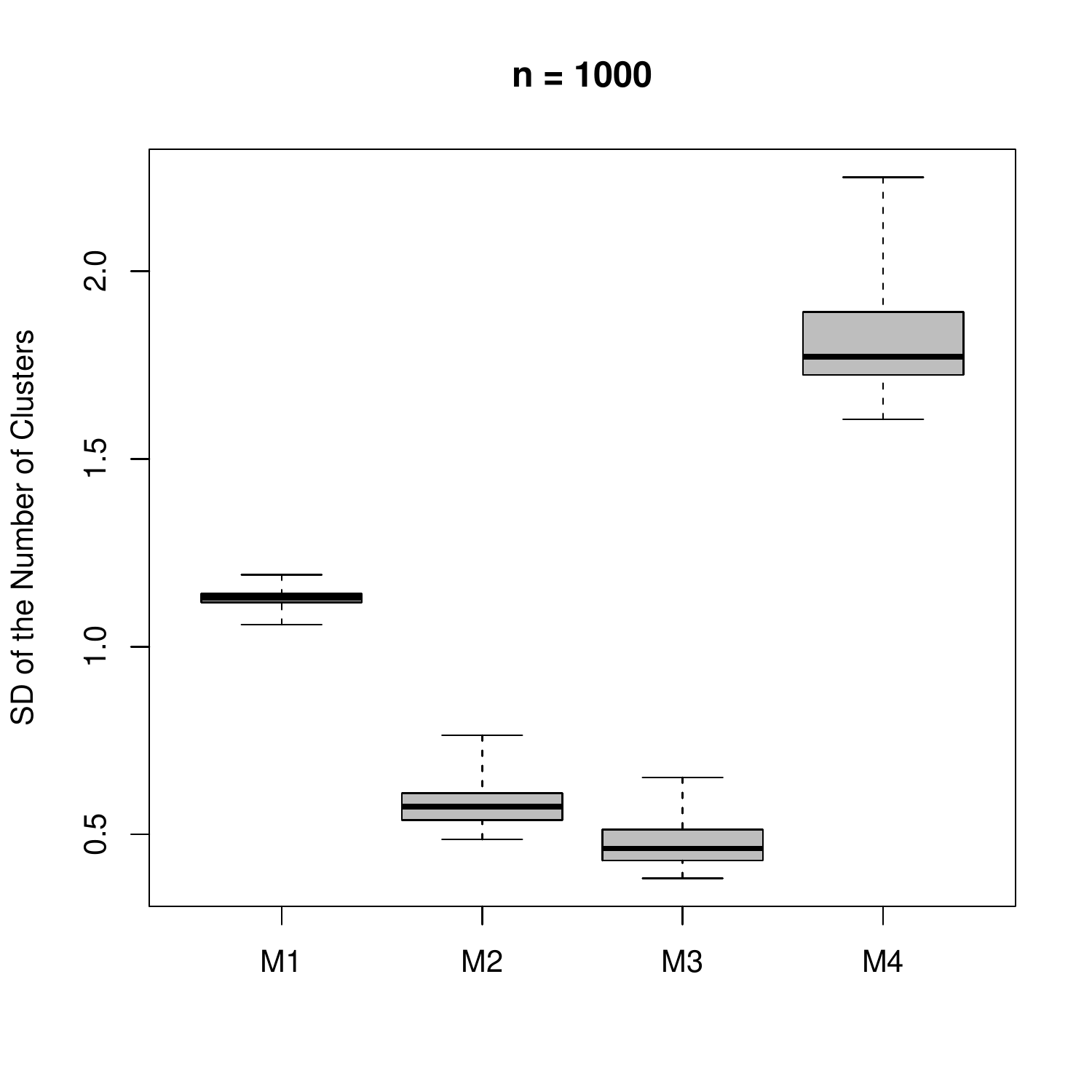} & \includegraphics[scale=0.40]{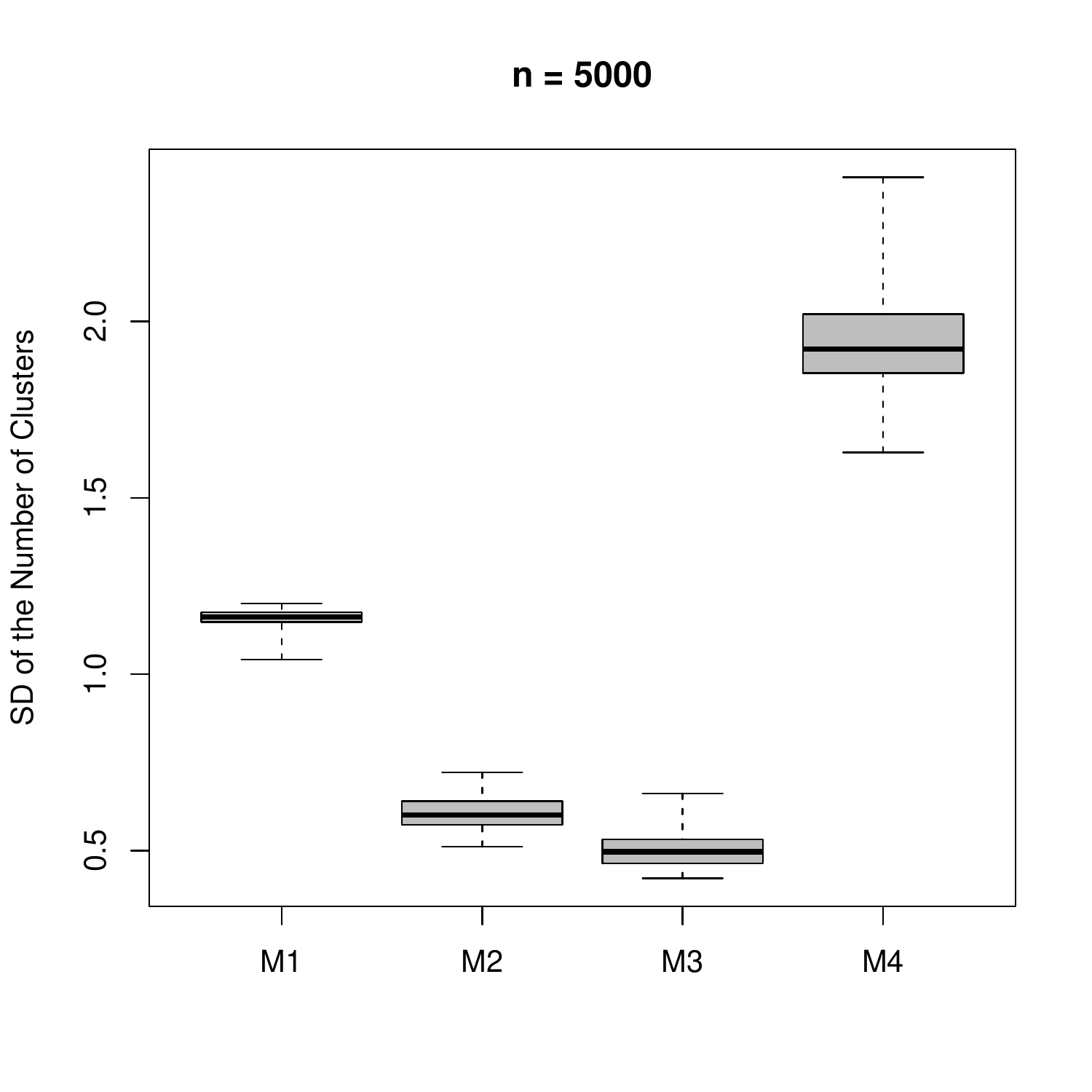}
  \end{tabular}
  \end{center}
  \caption{Side-by-side boxplots that display the average standard deviation associated with the posterior distribution of occupied mixture components for each of the four 
  procedures.}
  \label{Boxplots.Simulation.SDPANOC}
\end{figure}

Figures \ref{Boxplots.Simulation.PANOC} and \ref{Boxplots.Simulation.SDPANOC} show that the average number of occupied mixture components is much smaller for M2 and M3 
relative to M1 and M4. This pattern persists (possibly becomes more obvious) as the number of observations grows. The number of occupied mixture components for M2 and M3 are 
also highly concentrated around 3, 4 and 5 (recall that the data were generated using a mixture of four components). Conversely, M1 and M4 require many more occupied 
mixture components to achieve the same goodness-of-fit, a trend that persists when the sample size grows.


\section{Data Illustrations}\label{Data.Illustrations}

We now turn our attention to two well known data sets. The first is the \textit{Galaxy} data set \citep{Roeder:1990}, and the second is bivariate \textit{Air Quality} 
\citep{Chambers:1983}. Both are publicly available in \verb"R". For the second data set we removed 42 observations that were incomplete. We compare density estimates 
available from the DPMM to those from the RGMM. For each procedure we report the LPML as a measure of goodness-of-fit, a brief summary regarding the average number of 
occupied components, and posterior distribution associated with the number of clusters. It is worth noting that both data sets were standardized prior to model fit. We now 
provide more details on the two model specifications.

\begin{enumerate}
  \item DPMM: We employed the \verb"R" function \verb"DPdensity" available in \verb"DPpackage" (\citealt{DPpackage}). Decisions on hyperprior parameter values for both data 
  sets were again guided by \cite{escobar&west:95}. In both cases the model is specified by \eqref{DPM.Response.given.Cluster}--\eqref{DPM.Hyperprior.Scale.Scales}. We 
  collected 10000 MCMC iterates after discarding the first 1000 (5000) as burn-in for Galaxy (Air Quality) data and thinning by 10. Specific details associated with model 
  prior parameter values are now provided:
  \begin{enumerate}
    \item Galaxy: $d=1$, $a_{0}=2$, $b_{0}=2$, $\nu_{1}=4$, $\nu_{2}=4$, $\boldsymbol{m}_{2}=0$, $\mathbf{S}_{2}=1$, $\boldsymbol{\Psi}_{2}=0.15$, $\tau_{1}=2.01$ and
    $\tau_{2}=1.01$.

    \item Air Quality: $d=2$, $a_{0}=1$, $b_{0}=3$, $\nu_{1}=4$, $\nu_{2}=4$, $\boldsymbol{m}_{2}=\mathbf{0}_{2}$, $\mathbf{S}_{2}=\mathbf{I}_{2}$, 
    $\boldsymbol{\Psi}_{2}=\mathbf{I}_{2}$, $\tau_{1}=2.01$ and $\tau_{2}=1.01$.
  \end{enumerate}

  \item RGMM: We coded Algorithm RGMM in $\verb"Fortran"$ to generate posterior draws for this model. For both data sets, we collected 10000 MCMC iterates after discarding 
  the first 5000 as burn-in and thinning by 50. The values of $\tau$ were selected using the procedure outlined in Subsection \ref{Parameter.Calibration}: $(u,p)=(0.5,0.95)$ 
  and $(u,p)=(0.05,0.95)$ for Galaxy and Air Quality data respectively. Parameter selection for model components \eqref{Prior.Pi.NRep}--\eqref{Prior.Lambda.NRep} were 
  carried out according to the methods in Subsection~\ref{Parameter.Calibration}. Specific details now follow:
  \begin{enumerate}
    \item Galaxy: $k=10$, $d=1$, $\boldsymbol{\alpha}_{k,1}=10^{-1}\mathbf{1}_{10}$, $\boldsymbol{\mu}=0$, $\boldsymbol{\Sigma}=1$, $\tau=5.45$, $\boldsymbol{\Psi}=0.15$ and 
    $\nu=5$.

    \item Air Quality: $k=10$, $d=2$, $\boldsymbol{\alpha}_{k,1}=10^{-1}\mathbf{1}_{10}$, $\boldsymbol{\mu}=\mathbf{0}_{2}$, $\boldsymbol{\Sigma}=\mathbf{I}_{2}$, 
    $\tau=116.76$, $\boldsymbol{\Psi}=3\mathbf{I}_{2}$ and $\nu=6$.
  \end{enumerate}
\end{enumerate} 

\begin{figure}[htb!]
  \begin{center}
  \begin{tabular}{ccc}
    \hspace{-1cm}
    \includegraphics[width=8cm,height=8cm]{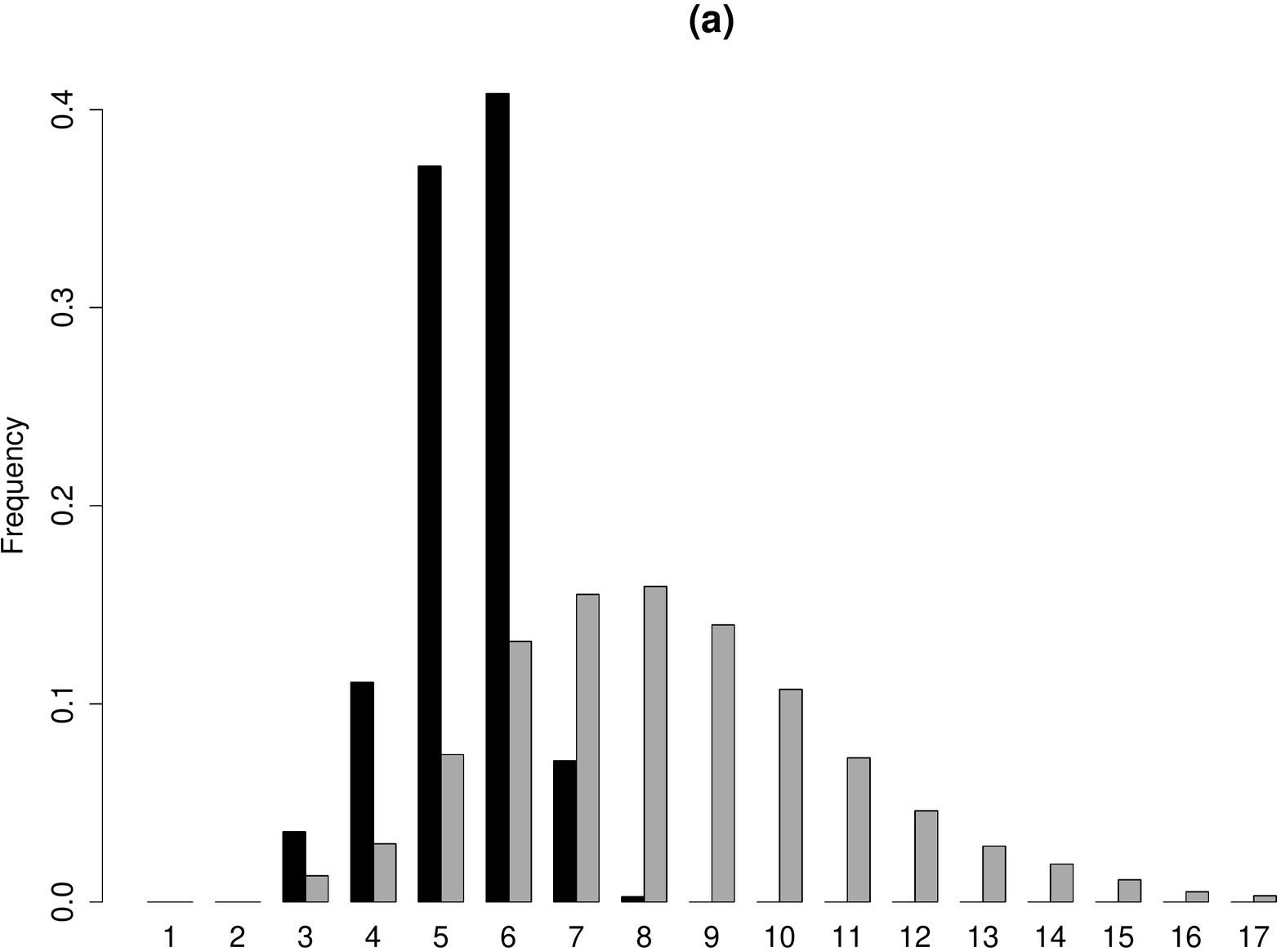} & \hspace{-1cm} & \includegraphics[width=8cm,height=8cm]{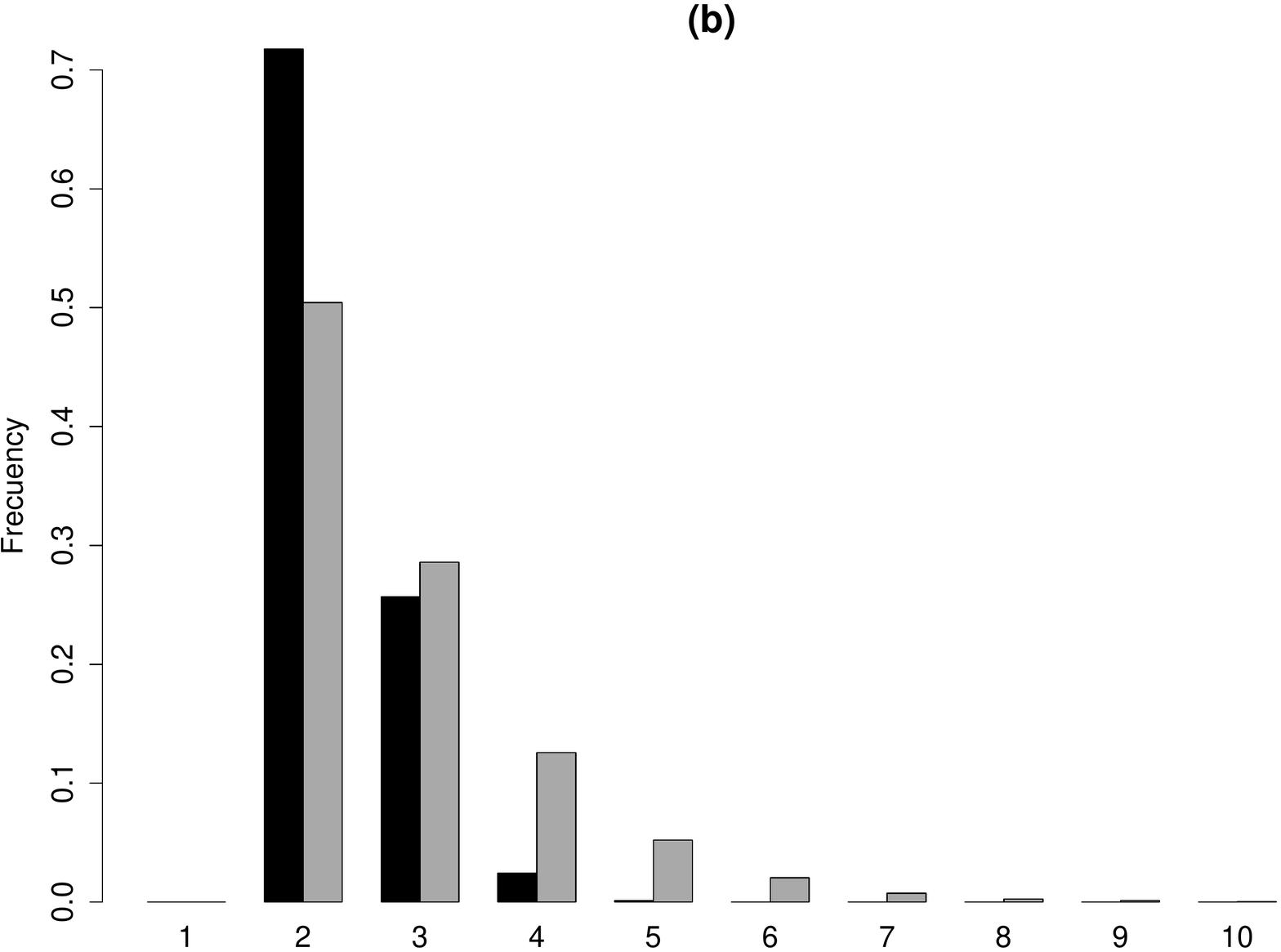}
  \end{tabular}
  \end{center}
  \caption{Posterior distribution for the active number of clusters in (a) Galaxy and (b) Air Quality data. Black (gray) bars correspond to RGMM (DPMM).}
  \label{Posterior.Number.Clusters}
\end{figure}

\begin{figure}[htb!]
  \begin{center}
  \begin{tabular}{ccc}
    \hspace{-1cm}
    \includegraphics[width=8cm,height=8cm]{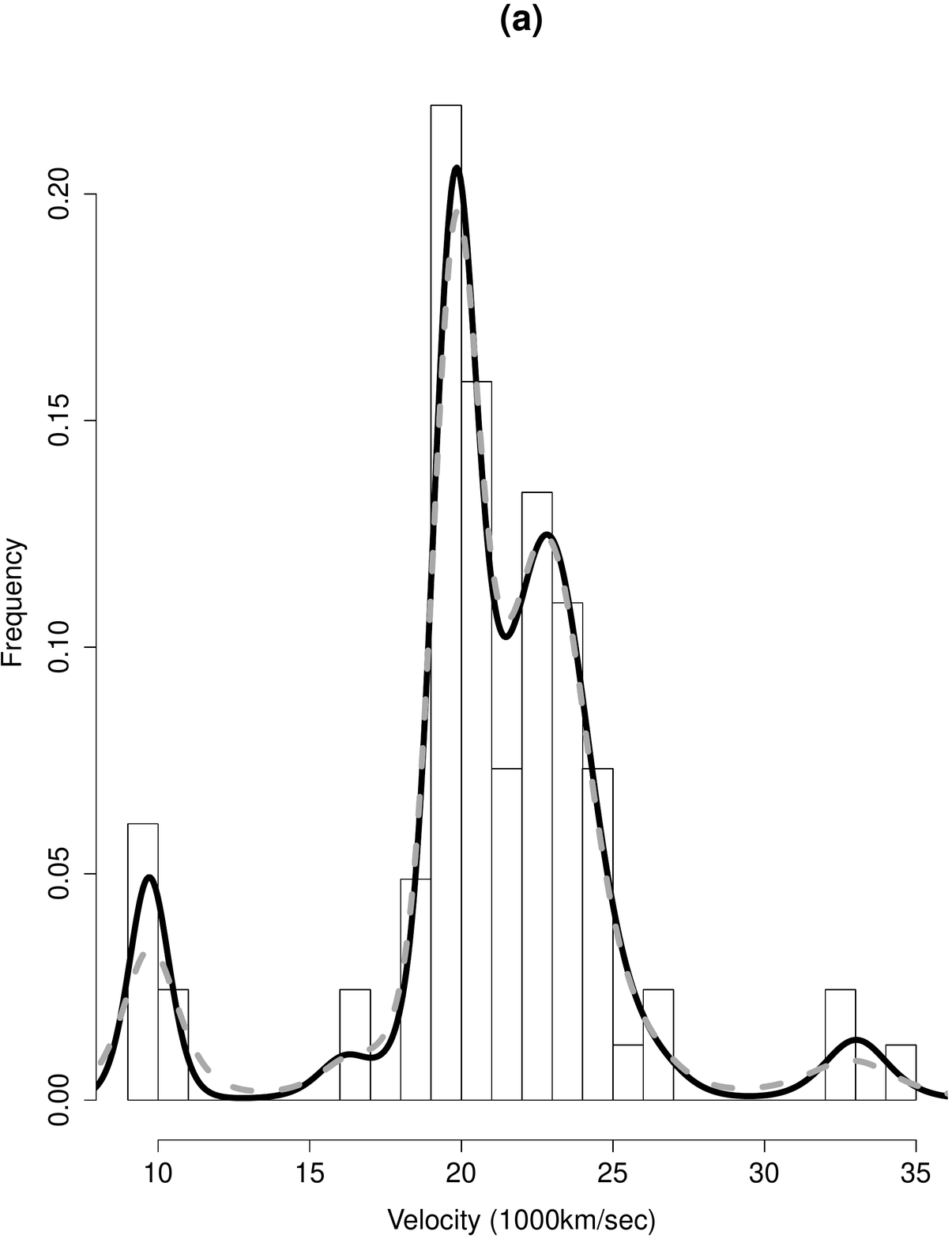} & \hspace{-1cm} & \includegraphics[width=8cm,height=8cm]{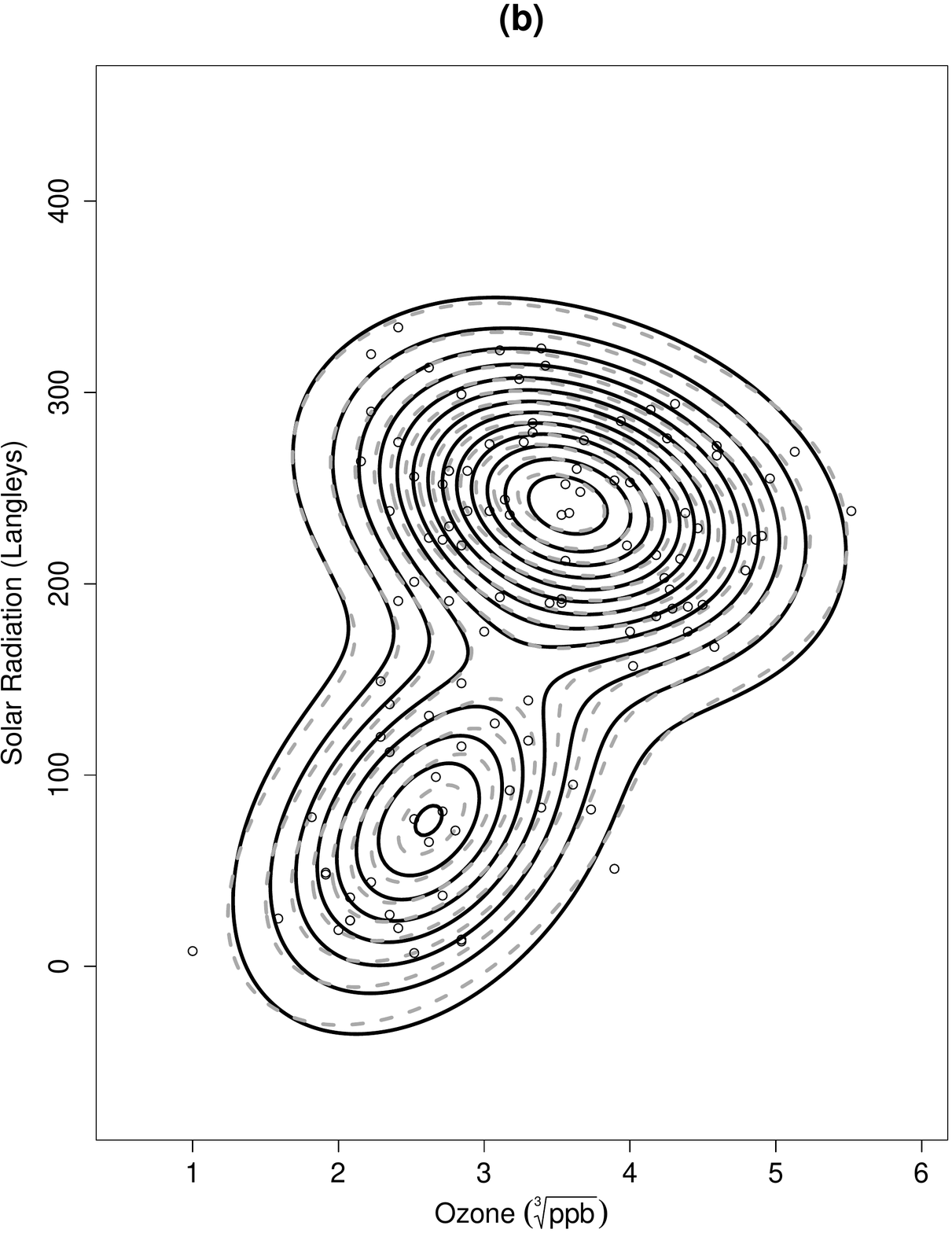}
  \end{tabular}
  \end{center}
  \caption{Posterior predictive densities for (a) Galaxy and (b) Air Quality data. Black solid (gray dashed) curves correspond to RGMM (DPMM).}
  \label{Posterior.Predictive.Densities}
\end{figure}

Results of the fits are provided in Table~\ref{Summary.Real.Data}. Notice that the fit associated with RGMM is better relative to the DPMM, which corroborates the argument 
that RGMM sacrifices no appreciable model fit for the sake of model parsimony. Figure~\ref{Posterior.Number.Clusters} further reinforces the idea that RGMM is more 
parsimonious relative to DPMM. This can be seen as the posterior distribution of the number of clusters (or non-empty components) for RGMM concentrates on values that are 
smaller relative to the DPMM. Graphs of the estimated densities (provided in Figure~\ref{Posterior.Predictive.Densities}) show that the cost of parsimony is negligible as 
density estimates are practically the same.

\begin{table}[htb!]
  \begin{center}
  \begin{tabular}{|c|c|c|c|}
    \hline
    \textrm{Data} & \textrm{LPML} & \textrm{Mean (Clusters)} & \textrm{SD (Clusters)} \\
    \hline
    \textrm{Galaxy (DPMM)} & -48.16 & 8.38 & 2.64 \\
    \textrm{Galaxy (RGMM)} & -36.68 & 5.37 & 0.91 \\
    \hline
    \textrm{Air Quality (DPMM)} & -274.82 & 2.83 & 1.11 \\
    \textrm{Air Quality (RGMM)} & -274.58 & 2.30 & 0.51 \\
    \hline
  \end{tabular}
  \end{center}
  \caption{Summary statistics related to model fit and the number of clusters for Galaxy and Air Quality data based on DPMM and RGMM.}
  \label{Summary.Real.Data}
\end{table}

\section{Discussion and Future Work}\label{Discussion}

We have created a class of probability models that explicitly parametrizes repulsion in a smooth way. In addition to providing pertinent theoretical properties, we 
demonstrated how this class of repulsive distributions can be employed to make hierarchical mixture models more parsimonious. Acompelling result is that this added parsimony 
comes at essentially no goodness-of-fit cost. We studied properties of the models, adapting the theory developed in \cite{NIPS2012_4589} to accommodate the potential 
function we considered. Moreover, we generalized the results to include not only Gaussian Mixtures of location but of also of scale (though the scale is constrained to be 
equal in each mixture component).

Our approach shares the same modeling spirit (presence of repulsion) as in \cite{NIPS2012_4589}, \cite{XuMullerTelesca:2016} and \cite{Steel:2016}. However, the specific 
mechanism we propose to model repulsion differs from these works. \cite{NIPS2012_4589} employ a potential (based on Lennard-Jones type potential) that introduces a stronger 
repulsion than our case, in the sense that in their model, locations are encouraged to be further apart. \cite{XuMullerTelesca:2016} is based on Determinantal Point 
Processes, which introduces repulsion through the determinant of a matrix driven by a Gaussian covariance kernel. By nature of the point process, this approach allows a 
random number of mixture components (similar to DPM models) something that our approach lacks. However, our approach allows a direct modeling of the repulsion that is easier 
to conceptualize. Finally, the work by \cite{Steel:2016} defines a family of probability densities that promotes well-separated location parameters through a penalization 
function, that cannot be re-expressed as a (pure) repulsive potential. However, for small relative distances, the penalization function can be identified as an interaction 
potential that produces repulsion similar to that found in \cite{NIPS2012_4589}. 

Presently we are pursuing a few directions of continued research. First, Propositions \ref{Kullback.Leibler.Support} and \ref{Posterior.Convergence.Rate} were established 
for Gaussian mixtures of dimension $d=1$ with mixture components sharing the same variance. Extending results to the general $d$ dimensional case would be a natural 
progression. Additionally, we are exploring the possibility of relaxing the assumption of common variance between mixture components and adapting the mentioned theoretical 
results to a larger class of potential functions. Studying the influence of the metric on the repulsive component in Definition~\ref{Repulsive.Distribution} and allowing the 
number of mixture components to be random are also topics of future research. \cite{RousseauMengersen:2011} developed some very interesting results that explore statistical 
properties associated with mixtures when $k$ is chosen to be conservatively large (overfitted mixtures) with decaying weights associated with these extra mixture components. 
They did so using a framework that is an alternative to what we developed here. Under some restrictions on the prior and regularity conditions for the mixture component 
densities, the asymptotic behavior of the posterior distribution on the weights tends to empty the extra mixture components. We are currently exploring connections between 
these two approaches.


{\bf Acknowledgments}:  We would like to thank Gregorio Moreno and Duvan Henao for helpful conversations and comments. Jos\'e Quinlan gratefully recognizes the financial 
support provided by CONICYT through Doctoral Scholarship Grant 21120153 and Fondecyt Grant 11121131. Fernando A. Quintana was supported by Fondecyt Grant 1141057 and Garritt 
L. Page was partially supported by Fondecyt Grant 11121131.


\singlespace
\bibliographystyle{ASA}
\bibliography{Reference}


\begin{appendices}

\section{Algorithm RGMM}\label{Algorith.RGMM}
In what follows we describe the Gibbs Sampler for the RGMM in its entirety. Let $B,S,T\in\N$ be the total number of iterations during the burn-in, the number of collected 
iterates, and the thinning, respectively.
\begin{itemize}
  \item [$\bullet$] (Start) Choose initial values $z^{(0)}_{i}:i\in[n]$, $\boldsymbol{\pi}^{(0)}_{k,1}$ and 
  $\boldsymbol{\theta}^{(0)}_{j},\boldsymbol{\Lambda}^{(0)}_{j}:j\in[k]$. Set $\boldsymbol{\Gamma}_{j}=\mathbf{O}_{d}:j\in[k]$, where $\mathbf{O}_{d}$ is the null matrix of 
  dimension $d\times d$.

  \item [$\bullet$] (Burn-in phase) For $t=0,\ldots,B-1$:
  \begin{enumerate}
    \item $(z_{i}^{(t+1)}\mid\cdots)\sim\Pb(z_{i}^{(t+1)}=j)=\pi^{(t,i)}_{j}$ independently for each $i\in[n]$, where
    \begin{equation*}
      \pi^{(t,i)}_{j}=
      \Bigg\{\sum_{l=1}^{k}\pi^{(t)}_{l}\mathrm{N}_{d}(\boldsymbol{y}_{i};\boldsymbol{\theta}^{(t)}_{l},\boldsymbol{\Lambda}^{(t)}_{l})\Bigg\}^{-1}
      \pi^{(t)}_{j}\mathrm{N}_{d}(\boldsymbol{y}_{i};\boldsymbol{\theta}^{(t)}_{j},\boldsymbol{\Lambda}^{(t)}_{j}):j\in[k].
    \end{equation*}

    \item $(\boldsymbol{\pi}^{(t+1)}_{k,1}\mid\cdots)\sim\mathrm{Dir}(\boldsymbol{\alpha}^{(t)}_{k,1})$, where
    \begin{align*}
      &\boldsymbol{\alpha}^{(t)}_{k,1}=(\alpha_{1}+n^{(t+1)}_{1},\ldots,\alpha_{k}+n^{(t+1)}_{k}) \\
      &n^{(t+1)}_{j}=\card(i\in[n]:z^{(t+1)}_{i}=j):j\in[k].
    \end{align*}

    \item For $j=1,\ldots,k$:
    \begin{itemize}
      \item [3.1.] Generate a candidate $\boldsymbol{\theta}^{(\star)}_{j}$ from $\mathrm{N}_{d}(\boldsymbol{\theta}^{(t)}_{j},\boldsymbol{\Omega}^{(t)}_{j})$, where
      \begin{equation*}
        \boldsymbol{\Omega}^{(t)}_{j}=\{\boldsymbol{\Sigma}^{-1}+n^{(t+1)}_{j}(\boldsymbol{\Lambda}^{(t)}_{j})^{-1}\}^{-1}.
      \end{equation*}

      \item [3.2.] Update $\boldsymbol{\theta}^{(t)}_{j}\to\boldsymbol{\theta}^{(t+1)}_{j}=\boldsymbol{\theta}^{(\star)}_{j}$ with probability $\min(1,\beta_{j})$, where
      \begin{equation*}
        \beta_{j}=
        \frac
        {\mathrm{N}_{d}(\boldsymbol{\theta}^{(\star)}_{j};\boldsymbol{\mu}^{(t)}_{j},\boldsymbol{\Sigma}^{(t)}_{j})}
        {\mathrm{N}_{d}(\boldsymbol{\theta}^{(t)}_{j};\boldsymbol{\mu}^{(t)}_{j},\boldsymbol{\Sigma}^{(t)}_{j})}
        \prod_{l\neq j}^{k}
        \Bigg[
        \frac
        {1-\exp\{-0.5\tau^{-1}(\boldsymbol{\theta}^{(\star)}_{j}-\boldsymbol{\theta}^{(t)}_{l})^{\top}
        \boldsymbol{\Sigma}^{-1}
        (\boldsymbol{\theta}^{(\star)}_{j}-\boldsymbol{\theta}^{(t)}_{l})\}}
        {1-\exp\{-0.5\tau^{-1}(\boldsymbol{\theta}^{(t)}_{j}-\boldsymbol{\theta}^{(t)}_{l})^{\top}
        \boldsymbol{\Sigma}^{-1}
        (\boldsymbol{\theta}^{(t)}_{j}-\boldsymbol{\theta}^{(t)}_{l})\}}
        \Bigg].
      \end{equation*}
      In the above expression for $\beta_{j}$ 
      \begin{align*}
        &\boldsymbol{\Sigma}^{(t)}_{j}=\{\boldsymbol{\Sigma}^{-1}+n^{(t+1)}_{j}(\boldsymbol{\Lambda}^{(t)}_{j})^{-1}\}^{-1} \\
        &\boldsymbol{\mu}^{(t)}_{j}=\boldsymbol{\Sigma}^{(t)}_{j}\{\boldsymbol{\Sigma}^{-1}\boldsymbol{\mu}+(\boldsymbol{\Lambda}^{(t)}_{j})^{-1}\boldsymbol{s}^{(t)}_{j}\}:
        \boldsymbol{s}^{(t)}_{j}=\sum_{i=1}^{n}\I_{\{j\}}(z^{(t+1)}_{i})\boldsymbol{y}_{i}.
      \end{align*}
      Otherwise, set $\boldsymbol{\theta}^{(t+1)}_{j}=\boldsymbol{\theta}^{(t)}_{j}$.

      \item [3.3.] Update $\boldsymbol{\Gamma}_{j}\to\boldsymbol{\Gamma}_{j}+B^{-1}\boldsymbol{\Omega}^{(t)}_{j}$.
    \end{itemize}

    \item $(\boldsymbol{\Lambda}^{(t+1)}_{j}\mid\cdots)\sim\mathrm{IW}_{d}(\boldsymbol{\Psi}^{(t)}_{j},\nu^{(t)}_{j})$ independently for each $j\in[k]$, where 
    $\nu^{(t)}_{j}=\nu+n^{(t+1)}_{j}$ and
    \begin{equation*}
      \boldsymbol{\Psi}^{(t)}_{j}=
      \boldsymbol{\Psi}+
      \sum_{i=1}^{n}\I_{\{j\}}(z^{(t+1)}_{i})(\boldsymbol{y}_{i}-\boldsymbol{\theta}^{(t+1)}_{j})(\boldsymbol{y}_{i}-\boldsymbol{\theta}^{(t+1)}_{j})^{\top}.
    \end{equation*}
  \end{enumerate}

  \item [$\bullet$] (Save samples) For $t=B,\ldots,ST+B-1$: Repeat steps 1, 2 and 4 of the burn-in phase. As for step 3 ignore 3.3, maintain 3.2 and replace 3.1 with
  \begin{itemize}
    \item [3.1a.] Generate a candidate $\boldsymbol{\theta}^{(\star)}_{j}$ from $\mathrm{N}_{d}(\boldsymbol{\theta}^{(t)}_{j},\boldsymbol{\Gamma}_{j})$.
  \end{itemize}
  Finally, save the generated samples every $T$th iteration.
  
  \item [$\bullet$] (Posterior predictive estimate) With the $T$ saved samples, compute
  \begin{equation*}
    f(\boldsymbol{y}\mid\boldsymbol{y}_{1},\ldots,\boldsymbol{y}_{n})\approx\frac{1}{T}\sum_{t=1}^{T}
    \Bigg\{
    \sum_{j=1}^{k}\pi^{(t)}_{j}\mathrm{N}_{d}(\boldsymbol{y};\boldsymbol{\theta}^{(t)}_{j},\boldsymbol{\Lambda}^{(t)}_{j})
    \Bigg\}.
  \end{equation*}

\end{itemize}

\section{Proof of Lemma \ref{Smoothness.Repulsive.Component}.}\label{Proof.Smoothness.Repulsive.Component}
Assign to $\R^{d}_{k}$ and $[0,1)$ the metrics $d_{1}(\boldsymbol{x}_{k,d},\boldsymbol{y}_{k,d})=\max\{\rho(\boldsymbol{x}_{i},\boldsymbol{y}_{i}):i\in[k]\}$ and 
$d_{2}(x,y)=|x-y|$, respectively. Continuity of $\mathrm{R}_{\mathrm{C}}:\R^{d}_{k}\to[0,1)$ follows from condition A1 of $C_{0}$-properties and the following inequality:
\begin{equation*}
  |\rho(\boldsymbol{x}_{r},\boldsymbol{x}_{s})-\rho(\boldsymbol{y}_{r},\boldsymbol{y}_{s})|<2d_{1}(\boldsymbol{x}_{k,d},\boldsymbol{y}_{k,d}).
\end{equation*}

\section{Proof of Proposition \ref{Existence.Repulsive.Distribution}.}\label{Proof.Existence.Repulsive.Distribution}
Notice that $g_{k,d}\in C(\R^{d}_{k};(0,\infty))$ by construction (see Lemma~\ref{Smoothness.Repulsive.Component}). Because of the continuity, measurability follows. Using 
conditions A1--A4 of $C_{0}$-properties it follows that for all $x\in[0,\infty)$, $\{1-C_{0}(x)\}\in[0,1)$. By Tonelli's Theorem
\begin{equation*}
  \int_{\R^{d}_{k}}g_{k,d}(\boldsymbol{x}_{k,d})\diff\boldsymbol{x}_{k,d}\leq\Bigg(\int_{\R^{d}}f_{0}(\boldsymbol{x})\diff\boldsymbol{x}\Bigg)^{k}=1.
\end{equation*}
The upper bound only proves that $g_{k,d}$ is integrable. However, this does not guarantee that $g_{k,d}$ is well defined, i.e. $\lambda^{k}_{d}(g_{k,d}>0)=0$. For this, it 
is sufficient to show that
\begin{equation*}
  \int_{\R^{d}_{k}}g_{k,d}(\boldsymbol{x}_{k,d})\diff\boldsymbol{x}_{k,d}>0
\end{equation*}
because for all $\boldsymbol{x}_{k,d}\in\R^{d}_{k}$, $g_{k,d}(\boldsymbol{x}_{k,d})\geq0$ by construction. To prove the above inequality, fix 
$\boldsymbol{x}^{0}_{k,d}\in\R^{d}_{k}$ such that $\boldsymbol{x}^{0}_{r}\neq\boldsymbol{x}^{0}_{s}$ for $r\neq s\in[k]$. Then $g_{k,d}(\boldsymbol{x}^{0}_{k,d})>0$. Because 
$g_{k,d}$ is a continuous function on $\R^{d}_{k}$, there exists $r_{0}\in(0,\infty)$ such that for all $\boldsymbol{x}_{k,d}\in B(\boldsymbol{x}^{0}_{k,d},r_{0})$
\begin{equation*}
  g_{k,d}(\boldsymbol{x}_{k,d})>0,
\end{equation*}
where $B(\boldsymbol{x}^{0}_{k,d},r_{0})$ is the cartesian product of $B_{2}(\boldsymbol{x}^{0}_{1},r_{0}),\ldots,B_{2}(\boldsymbol{x}^{0}_{k},r_{0})$. Further, 
$B(\boldsymbol{x}^{0}_{k,d},r_{0})\in\B(\R^{d}_{k})$ and $\lambda^{k}_{d}\{B(\boldsymbol{x}^{0}_{k,d},r_{0})\}=(\pi^{kd/2}r_{0}^{kd})\Gamma(1+d/2)^{-k}\in(0,\infty)$ by the 
Volume Formula, where $\Gamma(\,\cdot\,)$ is the Gamma function. Thus
\begin{equation*}
  \int_{\R^{d}_{k}}g_{k,d}(\boldsymbol{x}_{k,d})\diff\boldsymbol{x}_{k,d}
  \geq
  \int_{B(\boldsymbol{x}^{0}_{k,d},r_{0})}g_{k,d}(\boldsymbol{x}_{k,d})\diff\boldsymbol{x}_{k,d}>0.
\end{equation*}

\section{Proof of Lemma \ref{Dominated.Convergence}.}\label{Proof.Dominated.Convergence}
For any $x\in\R$ we have that
\begin{equation*}
  |f_{0}(x;\boldsymbol{\xi}^{0}_{k_{0}})-f(x;\boldsymbol{\xi}_{k_{0}})|\leq
  \frac{||\boldsymbol{\pi}^{0}_{k_{0},1}-\boldsymbol{\pi}_{k_{0},1}||_{1}}{(2\pi\lambda_{0})^{1/2}}
  +
  \frac{||\boldsymbol{\theta}^{0}_{k_{0},1}-\boldsymbol{\theta}_{k_{0},1}||_{1}}{\{2\pi\exp(1)\}^{1/2}\lambda_{0}}
  +
  u(\lambda,\boldsymbol{\theta}_{k_{0},1};x,\lambda_{0})|\lambda-\lambda_{0}|
\end{equation*}
and
\begin{equation*}
  u(\lambda,\boldsymbol{\theta}_{k_{0,1}};x,\lambda_{0})=\frac{1}{(2\pi)^{1/2}}
  \Bigg[
  \frac{k_{0}}{\lambda\lambda_{0}^{1/2}+\lambda_{0}\lambda^{1/2}}
  +
  \frac{\lambda_{0}^{1/2}}{2\lambda\lambda_{0}^{2}}\sum_{j=1}^{k_{0}}(x-\theta_{j})^{2}
  \Bigg]
\end{equation*}
for all $(\lambda,\boldsymbol{\theta}_{k_{0},1})\in(0,\infty)\times\R^{1}_{k_{0}}$, with $||\,\cdot\,||_{1}$ being the Euclidean $L_{1}$-norm in $\R^{1}_{k_{0}}$. Because 
$u(\lambda,\boldsymbol{\theta}_{k_{0},1};x,\lambda_{0})$ is continuous at $(\lambda_{0},\boldsymbol{\theta}^{0}_{k_{0},1})$,
\begin{equation*}
  f(x;\boldsymbol{\xi}_{k_{0}})\to f_{0}(x;\boldsymbol{\xi}^{0}_{k_{0}})
\end{equation*}
point-wise in $x$ when $\boldsymbol{\xi}_{k_{0}}\to\boldsymbol{\xi}^{0}_{k_{0}}$. The last statement is equivalent to the condition that
\begin{equation*}
  |\log\{f(x;\boldsymbol{\xi}_{k_{0}}\}-\log\{f_{0}(x;\boldsymbol{\xi}^{0}_{k_{0}})\}|f_{0}(x;\boldsymbol{\xi}^{0}_{k_{0}})\to0
\end{equation*}
point-wise in $x$ when $\boldsymbol{\xi}_{k_{0}}\to\boldsymbol{\xi}^{0}_{k_{0}}$.

By condition B2, we can assume that $\theta^{0}_{1}<\cdots<\theta^{0}_{k_{0}}$ (possibly after an appropriate relabeling). Choose $t^{0}_{1},t^{0}_{2}\in\R$ and 
$l^{0}_{1},l^{0}_{2}\in(0,\infty)$ such that $\lambda_{0}\in[l^{0}_{1},l^{0}_{2}]$ and, for all $x\in(-\infty,t^{0}_{1})\cup(t^{0}_{2},\infty)$
\begin{equation*}
  f_{0}(x;\boldsymbol{\xi}^{0}_{k_{0}})<1,\qquad\theta_{j}\in(t^{0}_{1},t^{0}_{2}):j\in[k_{0}].
\end{equation*}
Since $|\log\{f_{0}(x;\boldsymbol{\xi}^{0}_{k_{0}})\}|f_{0}(x;\boldsymbol{\xi}^{0}_{k_{0}})$ is uniformly continuous for $x\in[t^{0}_{1},t^{0}_{2}]$,
\begin{equation*}
  I_{1}=\int_{[t^{0}_{1},t^{0}_{2}]}|\log\{f_{0}(x;\boldsymbol{\xi}^{0}_{k_{0}})\}|f_{0}(x;\boldsymbol{\xi}^{0}_{k_{0}})\diff x\in(0,\infty).
\end{equation*}
Fix $\delta_{1}\in(0,1)$, $\delta_{2}=0.5\min(t^{0}_{2}-\theta^{0}_{k_{0}},\theta^{0}_{1}-t^{0}_{1})$ and define 
$V_{0}=D_{1}(\boldsymbol{\pi}^{0}_{k_{0},1},\delta_{1})\times D_{1}(\boldsymbol{\theta}^{0}_{k_{0},1},\delta_{2})\times[l^{0}_{1},l^{0}_{2}]$. Notice that 
$M(x,\boldsymbol{\xi}_{k_{0}})=|\log\{f(x;\boldsymbol{\xi}_{k_{0}})\}|$ is uniformly continuous for $(x,\boldsymbol{\xi}_{k_{0}})\in[t^{0}_{1},t^{0}_{2}]\times V_{0}$. Then 
$M_{0}=\max(M(x,\boldsymbol{\xi}_{k_{0}}):(x,\boldsymbol{\xi}_{k_{0}})\in[t^{0}_{1},t^{0}_{2}]\times V_{0})\in(0,\infty)$ and
\begin{equation*}
  \int_{[t^{0}_{1},t^{0}_{2}]}|\log\{f(x;\boldsymbol{\xi}_{k_{0}})\}|f_{0}(x;\boldsymbol{\xi}^{0}_{k_{0}})\diff x\leq I_{2}=
  \int_{[t^{0}_{1},t^{0}_{2}]}M_{0}f_{0}(x;\boldsymbol{\xi}^{0}_{k_{0}})\diff x\in(0,\infty).
\end{equation*}
By the Triangle Inequality
\begin{equation*}
  \int_{[t^{0}_{1},t^{0}_{2}]}|\log\{f_{0}(x;\boldsymbol{\xi}^{0}_{k_{0}})\}-\log\{f(x;\boldsymbol{\xi}_{k_{0}})\}|f_{0}(x;\boldsymbol{\xi}^{0}_{k_{0}})\diff x
  \leq I_{1}+I_{2}\in(0,\infty).
\end{equation*}
On the other hand, define the following continuous functions:
\begin{align*}
  &h_{1}(x)=0.5|\log(2\pi\lambda_{0})|+0.5\lambda_{0}^{-1}(x-\theta^{0}_{k_{0}})^{2}:x\in(-\infty,t^{0}_{1}) \\
  &h_{2}(x)=0.5|\log(2\pi l^{0}_{1})|+(2l^{0}_{1})^{-1}(x-\delta_{2}-\theta^{0}_{k_{0}})^{2}:x\in(-\infty,t^{0}_{1}) \\
  &h_{3}(x)=0.5|\log(2\pi\lambda_{0})|+0.5\lambda_{0}^{-1}(x-\theta^{0}_{1})^{2}:x\in(t^{0}_{2},\infty) \\
  &h_{4}(x)=0.5|\log(2\pi l^{0}_{1})|+(2l^{0}_{1})^{-1}(x+\delta_{2}-\theta^{0}_{1})^{2}:x\in(t^{0}_{2},\infty).
\end{align*}
Using the initial assumptions
\begin{align*}
  &|\log\{f(x;\boldsymbol{\xi}^{0}_{k_{0}})\}|\leq h_{1}(x):x\in(-\infty,t^{0}_{1}) \\
  &|\log\{f(x;\boldsymbol{\xi}_{k_{0}})\}|\leq h_{2}(x):(x,\boldsymbol{\xi}_{k_{0}})\in(-\infty,t^{0}_{1})\times V_{0} \\
  &|\log\{f(x;\boldsymbol{\xi}^{0}_{k_{0}})\}|\leq h_{3}(x):x\in(t^{0}_{2},\infty) \\
  &|\log\{f(x;\boldsymbol{\xi}_{k_{0}})\}|\leq h_{4}(x):(x,\boldsymbol{\xi}_{k_{0}})\in(t^{0}_{2},\infty)\times V_{0}.
\end{align*}
Taking into account the existence of second order moments of a Gaussian distribution
\begin{align*}
  &I_{3}=\int_{(-\infty,t^{0}_{1})}\{h_{1}(x)+h_{2}(x)\}f_{0}(x;\boldsymbol{\xi}^{0}_{k_{0}})\diff x\in(0,\infty) \\
  &I_{4}=\int_{(t^{0}_{2},\infty)}\{h_{3}(x)+h_{4}(x)\}f_{0}(x;\boldsymbol{\xi}^{0}_{k_{0}})\diff x\in(0,\infty).
\end{align*}
Again, using the Triangle Inequality
\begin{equation*}
  \int_{(-\infty,t^{0}_{1})\cup(t^{0}_{2},\infty)}
  |\log\{f_{0}(x;\boldsymbol{\xi}^{0}_{k_{0}})\}-\log\{f(x;\boldsymbol{\xi}_{k_{0}})\}|f_{0}(x;\boldsymbol{\xi}^{0}_{k_{0}})\diff x
  \leq I_{3}+I_{4}\in(0,\infty).
\end{equation*}
The previous arguments show that $|\log\{f_{0}(x;\boldsymbol{\xi}^{0}_{k_{0}})\}-\log\{f(x;\boldsymbol{\xi}_{k_{0}})\}|f_{0}(x;\boldsymbol{\xi}^{0}_{k_{0}})$ for all 
$(x,\boldsymbol{\xi}_{k_{0}})\in\R\times V_{0}$ is bounded above by a positive and integrable function that depends only in $x\in\R$. As a consequence of Lebegue's
Dominated Convergence Theorem
\begin{equation*}
  \int_{\R}\log\Bigg\{\frac{f_{0}(x;\boldsymbol{\xi}^{0}_{k_{0}})}{f(x;\boldsymbol{\xi}_{k_{0}})}\Bigg\}f_{0}(x;\boldsymbol{\xi}^{0}_{k_{0}})\diff x\to0
\end{equation*}
as $\boldsymbol{\xi}_{k_{0}}\to\boldsymbol{\xi}^{0}_{k_{0}}$. In other words, for all $\varepsilon>0$ there exists $\delta>0$ such that
\begin{equation*}
  \int_{\R}\log\Bigg\{\frac{f_{0}(x;\boldsymbol{\xi}^{0}_{k_{0}})}{f(x;\boldsymbol{\xi}_{k_{0}})}\Bigg\}f_{0}(x;\boldsymbol{\xi}^{0}_{k_{0}})\diff x<\varepsilon
\end{equation*}
provided that 
$\boldsymbol{\xi}_{k_{0}}\in B_{1}(\boldsymbol{\theta}^{0}_{k_{0},1},\delta)\times B_{1}(\boldsymbol{\pi}^{0}_{k_{0},1},\delta)\times(\lambda_{0}-\delta,\lambda_{0}+\delta)$.

\section{Proof of Lemma \ref{Repulsive.Probability.Property}.}\label{Proof.Repulsive.Probability.Property}
Set $\delta_{00}=0.25vk_{0}$ with $v>0$ specified by condition B2. Notice that
\begin{equation*}
   \boldsymbol{\theta}_{k_{0},1}\in B_{\delta}=
   \prod_{i=1}^{k_{0}}\Bigg(\theta^{0}_{i}-\frac{\delta}{k_{0}},\theta^{0}_{i}+\frac{\delta}{k_{0}}\Bigg)
   \subseteq B_{1}(\boldsymbol{\theta}^{0}_{k_{0},1},\delta).
\end{equation*}
for all $\delta\in(0,\delta_{00}]$. Using the definition of $\NRep_{k_{0},1}(\mu,\sigma^{2},\tau)$ and denoting $c_{k_{0}}=c_{k_{0},1}$ the associated normalizing constant, 
we have that
\begin{equation*}
  \Pb\{\boldsymbol{\theta}_{k_{0},1}\in B_{1}(\boldsymbol{\theta}^{0}_{k_{0},1},\delta)\}\geq
  \frac{1}{c_{k_{0}}}
  \int_{B_{\delta}}
  \Bigg\{\prod_{i=1}^{k_{0}}\mathrm{N}(\theta_{i};\mu,\sigma^{2})\Bigg\}
  \prod_{r<s}^{k_{0}}\Bigg[1-\exp\Bigg\{-\frac{(\theta_{r}-\theta_{s})^{2}}{2\tau\sigma^{2}}\Bigg\}\Bigg]\diff\boldsymbol{\theta}_{k_{0},1}.
\end{equation*}
for all $\delta\in(0,\delta_{00}]$. Now
\begin{equation*}
  \prod_{r<s}^{k_{0}}\Bigg[1-\exp\Bigg\{-\frac{(\theta_{r}-\theta_{s})^{2}}{2\tau\sigma^{2}}\Bigg\}\Bigg]
  \geq
  \Bigg[1-\exp\Bigg\{-\frac{v_{0}}{2\tau\sigma^{2}}\Bigg\}\Bigg]^{\ell_{k_{0}}}=R_{0}\in(0,\infty)
\end{equation*}
for all $\boldsymbol{\theta}_{k_{0},1}\in B_{\delta}$, with $v_{0}=(v-2\delta_{00}k_{0}^{-1})^{2}$ and $\ell_{k_{0}}=0.5k_{0}(k_{0}-1)$. Using this information and Fubini's 
Theorem
\begin{equation*}
  \Pb\{\boldsymbol{\theta}_{k_{0},1}\in B_{1}(\boldsymbol{\theta}^{0}_{k_{0},1},\delta)\}
  \geq
  \frac{R_{0}}{c_{k_{0}}}
  \prod_{i=1}^{k_{0}}
  \Bigg\{\Phi\Bigg(\frac{\theta^{0}_{i}-\mu}{\sigma}+\frac{\delta}{k_{0}\sigma}\Bigg)-\Phi\Bigg(\frac{\theta^{0}_{i}-\mu}{\sigma}-\frac{\delta}{k_{0}\sigma}\Bigg)\Bigg\}
\end{equation*}
for all $\delta\in(0,\delta_{00}]$. Because for each $i\in[k_{0}]$
\begin{equation*}
  \frac{1}{\delta}
  \Bigg\{
  \Phi\Bigg(\frac{\theta^{0}_{i}-\mu}{\sigma}+\frac{\delta}{k_{0}\sigma}\Bigg)
  -
  \Phi\Bigg(\frac{\theta^{0}_{i}-\mu}{\sigma}-\frac{\delta}{k_{0}\sigma}\Bigg)
  \Bigg\}
  \to
  \frac{2}{k_{0}\sigma}\mathrm{N}\Bigg(\frac{\theta^{0}_{i}-\mu}{\sigma};0,1\Bigg)=S^{0}_{i}\in(0,\infty)
\end{equation*}
as $\delta\to0$ (right-side limit), there exists $\delta_{0i}>0$ such that
\begin{equation*}
  \Bigg\{
  \Phi\Bigg(\frac{\theta^{0}_{i}-\mu}{\sigma}+\frac{\delta}{k_{0}\sigma}\Bigg)
  -
  \Phi\Bigg(\frac{\theta^{0}_{i}-\mu}{\sigma}-\frac{\delta}{k_{0}\sigma}\Bigg)
  \Bigg\}
  \geq\frac{S^{0}_{i}}{2}.
\end{equation*}
for all $\delta\in(0,\delta_{0i}]$. Finally, choose $\delta_{0}=\min(\delta_{0j}:j\in\{0\}\cup[k_{0}])$ to conclude that
\begin{equation*}
  \Pb\{\boldsymbol{\theta}_{k_{0},1}\in B_{1}(\boldsymbol{\theta}^{0}_{k_{0},1},\delta)\}
  \geq
  \frac{R_{0}}{c_{k_{0}}}
  \Bigg(
  \prod_{i=1}^{k_{0}}\frac{S^{0}_{i}}{2}
  \Bigg)\exp\{-k_{0}\log(1/\delta)\}\in(0,\infty).
\end{equation*}
for all $\delta\in(0,\delta_{0}]$.
\newline
\textbf{Remark:} The previous inequality also applies replacing $B_{1}(\boldsymbol{\theta}^{0}_{k_{0},1},\delta)$ by $D_{1}(\boldsymbol{\theta}^{0}_{k_{0},1},\delta)$.

\section{Proof of Proposition \ref{Kullback.Leibler.Support}.}\label{Proof.Kullback.Leibler.Support}
We will follow the proof of Lemma 1 in \cite{NIPS2012_4589} with a few variations. For this, let $\varepsilon>0$ and define
\begin{equation*}
  B_{\mathrm{KL}}(f_{0},\varepsilon)=
  \Bigg\{f\in\F:\int_{\R}\log\Bigg\{\frac{f_{0}(x;\boldsymbol{\xi}^{0}_{k_{0}})}{f(x;\boldsymbol{\xi}_{\star})}\Bigg\}f_{0}(x;\boldsymbol{\xi}^{0}_{k_{0}})\diff x
  <\varepsilon\Bigg\}
\end{equation*}
with $\boldsymbol{\xi}_{\star}\in\bigcup_{k=1}^{\infty}\boldsymbol{\Theta}_{k}$. Using the stochastic representation \eqref{Distribution.Xi.given.K},
\begin{equation*}
  \Pi\{B_{\mathrm{KL}}(f_{0},\varepsilon)\}
  \geq
  \kappa(k_{0})
  \Pb\Bigg(\boldsymbol{\xi}_{k_{0}}\in\boldsymbol{\Theta}_{k_{0}}:
  \int_{\R}\log\Bigg\{\frac{f_{0}(x;\boldsymbol{\xi}^{0}_{k_{0}})}{f(x;\boldsymbol{\xi}_{k_{0}})}\Bigg\}f_{0}(x;\boldsymbol{\xi}^{0}_{k_{0}})\diff x
  <\varepsilon\Bigg).
\end{equation*}
By condition B3, $\kappa(k_{0})>0$. In this case, to guarantee \eqref{Definition.Kullback.Leibler.Support} it is sufficient to show that
\begin{equation*}
  \Pb\Bigg(\boldsymbol{\xi}_{k_{0}}\in\boldsymbol{\Theta}_{k_{0}}:
  \int_{\R}\log\Bigg\{\frac{f_{0}(x;\boldsymbol{\xi}^{0}_{k_{0}})}{f(x;\boldsymbol{\xi}_{k_{0}})}\Bigg\}f_{0}(x;\boldsymbol{\xi}^{0}_{k_{0}})\diff x
  <\varepsilon\Bigg)>0.
\end{equation*}
Lemma~\ref{Dominated.Convergence} guaranties the existence of $\delta_{1}>0$ such that for all $\boldsymbol{\xi}_{k_{0}}\in 
B_{1}(\boldsymbol{\theta}^{0}_{k_{0},1},\delta_{1})\times B_{1}(\boldsymbol{\pi}^{0}_{k_{0},1},\delta_{1})\times(\lambda_{0}-\delta_{1},\lambda_{0}+\delta_{1})$
\begin{equation*}
  \int_{\R}\log\Bigg\{\frac{f_{0}(x;\boldsymbol{\xi}^{0}_{k_{0}})}{f(x;\boldsymbol{\xi}_{k_{0}})}\Bigg\}f_{0}(x;\boldsymbol{\xi}^{0}_{k_{0}})\diff x<\varepsilon.
\end{equation*}
Choose $\delta=\min(\delta_{0},\delta_{1})$ where $\delta_{0}>0$ is given by Lemma~\ref{Repulsive.Probability.Property}. Now 
$p_{1}=\Pb\{\boldsymbol{\theta}_{k_{0},1}\in B_{1}(\boldsymbol{\theta}^{0}_{k_{0},1},\delta)\}>0$. The same holds for 
$p_{2}=\Pb\{\boldsymbol{\pi}_{k_{0},1}\in B_{1}(\boldsymbol{\pi}^{0}_{k_{0},1},\delta))$ and $p_{3}=\Pb\{\lambda\in(\lambda_{0}-\delta,\lambda_{0}+\delta)\}$. Thus, 
independence between $\boldsymbol{\pi}_{k_{0},1}$, $\boldsymbol{\theta}_{k_{0},1}$ and $\lambda$ implies
\begin{equation*}
  \Pb\Bigg(\boldsymbol{\xi}_{k_{0}}\in\boldsymbol{\Theta}_{k_{0}}:
  \int_{\R}\log\Bigg\{\frac{f_{0}(x;\boldsymbol{\xi}^{0}_{k_{0}})}{f(x;\boldsymbol{\xi}_{k_{0}})}\Bigg\}f_{0}(x;\boldsymbol{\xi}^{0}_{k_{0}})\diff x
  <\varepsilon\Bigg)
  \geq p_{1}p_{2}p_{3}>0.
\end{equation*}

\section{Proof of Lemma \ref{Repulsive.Probability.Inequality}.}\label{Proof.Repulsive.Probability.Inequality}
As already mentioned at the beginning of Subsection~\ref{Rep.Properties}, $\boldsymbol{\theta}_{k,1}\sim\NRep_{k,1}(\mu,\sigma^{2},\tau)$ is an exchangeable distribution in 
$\theta_{1},\ldots,\theta_{k}$ for $k\geq2$. This implies that the probability laws of each $\theta_{i}:i\in[k]$ are the same. To prove the desired inequality, observe that
for all $t\in(0,\infty)$
\begin{equation*}
  \Pb(|\theta_{i}|>t)\leq\frac{c_{k-1}}{c_{k}}\int_{A_{t}}\mathrm{N}(x;\mu,\sigma^{2})\diff x=\frac{c_{k-1}}{c_{k}}\int_{B_{t}}\mathrm{N}(s;0,1)\diff s.
\end{equation*}
where $A_{t}=\{x\in\R:|x|>t\}$ and $B_{t}=\{s\in\R:|\mu+\sigma s|>t\}$. Now
\begin{equation*}
  B_{t}\subseteq\{s\in\R:|\mu|+\sigma|s|>t\}=\Bigg\{s\in\R:|s|>\frac{t-|\mu|}{\sigma}\Bigg\}=C_{t}.
\end{equation*}
Set $\gamma=\max\{2|\mu|+1,(2+\sqrt{2})|\mu|\}\in(0,\infty)$. By Mill's Inequality, for all $t\in[\gamma,\infty)$
\begin{align*}
  \int_{C_{t}}\mathrm{N}(s;0,1)\diff s&\leq\frac{2}{(2\pi)^{1/2}}\sigma(t-|\mu|)^{-1}\exp\{-(2\sigma^{2})^{-1}(t-|\mu|)^{2}\} \\
  &\leq\frac{2}{(2\pi)^{1/2}}\sigma(|\mu|+1)^{-1}\exp\{-(4\sigma^{2})^{-1}t^{2}\}.
\end{align*}
Using the previous information
\begin{equation*}
  \Pb(|\theta_{i}|>t)\leq\frac{2}{(2\pi)^{1/2}}\sigma(|\mu|+1)^{-1}\exp\{-(4\sigma^{2})^{-1}t^{2}\}
\end{equation*}
for all $t\in[\gamma,\infty)$ and $i\in[k]$.

\section{Proof of Lemma \ref{Repulsive.Constant.Property}.}\label{Proof.Repulsive.Constant.Property}
By the Change of Variables Theorem and Fubini's Theorem, it can be shown that for all $k\geq2$ $(k\in\N)$
\begin{equation*}
  c_{k}=\int_{\R^{1}_{k-1}}
  F_{k-1}(\boldsymbol{\theta}_{-1,1})
  \Bigg\{\prod_{i=2}^{k}\mathrm{N}(\theta_{i};0,1)\Bigg\}
  \prod_{2\leq r<s}^{k}\Bigg[1-\exp\Bigg\{-\frac{(\theta_{r}-\theta_{s})^{2}}{2\tau}\Bigg\}\Bigg]\diff\boldsymbol{\theta}_{-1,1}
\end{equation*}
where $\boldsymbol{\theta}_{-1,1}=(\theta_{i}:i\neq 1)\in\R^{1}_{k-1}$ and $F_{k-1}:\R^{1}_{k-1}\to(0,1)$ is given by
\begin{equation*}
  F_{k-1}(\boldsymbol{\theta}_{-1,1})=
  \int_{\R}\mathrm{N}(\theta_{1};0,1)\prod_{j=2}^{k}\Bigg[1-\exp\Bigg\{-\frac{(\theta_{1}-\theta_{j})^{2}}{2\tau}\Bigg\}\Bigg]\diff\theta_{1}.
\end{equation*}
Notice that $F_{k-1}\in C(\R^{1}_{k-1};(0,1))$ (as a consequence of Lebesgue's Dominated Convergence Theorem) and $F_{k-1}(\boldsymbol{\theta}_{-1,1})\to1$ as 
$||\boldsymbol{\theta}_{-1,1}||\to\infty$. By Jensen's Inequality, for all $\boldsymbol{\theta}_{-1,1}\in\R^{1}_{k-1}$
\begin{equation*}
  \log\{F_{k-1}(\boldsymbol{\theta}_{-1,1})\}
  \geq
  \sum_{j=2}^{k}\int_{\R}\mathrm{N}(\theta_{1};0,1)\log\Bigg[1-\exp\Bigg\{-\frac{(\theta_{1}-\theta_{j})^{2}}{2\tau}\Bigg\}\Bigg]\diff\theta_{1}.
\end{equation*}
Now
\begin{equation*}
  \Bigg|\int_{\R}\mathrm{N}(\theta_{1};0,1)\log\Bigg[1-\exp\Bigg\{-\frac{(\theta_{1}-\theta_{j})^{2}}{2\tau}\Bigg\}\Bigg]\diff\theta_{1}\Bigg|
  \leq
  -2\frac{\tau^{1/2}}{\pi^{1/2}}\int_{0}^{\infty}\log\{1-\exp(-\theta_{1}^{2})\}\diff\theta_{1}.
\end{equation*}
Using the substitution $\theta_{1}(z)=z^{1/2}:z\in(0,\infty)$ and then integrating by parts
\begin{equation*}
  \int_{0}^{\infty}\log\{1-\exp(-\theta_{1}^{2})\}\diff\theta_{1}=-\int_{0}^{\infty}\frac{z^{3/2-1}}{\exp(z)-1}\diff z=
  -\Gamma(3/2)\zeta(3/2)\in(-\infty,0)
\end{equation*}
where $\Gamma(\,\cdot\,)$ and $\zeta(\,\cdot\,)$ are the Gamma and Riemann Zeta functions, respectively. The previous information implies that
\begin{equation*}
  \Bigg|\int_{\R}\mathrm{N}(\theta_{1};0,1)\log\Bigg[1-\exp\Bigg\{-\frac{(\theta_{1}-\theta_{j})^{2}}{2\tau}\Bigg\}\Bigg]\diff\theta_{1}\Bigg|
  \leq2.6124\tau^{1/2}\in(0,\infty).
\end{equation*}
With this bound, defining $A_{2}=2.6124\tau^{1/2}$ and $A_{1}^{-1}=\exp(A_{2})$ the following holds: for all $\boldsymbol{\theta}_{-1,1}\in\R^{1}_{k-1}$
\begin{equation*}
  \log\{F_{k-1}(\boldsymbol{\theta}_{-1,1})\}\geq-(k-1)A_{2}
\end{equation*}
which implies
\begin{equation*}
  F_{k-1}(\boldsymbol{\theta}_{-1,1})\geq A_{1}^{-1}\exp(-A_{2}k).
\end{equation*}
To conclude the proof, notice that
\begin{equation*}
  c_{k-1}=\int_{\R^{1}_{k-1}}\Bigg\{\prod_{i=2}^{k}\mathrm{N}(\theta_{i};0,1)\Bigg\}
  \prod_{2\leq r<s}^{k}\Bigg[1-\exp\Bigg\{-\frac{(\theta_{r}-\theta_{s})^{2}}{2\tau}\Bigg\}\Bigg]\diff\boldsymbol{\theta}_{-1,1}.
\end{equation*}
Using the previous equation it follows that for all $k\geq2$ $(k\in\N)$
\begin{equation*}
  c_{k}\geq A_{1}^{-1}\exp(-A_{2}k)c_{k-1}>0,
\end{equation*}
the above being equivalent to
\begin{equation*}
  0<\frac{c_{k-1}}{c_{k}}\leq A_{1}\exp(A_{2}k).
\end{equation*}

\section{Proof of Proposition \ref{Posterior.Convergence.Rate}.}\label{Proof.Posterior.Convergence.Rate}
Following Theorem 3.1 in \cite{Scricciolo:2011} $p=2$ induce a (finite) Gaussian Mixture Model, $\lambda\sim\mathrm{IG}(a,b):a,b\in(0,\infty)$ satisfy $(i)$ and 
$\boldsymbol{\pi}_{k,1}\sim\mathrm{Dir}(k^{-1}\mathbf{1}_{k})$ satisfy $(iii)$. Condition $\mathrm{B3'}$ is equivalent to $(ii)$. However, $(iv)$ does not apply because the 
cluster-location parameters are not i.i.d. in our framework.

Along the proof of Theorem 3.1 we identified those steps that can be adapted by the assumption $\boldsymbol{\theta}_{k,1}\sim\NRep_{k,1}(\mu,\sigma^{2},\tau)$. It is
important to mention that Theorem 3.1 appeals to conditions (A.1), (A.2) and (A.3) in Theorem A.1 (Appendix of Scricciolo's paper) which is a powerful result given by 
\cite{GhosalVanDerVaart:2001}. We will check that (A.1) to (A.3) are satisfied:
\begin{itemize}
  \item [(A.1)] The proof is the same as the arguments presented at page 277 and the first paragraph in page 278. The reason for this is that it only depends on the 
  structure of the mixture, leaving aside the prior distributions for all the involved parameters.

  \item [(A.2)] What needs to be modified on the first inequality found on page 278 is the term $E(K)\Pi([-a_{n},a_{n}]^{c})$. This quantity is part of the chain of 
  inequalities
  \begin{equation*}
    \sum_{i=1}^{k_{n}}\rho(i)\sum_{j=1}^{i}\Pb(|\theta_{j}|>a_{n})=\sum_{i=1}^{k_{n}}i\rho(i)\Pi([-a_{n},a_{n}]^{c})\leq E(K)\Pi([-a_{n},a_{n}]^{c})\lesssim
    \exp\{-ca_{n}^{\vartheta}\}
  \end{equation*}
  under the conditions $(ii)$ and $(iv)$. In our case, $\rho(i)=\kappa(i)$ for $i\in\N$. By way of Lemma~\ref{Repulsive.Probability.Inequality}
  \begin{equation*}
    \sum_{j=1}^{i}\Pb(|\theta_{j}|>a_{n})\leq\frac{2i}{(2\pi)^{1/2}}\frac{c_{i-1}}{c_{i}}\sigma(|\mu|+1)^{-1}\exp\{-(4\sigma^{2})^{-1}a_{n}^{2}\}
  \end{equation*}
  under the convention that $c_{0}=1$ and $n\in\N$ is big enough. Thus,
  \begin{equation*}
    \sum_{i=1}^{k_{n}}\rho(i)\sum_{j=1}^{i}\Pb(|\theta_{j}|>a_{n})
    \leq
    \frac{2}{(2\pi)^{1/2}}\sigma(|\mu|+1)^{-1}\exp\{-(4\sigma^{2})^{-1}a_{n}^{2}\}\sum_{i=1}^{k_{n}}i\rho(i)\frac{c_{i-1}}{c_{i}}
  \end{equation*}
  and by Lemma~\ref{Repulsive.Constant.Property}
  \begin{equation*}
    \sum_{i=1}^{k_{n}}i\rho(i)\frac{c_{i-1}}{c_{i}}\leq A_{1}B_{1}\sum_{i=1}^{\infty}i\exp\{-(B_{2}-A_{2})i\}\in(0,\infty).
  \end{equation*}
  Finally, we obtain the following upper bound (in order), which is analogous to that obtain in \cite{Scricciolo:2011}:
  \begin{equation*}
    \sum_{i=1}^{k_{n}}\rho(i)\sum_{j=1}^{i}\Pb(|\theta_{j}|>a_{n})\lesssim\exp\{-(4\sigma^{2})^{-1}a_{n}^{2}\}.
  \end{equation*}

  \item [(A.3)] We only need to adapt the following inequality found on page 279, whose validity is deduced from $(iv)$:
  \begin{equation*}
    \Pb\{\boldsymbol{\theta}_{k_{0}}\in B(\boldsymbol{\theta}^{0}_{k_{0}};\varepsilon)\}
    =\Pi^{\otimes k_{0}}\{B(\boldsymbol{\theta}^{0}_{k_{0}};\varepsilon)\}
    \gtrsim
    \exp\{-d_{1}k_{0}\log(1/\varepsilon)\}
  \end{equation*}
  In our case, $\boldsymbol{\theta}_{k_{0}}=\boldsymbol{\theta}_{k_{0},1}$, $\boldsymbol{\theta}^{0}_{k_{0}}=\boldsymbol{\theta}^{0}_{k_{0},1}$ and 
  $B(\boldsymbol{\theta}^{0}_{k_{0}};\varepsilon)=D_{1}(\boldsymbol{\theta}^{0}_{k_{0},1},\varepsilon)$. At the end of the proof of Lemma~\ref{Repulsive.Probability.Property} 
  it is shown that for every $\delta=\varepsilon\in(0,\delta_{0}]$
  \begin{equation*}
    \Pb\{\boldsymbol{\theta}_{k_{0},1}\in D_{1}(\boldsymbol{\theta}^{0}_{k_{0},1},\varepsilon)\}
    \geq
    \frac{R_{0}}{c_{k_{0}}}
    \Bigg(\prod_{i=1}^{k_{0}}\frac{S^{0}_{i}}{2}\Bigg)
    \exp\{-k_{0}\log(1/\varepsilon)\}.
  \end{equation*}
  With this information, we obtain a lower bound (in order) analogous to that obtained in \cite{Scricciolo:2011}:
  \begin{equation*}
    \Pb\{\boldsymbol{\theta}_{k_{0}}\in B(\boldsymbol{\theta}^{0}_{k_{0}};\varepsilon)\}\gtrsim\exp\{-k_{0}\log(1/\varepsilon)\}.
  \end{equation*}
\end{itemize}

\end{appendices}


\end{document}